\documentclass[journal]{IEEEtran}

\usepackage{amsmath}
\usepackage{amsfonts,amssymb}
\usepackage{caption}
\usepackage{graphicx}
\usepackage{subfigure}
\usepackage{booktabs}
\usepackage[table]{xcolor}
\usepackage{multirow}
\usepackage{makecell}
\usepackage{threeparttable}
\usepackage{algorithm}
\usepackage{algpseudocode}  
\usepackage{amsmath}
\usepackage{soul} 
\usepackage{color, xcolor}

\newtheorem{remark}{Remark}
\everymath{\displaystyle}

\begin{document}
\title{Topology-Aware Resilient Routing Protocol for FANETs: An Adaptive Q-Learning Approach}
\author{Yanpeng~Cui,~\IEEEmembership{Student Member,~IEEE,}
		    Qixun~Zhang,~\IEEEmembership{Member,~IEEE,}
		    Zhiyong~Feng,~\IEEEmembership{Senior Member,~IEEE,}
		    Zhiqing~Wei,~\IEEEmembership{Member,~IEEE,}
		    Ce~Shi,
		    and Heng~Yang,~\IEEEmembership{Student Member,~IEEE}
		      
\thanks{This work was partly supported by National Key Research and Development Project (Grant No. 2020YFA0711300) and National Natural Science Foundation of China (NSFC) (Grant No. 62022020, 61790553). \emph{(Corresponding author: Zhiyong Feng.)}}

\thanks{Yanpeng Cui, Qixun Zhang, Zhiyong Feng, Zhiqing Wei, Ce Shi and Heng Yang are with the Key Laboratory of Universal Wireless Communications, Ministry of Education, School of Information and Communication Engineering, Beijing University of Posts and Telecommunications, Beijing 100876, China, e-mail: \{cuiyanpeng94; zhangqixun; fengzy; weizhiqing; sc; yangheng\}@bupt.edu.cn.}
}
\markboth{ }
{Cui \MakeLowercase{\textit{et al.}}: Topology-Aware Resilient Routing Protocol for Flying Ad Hoc Networks: An Adaptive Q-Learning Approach}
\maketitle
\begin{abstract}
Flying ad hoc networks (FANETs) plays a crucial role in numerous military and civil applications since it shortens mission duration and enhances coverage significantly compared with a single unmanned aerial vehicle (UAV).
Whereas, designing an energy-efficient FANETs routing protocol with a high packet delivery rate (PDR) and low delay is challenging owing to the dynamic topology changes.
In this paper, we propose a topology-aware resilient routing strategy based on adaptive Q-learning (TARRAQ) to accurately capture topology changes with low overhead and make routing decisions in a distributed and autonomous way. First, we analyze the dynamic behavior of UAVs nodes via queuing theory, then the closed-form solutions of neighbors' change rate (NCR) and neighbors' change inter-arrival time (NCIT) distribution are derived. Based on the real-time NCR and NCIT, a resilient sensing interval is determined by defining the expected sensing delay of network events. Besides, we also present an adaptive Q-learning approach that enables UAVs to make distributed, autonomous and adaptive routing decisions, where the above sensing interval ensures that the action space can be updated in time with low cost. The simulation results verify the accuracy of the topology dynamic analysis model, and also prove that our TARRAQ outperforms the Q-learning-based topology-aware routing, mobility prediction-based virtual routing and greedy perimeter stateless routing based on energy-efficient Hello in terms of 25.23\%, 20.24\% and 13.73\% lower overhead, 9.41\%, 14.77\% and 16.70\% higher PDR and 5.12\%, 15.65\% and 11.31\% lower energy consumption, respectively.

\end{abstract}


\section{Introduction}

\IEEEPARstart{T}{he} paradigm of the Internet of Things (IoT) enables highly integrated smart machines and devices to access and process information from the physical environment without human interaction \cite{Cui}. As the emerging device of IoT and the carrier of aerial technology, Unmanned Aerial Vehicle (UAV) plays a crucial role in military and civilian IoT applications such as coverage monitoring, emergency communication, remote sensing and disaster rescue (e.g. the recent Australian wildfire) \cite{WildFire}, thanks to its flexibility, small volume, low cost, concealment, rapid deployment \cite{Arafat}, etc. Compared with the application of a single UAV node, the coordination of multi-UAVs can shorten the mission duration and improve efficiency. Thus, the Flying Ad-hoc Networks (FANETs) emerges to improve cooperation and scalability. However, there are challenges in cooperative communication and routing in FANETs due to the intermittent connectivity and the energy consumption of UAVs \cite{B. Alzahrani}. Recently, extensive research has been conducted on how to improve the mobile ad-hoc network routing protocols and implement them in FANETs, but the latter's sparse heterogeneity structure, dynamic mobility, limited energy and serious fragmentation are the major issues in designing an efficient routing strategy ensuring a robust data exchange between UAVs \cite{O. S. Oubbati}.

There are multiple categories of FANETs routing protocols for different Quality of Service (QoS), such as topology-based and geographic-based. The topology-based protocols include hierarchical and flat patterns. The former are designed for large-scale networks due to the strong scalability, whereas the cost and complexity of clustering might outweigh the pros in sparse FANETs \cite{Arafat_Survey_Cluster}. For flat routing methods, the proactive ones have a huge overhead for routing table while the reactive ones create the intolerable delay, and most of their hybrid ignore the standard attributes of UAVs, namely the Global Positioning System (GPS), which provides a precious opportunity for geographic-based protocols \cite{J. Jiang}.
Based on location information, the geographic-based protocols find appropriate relays towards the destination by greedy forwarding, hence nodes no longer need to explore the state of the entire network \cite{A. Bujari}. Although FANETs with high mobility and dynamic mission will benefit from geographic-based protocols, there are still some issues to be addressed.

First, the routing decisions should be made in a distributed and autonomous way for each UAV since the global state of the entire network is unavailable when they are configured with geographic-based protocols. Q-learning, which is one of the easiest and most practiced reinforcement learning (RL) techniques, is envisioned as a promising solution \cite{Reinforcement_Survey}. By adjusting their behavior according to the reward function of environmental feedback, the intelligent agent can achieve optimal decisions without prior knowledge. This makes it possible for adaptive autonomous routing decisions. Whereas, the shortest route may be the one with poor link duration (LD) and unbalanced energy consumption, and the routing holes will appear once there is no suitable relay next \cite{L. Lin}. At the time of writing, considering multiple metrics can work to a certain extent at the cost of detour. However, since the learning-based protocol works via the continuous and periodic interaction between agents and the environment, the relays will deviate from the best decision if there is a false hole owing to the inaccurate environment information.

Due to the time-varying topology of FANETs, the learning performance will deteriorate since nodes often lag in capturing rapid topology changes. Thanks to the Hello protocol, nodes can announce their existence and exchange status periodically with each other \cite{Lakew}. The Q-learning-based routing protocols use Hello messages to interactive with the dynamic environment, namely maintain the neighbor status, where the Hello timer and expiration timer (ET) play a significant role. More concretely, the former represents the periodic sensing interval (SI) while the latter reflects the validity period of neighbors \cite{Jovel}. However, both of them are deterministic in the routing protocol standard proposed by the Internet Engineering Task Force \cite{OLSR}, which often has a poor performance in the rapidly changing FANETs due to the following reasons.

On one hand, the network performance is affected by SI. A smaller SI can ensure the accuracy of topology change detection whereas the additional overhead is wasted if the frequency of perception is higher than the neighbors' change rate (NCR) or the traffics' arrival rate (TAR). Conversely, although the overhead and energy consumption can be reduced by a larger SI, there will be serious packet loss if the SI lags behind the arrival of traffics or is too large to perceive the change of key links \cite{Oliveira}. On the other hand, the SI is determined by the network performance demands. For example, UAVs are sensitive to delay when acting as an aerial relay \cite{Xiao}, and they will have more stringent requirements for overhead (energy efficiency) when performing coverage monitoring tasks \cite{Lin}. Besides, the stability and reliability of the links are more critical when UAVs are utilized to search and rescue \cite{Chen}. Thus, resilient SI should be designed to enable the network to meet various performance demands.

In this paper, we conduct thorough research on the above issues. We begin with constructing a queuing service framework for analyzing link establishment and disconnection between UAVs, then the closed-form expressions of LD, NCR and distribution of neighbors' change inter-arrival time (NCIT) are derived. Then we propose a protocol called Topology-Aware Resilient Routing based on Adaptive Q-learning (TARRAQ) that enables UAVs to make distributed, autonomous and adaptive routing decisions, which contains three phases: 1) neighbor discovery; 2) neighbor maintenance; 3) relay selection. During the first two phases, the residual LD predicted by Kalman Filter (KF) is working as an expiration timer for available neighbors, and the NCR and NCIT are used to calculate the resilient SI on demand, namely all UAVs can dynamically adjust their sensing scheme based on dynamic behavior and performance requirements, thus the accurate information of neighbors can be obtained accordingly with tolerable overhead. When it comes to the last phase, the neighbor information obtained previously is regarded as the finite states where agents attempt to take actions, and an adaptive Q-learning approach is presented to ensure the distributed and autonomous routing decisions. For clarity, we summarize our contribution as follows.

\begin{itemize}

	\item We analyze the dynamic topology changes of FANETs via queuing theory, which reveals the mapping relationship between LD and service duration, NCR and customer change rate, NCIT and distribution of customer change inter-arrival time, respectively. And the closed-form expressions of LD, NCR and NCIT are derived to describe the mobility behavior accurately. To the best of our knowledge, this is the first research to accurately characterize the topology changes of FANETs;

	\item We propose a novel resilient perception strategy based on the first contribution. The NCR with dynamic exponential weighted moving average (DEWMA) process is used to calculate the topology changes. By defining the expected sensing delay of the network events, the resilient SI is determined based on the real-time NCR and NCIT to satisfy the network's dynamic demands. Besides, with the KF method introduced, the residual LD is predicted to determine the validity period of neighbors;
	
	\item We present an adaptive Q-learning approach, which enables UAVs to make distributed and autonomous routing decisions. In dynamic FANETs, the action space is updated in time with the low cost through the second contribution. The reward function is designed based on the link quality, residual energy and distance of neighbors to find a stable path with a lower loss rate, fewer hops and energy consumption. Besides, the action selection, learning rate and discount factor are adjusted according to the residual LD, which achieves adaptive learning from the variable network environment.
\end{itemize}

\begin{table}
	\centering
	\caption{Main Abbreviations Used in This Paper.}
	\renewcommand{\arraystretch}{1}
	\begin{tabular}{m{2cm}<{\centering} m{6cm}<{\centering}}
		\toprule[1pt]\toprule[0.5pt]
		Abbreviations & Explanation \\
		\midrule[0.5pt]
		CQS & Cyclic Queuing System \\
		DEWMA & Dynamic Exponential Weighted Moving Average \\
		EE-Hello & Energy-Efficient Hello \\
		EIT & Events' Inter-arrival Time \\
		ET & Expiration Timer \\
		E2ED & End-to-End Delay \\
		FANETs & Flying Ad-hoc Networks \\
		KF & Kalman Filter \\
		LD & Link Duration \\
		MPVR & Mobility Prediction-based Virtual Routing \\
		NCIT & Neighbors' Changes Inter-arrival Time \\
		NCR & Neighbors' Change Rate \\
		NIT & Neighbors' Inter-arrival Time \\
		PDR & Packet Delivery Ratio \\
		PPP & Poisson Point Process \\
		QTAR & Q-Learning-based Topology-Aware Routing \\
		RWP & Random Waypoint \\
		SI & Sensing interval \\
		TAR & Traffics' Arrival Rate \\
		TARRAQ & Topology-Aware Resilient Routing Based on Adaptive Q-learning\\
		\bottomrule[0.5pt]\bottomrule[1pt]
	\end{tabular}
	\label{Abbreviations}
\end{table}

The remainder of this article is organized as follows. In section \uppercase\expandafter{\romannumeral2}, the related works are classified and analyzed. Section \uppercase\expandafter{\romannumeral3} introduces the motivation scenario, network model and Q-learning framework for FANETs routing. The queuing service system model and the TARRAQ protocol is proposed in sections \uppercase\expandafter{\romannumeral4} and \uppercase\expandafter{\romannumeral5}, respectively. Extensive simulations are performed in section \uppercase\expandafter{\romannumeral6} to verify the accuracy of the proposed model, and the performance of TARRAQ is discussed. Finally, we conclude this article in section \uppercase\expandafter{\romannumeral7}. The key and unique abbreviations are summarized in \textbf{Table \ref{Abbreviations}}.

\section{Related Works}

\subsection{Related Works on Perception Strategy}

\subsubsection{Mobility-Based Perception Strategy}

The mobility-based perception strategy tends to determine the optimal SI by studying the influence of dynamic behavior on topology changes. It includes two opposing camps: formula-based and intelligence-based perception strategy.

Qualitative and intuitive analysis of the correlation between network parameters (e.g. node's mobility, density and SI) and perception performance (e.g. accuracy and overhead) is a typical feature of the formula-based perception strategy. 
As one of the earliest typical perception strategies, an adaptive Hello protocol was proposed by Giruka \cite{Giruka}. It allows nodes to broadcast Hello messages once they move a certain distance. However, it only studies the influence of mobility rather than other factors. Besides, the linear relationship between the optimal SI and mobility is inaccurate. Park \textit{et al.} \cite{Park} defined the mobility factor by dividing transmission range by speed, and obtained the best mapping coefficient through experiments with the goal of maximizing throughput. 
Mahmud \textit{et al.} \cite{Mahmud} defined the best SI based on density, airspace size, node's speed and transmission range, and proposed the energy-efficient Hello (EE-Hello) protocol. Although the overhead efficiency was improved, the end-to-end delay (E2ED) has not been reduced.

Considering that both the interaction between parameters and their influence on network performance are difficult to understand and explain, the intelligence-based perception strategy prefers to construct a correlation between key parameters and the best SI using intelligent tools. Shah \textit{et al.} \cite{Shah} proposed a scheme to determine SI based on an artificial neural network (ANN), where the input is the node's transmitting power and mobility while the output is the Hello interval. The network throughput and packet delivery ratio (PDR) are improved at the expense of network delay. However, their ANN architecture with only 10 nodes in a single hidden layer can complete the learning process in 20 epochs, which means the relationship between input and output is not complicated. In fact, it can be determined by analyzing the topology changes, and we will elaborate on it in section \uppercase\expandafter{\romannumeral4}. 
An energy-efficient routing approach through an adaptive neuro-fuzzy inference system was proposed in \cite{Bisen}. The nodes with more residual energy and higher mobility are set with a smaller SI, which provides excellent performance in high mobility and dense FANETs. However, the influence of network density, traffic flow and node transmission range on the optimal SI has been ignored. According to the factors considered in the EE-Hello scheme \cite{Mahmud}, a new energy-efficient ad-hoc on-demand distance vector based on the response surface methodology was proposed by Mohamed \cite{Daas}. 
However, its adaptability needs to be improved once the actual scenario and the model are quite different.

\subsubsection{Event-Based Perception Strategy}

Unlike the mobility-based perception strategy, the event-based one tends to proceed from the original intention of the perception \cite{Liu}, that is, to provide accurate network information for routing decisions \cite{Ernst}. Nelson \textit{et al.} \cite{Hernandez-Cons} dynamically adjust SI by monitoring the NCR and packet loss rate of nodes. Specifically, the SI will decrease if the NCR exceeds the threshold or packet loss occurs, and vice versa. However, the optimal threshold in multiple scenarios is difficult to determine, and the influence of traffics has not been discussed. Considering the situation where traffic inter-arrival time follows the Poisson distribution, an event-driven adaptive Hello protocol was proposed in \cite{Han}. 
But it is counterproductive in low-speed sparse networks since the links established within 10 s are probably available even if new traffic is frequently generated.

\subsection{Related Works on Routing Protocols}

\subsubsection{Topology-based Routing Protocols}
In this category, an appropriate routing path from the source to the destination is required before data transmission begins. According to how the routing is maintained, it can be further categorized as proactive, reactive and hierarchical schemes.

In the proactive ones, UAVs preserve the latest route information in the network, regardless of whether they have data packets to send or not. The predictive Optimized link state routing protocol (OLSR), which combines the advantages of proactive routing and the GPS information available on board, was introduced in \cite{P-OLSR}. The expected transmission count metric is defined based on link quality and moving speed, thus the routing follows the topology change without interruptions. Note that for high-mobility FANETs, it is not sufficient to only exchange the control packets. The UAVs should have the capability to predict the status of neighbors to meet the critical reliability requirement. In \cite{OLSR_PMD}, an enhanced OLSR based on mobility and delay prediction was proposed. It employed the KF to choose stable neighbor nodes as relays and introduced a cross-layer queuing delay prediction model to achieve traffic load balance and reduce the E2ED. Whereas the proactive protocols are only suitable for real-time applications in coordinated formation architectures with low mobility degree since periodic routing table update require a large number of control packets in highly dynamic FANETs, which results in poor overhead and inefficiency.

The reactive routings are introduced to address the cons of proactive ones since it triggers the route discovery process only when the UAV has packets to send. Oubbati \textit{et al.} \cite{ECaD} proposed an energy-efficient connectivity-award date (ECaD) routing protocol, which is envisioned as a promising reactive solution. The key idea behind ECaD is to exploit UAVs to efficiently anticipate path failures before their occurrence and select alternative next hops. It also considered the link connectivity expiration time and residual energy of UAVs to ensure communication stability. Hong \textit{et al.} \cite{Hong} proposed a topology-aware routing scheme for UAV swarm network. Based on the smooth processing of the flight controller's output, it calculates LD and sets half of its minimum value as SI. However, the topology changes should be associated with link changes or traffic arrival intervals rather than LD since the latter is irrelevant to the network density \cite{Jian}. Besides, although it has a better perceived latency, the overhead is higher than that of the conventional protocols.

Hierarchical routing protocols are more advantageous in terms of flexibility and scalability for large-scale FANETs. In \cite{Arafat}, the authors proposed an energy-efficient swarm-intelligence-based clustering protocol where the particle fitness function is exploited for geographic location, residual energy and inter-cluster distance. Whereas in our view, it can be further improved by incorporating a mobility prediction mechanism to perform efficient routing in high-mobility FANETs, and it would be a promising protocol if the neighbor discovery process is further considered.

\subsubsection{Geographic-Based Routing Protocols}

Compared with the topology-based routing protocols, the geography-based ones consume less bandwidth and lower overhead. Jiang \textit{et al.} introduced a mobility prediction-based virtual routing (MPVR) strategy for FANETs \cite{MPVR}. The deviation degree, hop count and LD are considered when selecting the next hop. However, the residual energy and the local minimum problem is ignored. Besides, the specific method of obtaining neighbor information, which is exactly the prerequisite of MPVR, was not introduced. In \cite{QGEO}, a novel protocol named Q-learning-based geographic routing (QGeo) was proposed. Both transmission distance and link status were considered in the reward function, thus the reliable transmission was achieved with low overhead. Nevertheless, it is not sufficient to find relays without considering the link status. Note that QGeo utilizes a fixed SI, it may suffer from serious performance loss in applications that require high-accuracy link information, and leading to limited adaptability for high-mobility FANETs. Liu \textit{et al.} proposed a Q-learning-based multi-objective optimization routing (QMR) protocol \cite{QMR}, where both the delay and the energy consumption are considered. The QMR adjusted the Q-learning parameters and estimate neighbor relations in real time. Whereas, the QMR believes that close neighbors have more rewards than distant ones. In fact, the links of remote relays are not all inclined to break, which is related to the relative motion. Conversely, near relays are not typically optimal due to the sharp increase of hops and delays. The Q-learning-based topology-aware routing (QTAR) \cite{QTAR}, which improved the routing decision based on the two-hop neighbors' information and adaptive Q-learning technique, is envisioned as a promising solution. However, the authors ignored the value of LD, which may play a crucial role in the reward function to obtain stable links. In addition, the calculation methods for LD and SI inherited from \cite{Hong} are designed for swarm networks rather than FANETs, which may not be applicable owing to the rapid topology change.

\section{Preliminaries}

\subsection{Motivation Scenario}

In this study, we consider a FANETs scenario consisting of multiple UAVs and a BS for wild disaster monitoring. Aiming at detecting the region of interest (RoI) and transmitting data to the BS, the UAVs can make free movement and change their direction and altitude independently to achieve a trade-off between sensing range and accuracy for RoI. Thus their movement can be considered as 3D random waypoint (RWP) model owing to the random RoI.
With the BS as the destination, the UAV traffic source appears randomly as the concerned event occurred, and the remaining UAVs are considered as the relays to forward the traffic packets to BS, whose location is known by all UAVs before they takeoff. 

\begin{figure}
	\centering
	{\includegraphics[width=8.8cm]{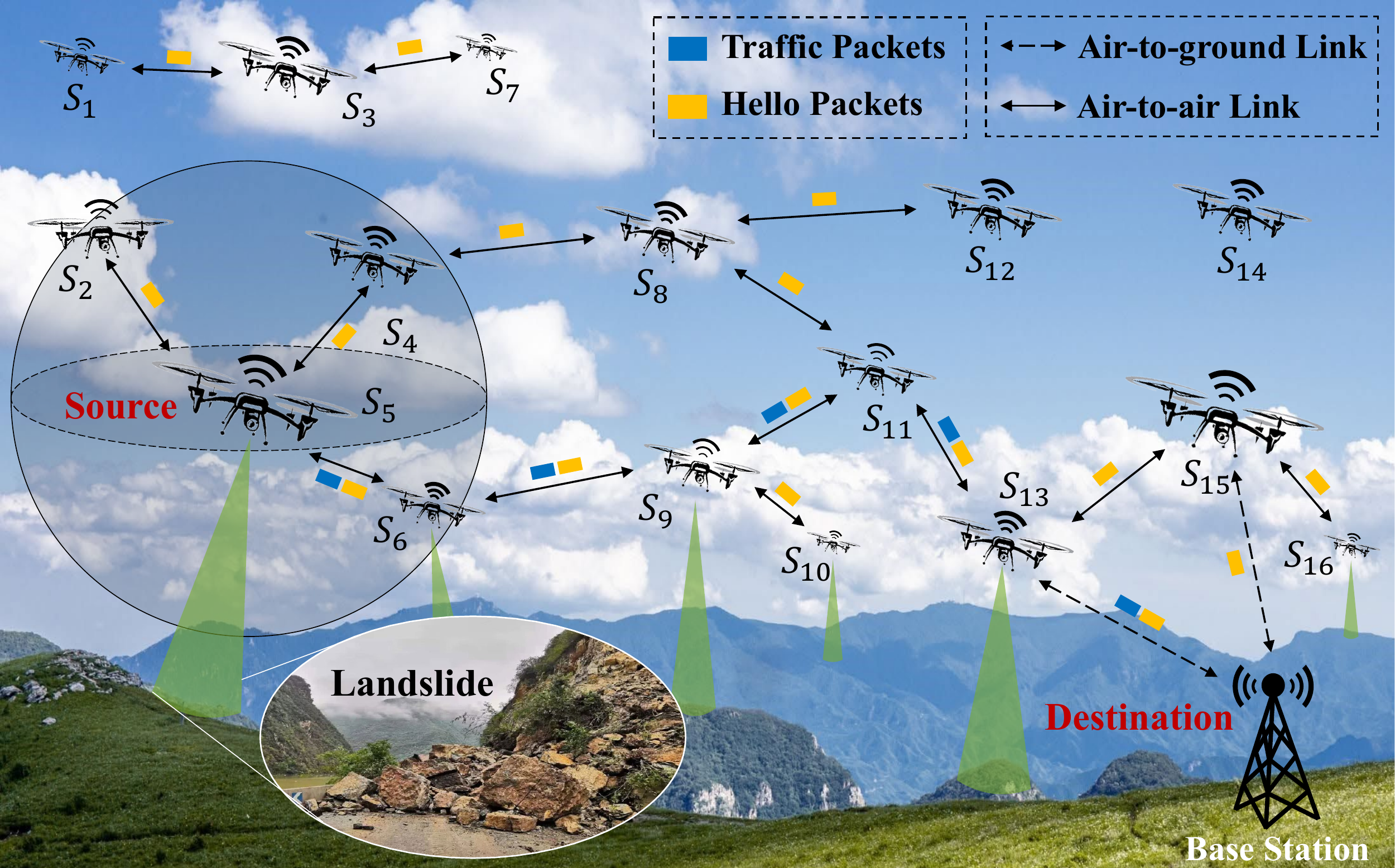}}
	\caption{Typical FANETs scenario for wild disaster monitoring.} \label{Scenario}
\end{figure}

As shown in \textbf{Fig. \ref{Scenario}}, the state of UAVs will be exchanged via Hello messages. When the traffic packets are generated in $S_2$, the location and other state information of next hop are utilized to forward packets, and finally construct a route path to a destination (e.g. $S_5$-$S_6$-$S_9$-$S_{11}$-$S_{13}$-BS). The routing errors will be sent back to find a new route once the forward link is broken. Besides, there may be a serious network fragmentation problem, namely a single UAV or sub-net of UAV groups (SUGs) that are disconnected from BS (e.g. $S_{14}$ and $S_1$-$S_3$-$S_7$). The traffic packets generated in SUGs will be forwarded to the edge of SUGs and stored. They will be forwarded to BS immediately once the path is available, or be dropped once the max cache time arrives. The aim of our research is to find the optimal route with low latency, high delivery rate and low overhead in the rapidly changing FANETs.

\subsection{Network Model}
Assume that all UAVs are aware of the location information using GPS within tolerable error, which means that the $i$ th UAV $S_i$ has a vector of 3D position $\textbf{p}_i\in\mathbb{R}^{3\times1}$. 
Considering that the UAV communication channels are mainly
dominated by the line of sight (LoS) links \cite{Z. Xiao}, we assume that the path loss model for link between $S_i$ and $S_j$ follows a proportion of $d_{i,j}^{-\alpha}$, where $\alpha$ is the path loss exponent and $d_{i,j}$ is the Euclidean distance calculated by
\begin{equation}\label{Euclidean Dis}
	d_{i,j}=\sqrt{\sum\nolimits_{k=1}^3\left|\,p_{i,k}-p_{j,k}\right|^2}.
\end{equation}
The Signal to Interference plus Noise Ratio (SINR) from $S_i$ to $S_j$ is denoted by
\begin{equation}\label{SINR}
	\gamma_{i,j}=\frac{P_{i,j}h_{i,j}d_{i,j}^{-\alpha}}{N_0+U_I},
\end{equation}
where $P_{i,j}$ is the transmission power from $S_i$ to $S_j$, $N_0$ is the ambient noise, and the power gain of small-scale fading channel $h_{i,j}$ is exponentially distributed with a unit mean. According to \cite{Srinivasa}, the interference $U_I$ can be calculated by 
\begin{equation}\label{interference}
	U_I=\frac{3N_e(D_m^{3-\alpha}-\epsilon^{3-\alpha})}{2D_m^3(3-\alpha)},
\end{equation}
where $\epsilon$ denotes the minimum distance between UAVs to avoid a collision, $D_m=\sqrt{3}L$ is the farthest distance between any two UAVs, and $N_e=4\sqrt{3}\pi N$ denotes the equivalent number of UAVs in a sphere with a radius of $D_m$. Thus, the successful transmission probability $P\left(\gamma_{i,j}\ge\gamma_{th}\right)$ is given by
\begin{equation}\label{eq2}
	\begin{split}
		Prob\left( {{\gamma _{i,j}}\ge\gamma_{th}} \right) &= Prob\left( {{h_{i,j}} \geqslant \frac{{\gamma_{th} {d_{i,j}}^{\alpha} \left({N_0}+{U_I}\right)}}{{{P_{i,j}}}}} \right) \\
		& = \exp \left( { - \frac{{\gamma_{th} {d_{i,j}}^\alpha \left({N_0}+{U_I}\right)}}{{{P_{i,j}}}}} \right),\\
	\end{split}
\end{equation}
where $\gamma_{th}$ is the SINR threshold. $Prob\left(\gamma_{i,j}\ge\gamma_{th}\right)\ge \phi$ must be satisfied to ensure the link quality, where $\phi$ is the constraint on SINR probability. Thus, the effective transmission range $R$ is defined as
\begin{equation}\label{R}
	R = {\left( { - \frac{{{P_{i,j}}\ln \left( \phi \right)}}{{\gamma_{th} ({N_0} + {U_I})}}} \right)^{\frac{1}{\alpha }}}.
\end{equation}

During the transition process of $l$-bit packet from $S_i$ to $S_j$, the energy consumption is calculated as
\begin{equation}\label{Energy Consumption}
	\left\{
	\begin{aligned}
		&E_{i,j}^r=l\times E_{elec}\\
		&E_{i,j}^t=l\times E_{elec}+l\times d_{i,j}^\alpha\times E_{fs}
	\end{aligned}\ ,
	\right.
\end{equation}
where $E_{i,j}^r$ and $E_{i,j}^t$ denote the energy consumed by receiving and sending $l$ bit data, respectively. $E_{elec}$ and $E_{fs}$ denote the coefficients of transmitter (receiver) and power amplifier \cite{Energy_Consumption}.

\subsection{Q-learning Framework for FANETs Routing}

Note that the purpose of routing is to determine an appropriate path that can transfer packets from the source to the destination, which consists of many single-hop decisions. It involves the probability of selecting the best relay from finite neighbors despite the lack of information about the entire network, which is exactly the core idea of Q-learning, namely finding the best action-selection strategy for finite Markov decision process even without the prior knowledge about the effect of actions on the environment.

Therefore, the routing process can be modeled by the Markov decision process and solved by the Q-learning algorithm as follows. By denoting the packet as an agent, the entire network can be regarded as the environment while the state's space is composed of all UAVs for forwarding packets. The selection of next hop is treated as an action, thus the action space represents the available neighbors. Accordingly, the state transition is equivalent to the packet forwarding decision among UAVs. 

In Q-learning, agents will get the reward $R(s_t,a_t)$ when taking action $a_t$ from state $s_t$, and the Q-table is updated by 
\begin{equation}\label{Q}
	\begin{aligned}
		&Q_{t+1}(s_t,a_t)\leftarrow (1-\alpha)Q_t(s_t,a_t)\\
		&+\alpha\left[R(s_t,a_t)+\gamma \max\limits_{a^{\prime}\in A_{t+1}} Q_t(s_{t+1},a^{\prime})-Q_t(s_t,a_t)\right],
	\end{aligned}
\end{equation}
where $Q_t(s_t,a_t)$ is the Q value when taking action $a_t$ from state $s_t$, $Q_{t+1}$ is the $Q$ value at time $t+1$, and $\alpha\in\left(0,1\right)$ is the learning rate while $\gamma\in\left(0,1\right)$ denotes the discount factor. Once the action $a_t$ is performed, the agent's state will change from $s_t$ to $s_{t+1}$. Note that $\max\limits_{a^{\prime}\in A_{t+1}} Q_t(s_{t+1},a^{\prime})$ is the maximum Q value of all possible action $a^{\prime}\in A_{t+1}$, where $A_{t+1}$ is the action space in state $s_{t+1}$. Aiming at different QoS metrics, the reward $R(s_t,a_t)$ can be defined and calculated through the information embedded in Hello messages.

\begin{figure}
	\centering
	{\includegraphics[width=8cm]{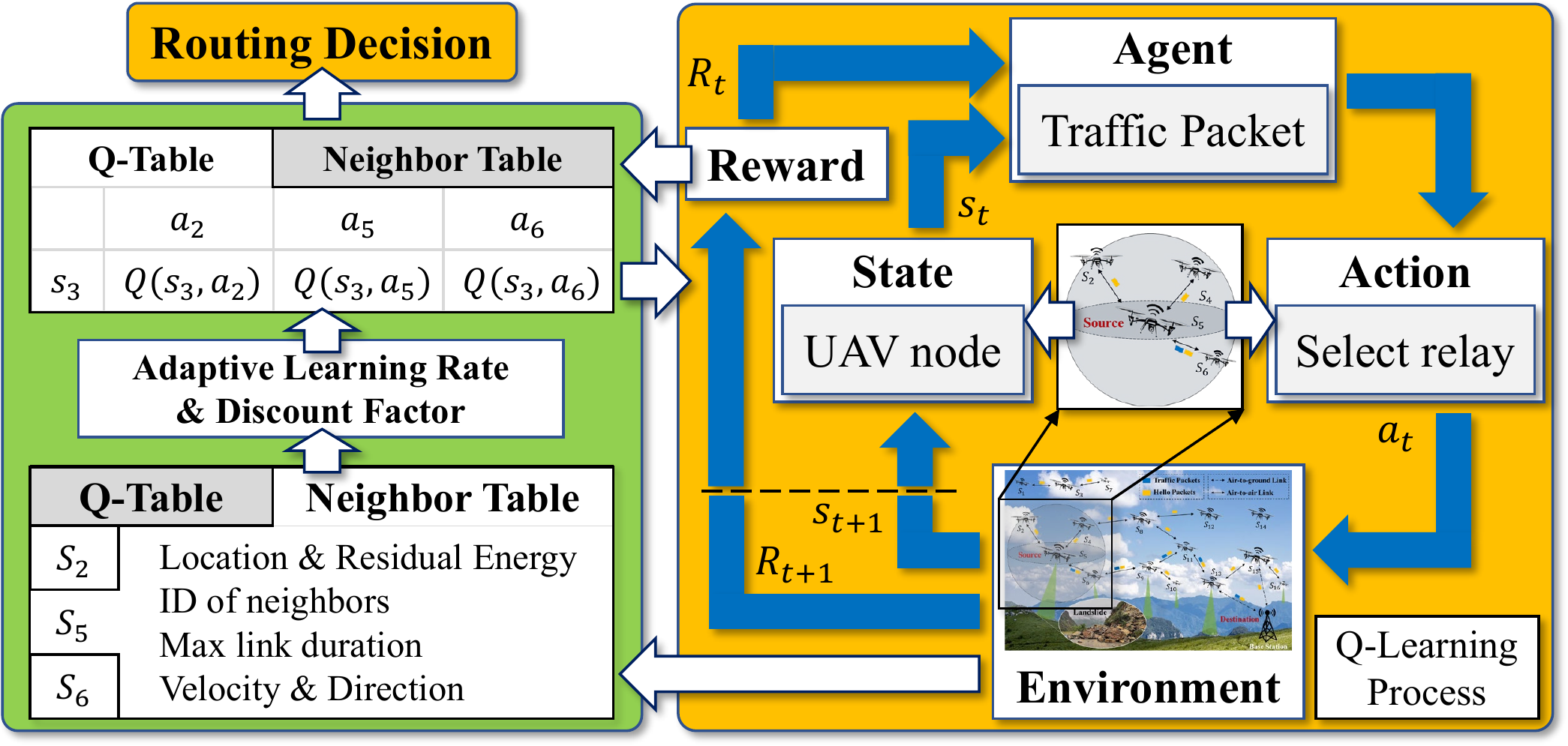}}
	\caption{Q-learning Framework for FANETs Routing.} \label{Framework}
\end{figure}

All UAVs will permanently maintain the neighbor table and Q-table by exchanging status information with neighbors regularly. The former is used to update the action space and calculate the reward, while the latter is used to determine the best action. As shown in \textbf{Fig. \ref{Framework}}, once the traffic packet is generated or received at the UAV node (e.g. UAV $S_5$), an agent will take action from the action space, that is, $S_5$ selects a relay from neighbors $S_2$, $S_4$ and $S_6$. Then the reward can be calculated based on QoS demands. Accordingly, the Q-table will be updated constantly via (\ref{Q}) as the agent learns from the environment. The best relay namely the optimal action can be determined via iterative Q-table, which means that a hop-by-hop traffic forwarding will be performed from $S_5$ to destination.

Although there are other advanced RL techniques that can be exploited to solve the routing problem, e.g., Deep Q Network (DQN) \cite{DQN} and Deep Deterministic Policy Gradient (DDPG) \cite{DDPG}, the Q-learning based scheme is envisioned as the best option in our motivation scenario. Note that the above process is performed by each UAV in a distributed manner since the global view of the network is impossible, which means each node only stores and retrieves a certain row of the Q-table. The slow convergence caused by an intractably large Q-table is no longer an issue for TARRAQ. The action space of each state will be very small even in a dense network due to the limited number of neighbors.

\section{Topology Dynamics Analysis Based on Queuing Theory}

The dynamic topology adaptation of the routing protocol is mainly reflected in the timely detection of network fragmentation and adjustment of routing decisions when topology changes, which is related to the mobility model, node density and transmission range. Thus, as the primary task of implementing a topology-aware routing protocol for FANETs, it is critical to study the characteristics of mobility behavior, such as the rate and the inter-arrival time distribution of neighbor's arrival, departure and change. 

When the establishment, maintenance and disconnection of links are regarded as queuing service process, there is obviously a one-to-one correspondence between the behavior of UAVs and customers. UAV plays a dual role during the queuing service process: a server when analyzing its topology changes and a customer when considering the impact of its behavior on the topology changes of other UAVs. The UAV's mobility model determines how customers enter and leave the service system. Besides, the UAV's speed range, communication range and network density determine the service duration, the rate and inter-arrival time distribution of customer's change. It provides a precious opportunity for applying the queuing model to analyze mobility behavior, and thus the existing queuing theory theorems can be directly used to obtain rigorous conclusions without complex derivation once the queuing model is determined. In this section, based on the queuing theory, the influence of mobility features on NCR is revealed, and the distribution of NCIT is deduced, which provides a theoretical basis for designing the resilient perception strategy and Q-learning routing protocol.
\begin{remark}
	To clarify the relationship between the CQS and the topology changes event, their connections are summarized in \textbf{Table \ref{Connections}}, which will be used alternately below unless otherwise specified.
\end{remark}

\subsection{Queuing Model} 

\begin{figure}
	\centering
	\subfigure[]{\label{a}\includegraphics[width=3.4cm]{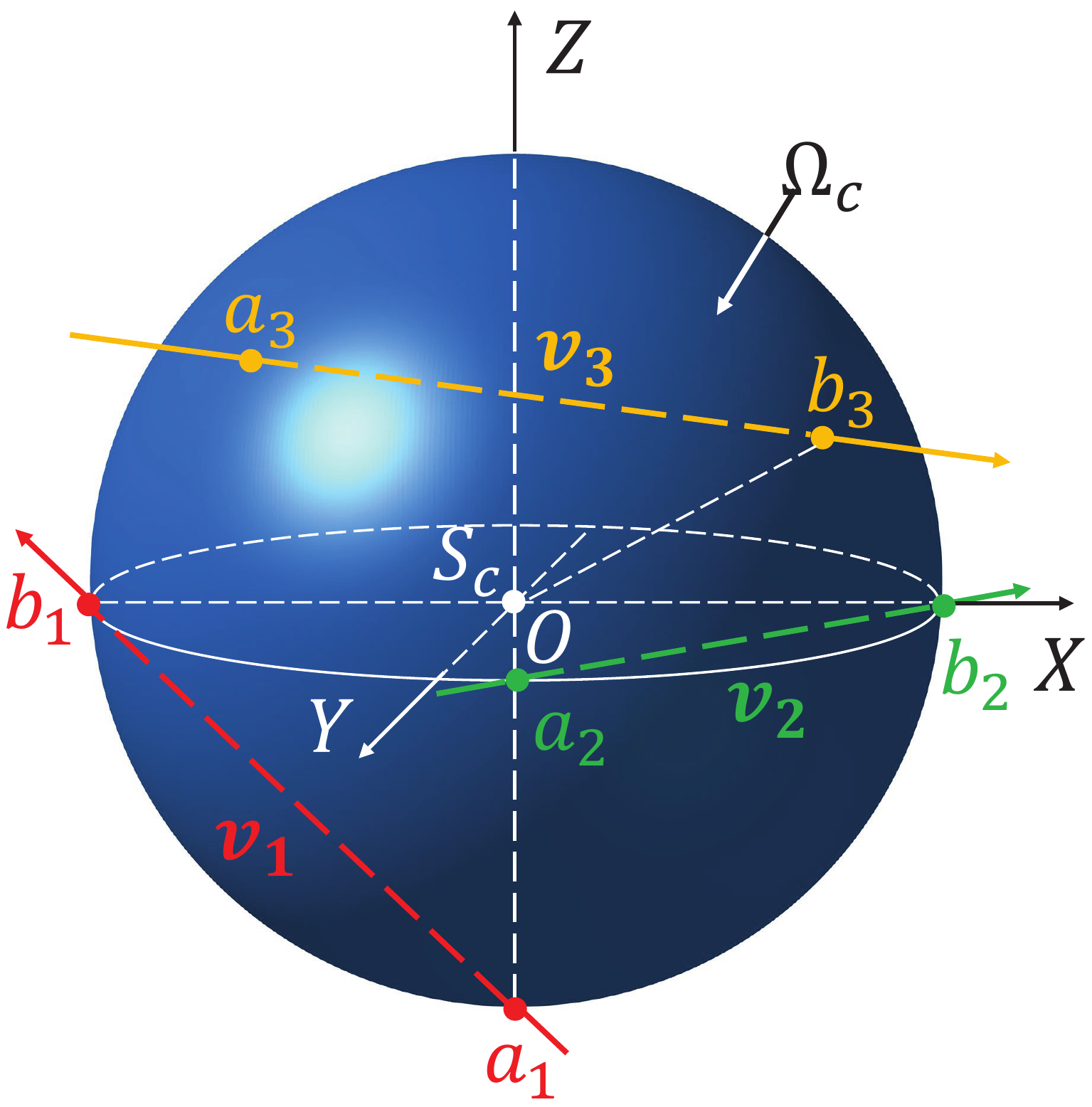}}
	\subfigure[]{\label{b}\includegraphics[width=2.2cm]{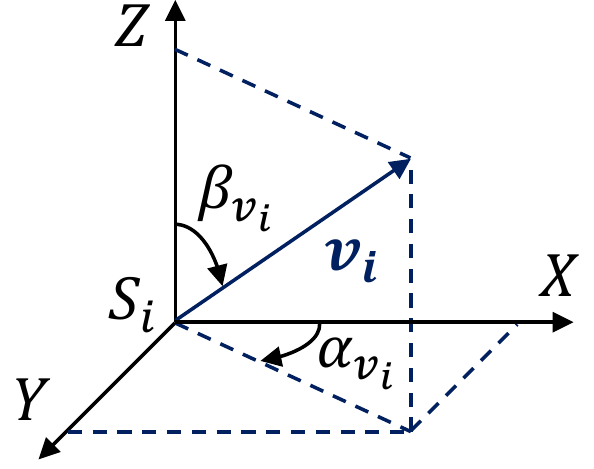}}\quad
	\subfigure[]{\label{c}\includegraphics[width=2.2cm]{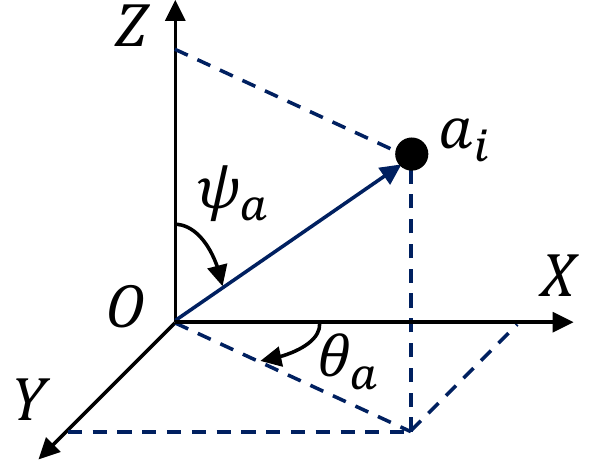}}
	\caption{(a) The communication range $\Omega_c$ of $S_c$. Three nodes enter $\Omega_c$ from $a_1$, $a_2$ and $a_3$, and leaving it from $b_1$, $b_2$ and $b_3$ with relative velocity $\boldsymbol{v_1}$, $\boldsymbol{v_2}$ and $\boldsymbol{v_3}$. (b) The moving direction of UAV $S_i$. (c) The azimuth and elevation of the entrance location $a_i$.}
	\label{Sphere}
\end{figure}

The communication links between UAVs will be established or disconnected frequently due to the topology changes. As shown in \textbf{Fig. \ref{Sphere}(a)}, taking node $S_c$ as an example, the process that other nodes entering its communication range $\Omega_c$, maintaining the link with $S_c$ and leaving $\Omega_c$, is regarded as a queuing service process. All nodes are divided into two categories, namely the nodes inside and outside $\Omega_c$, which are denoted as collections $M_1$ and $M_2$, respectively. Since UAV moves based on the 3D RWP mobility model, the nodes in $M_1$ will join $M_2$ immediately once their links established with $S_c$ are broken, namely, they will be regarded as the customer source once the service is over. Thus, the above process is essentially a cyclic queuing system (CQS).

\subsubsection{Queuing Rules}

The nodes can establish links with $S_c$ immediately once they enter $\Omega_c$, which means that customers can be served instantly without waiting, hence the number of servers and system capacity can be regarded as infinite.

\subsubsection{System Service Duration}

Assuming that the CQS is nonempty, namely there is at least one UAV within $M_1$, the service duration of CQS is equivalent to LD. Specifically, let's consider the case that there are two nodes $S_c$ and $S_o$ in 3-D space, where $S_o$ is an ordinary node that will be, being or has been served by the central node $S_c$. The velocity of $S_c$ and $S_o$ are defined as $\boldsymbol{v_c}=v_c\boldsymbol{k}$ and $\boldsymbol{v_o}=v_osin\beta_{\boldsymbol{v_o}}cos\alpha_{\boldsymbol{v_o}}\boldsymbol{i}+v_osin\beta_{\boldsymbol{v_o}}sin\alpha_{\boldsymbol{v_o}}\boldsymbol{j}+v_ocos\beta_{\boldsymbol{v_o}}\boldsymbol{k}$, and their relative velocity $\boldsymbol{v}$ and its magnitude $v$ can be given by
\begin{equation}\label{velocity}
	\begin{aligned}
		\boldsymbol{v} = \boldsymbol{v_o}-\boldsymbol{v_c}
		= &\ v_osin\beta_{\boldsymbol{v_o}}cos\alpha_{\boldsymbol{v_o}}\boldsymbol{i}
		+ v_osin\beta_{\boldsymbol{v_o}}sin\alpha_{\boldsymbol{v_o}}\boldsymbol{j}
		\\& + \left(v_ocos\beta_{\boldsymbol{v_o}}-v_c\right)\boldsymbol{k},
	\end{aligned}
\end{equation}
\begin{equation}\label{speed}
	v=|\boldsymbol{v}|=\sqrt{v_o^2+v_c^2-2v_cv_ocos\beta_{\boldsymbol{v_o}}\ }.
\end{equation}

As shown in \textbf{Fig. \ref{Sphere}(b)}, $\alpha_{\boldsymbol{v_k}}$ is regarded as the angle between the $X$-axis and the horizontal component of $\boldsymbol{v_k}$, while $\beta_{\boldsymbol{v_k}}$ is regarded as the angle between $\boldsymbol{v_k}$ and the $Z$-axis. The analysis process will be simplified significantly based on the relative motion of $S_c$ and $S_o$, that is, $S_c$ is regarded as a fixed node when $S_o$ enters $\Omega_c$ with relative velocity $\boldsymbol{v}$, whose direction can be calculated as $\alpha_{\boldsymbol{v}}=\alpha_{\boldsymbol{v_o}}$ and
\begin{equation}\label{beta_v}
	\beta_{\boldsymbol{v}} = arccos\frac{v_ocos\beta_{\boldsymbol{v_o}}-v_c}{\sqrt{v_o^2+v_c^2-2v_cv_ocos\beta_{\boldsymbol{v_o}}}}.
\end{equation}

Assuming that $S_o$ enters $\Omega_c$ from $a_3$ and leaves $\Omega_c$ from $b_3$ without changing the velocity, as shown in \textbf{Fig. \ref{Sphere}(a)}, we have $\boldsymbol{oa_3}=R(\cos\theta_{a_3}\sin\psi_{a_3}\cdot\boldsymbol{i}+\sin\theta_{a_3}\sin\psi_{a_3}\cdot\boldsymbol{j}+\cos\psi_{a_3}\cdot\boldsymbol{k})$, $\boldsymbol{a_3b}=|\boldsymbol{a_3b}|(\cos\alpha_{\boldsymbol{v}}\sin\beta_{\boldsymbol{v}}\cdot\boldsymbol{i}+\sin\alpha_{\boldsymbol{v}}\sin\beta_{\boldsymbol{v}}\cdot\boldsymbol{j}+\cos\beta_{\boldsymbol{v}}\cdot\boldsymbol{k})$, where $\theta_{a_3}$ and $\psi_{a_3}$ are the angles shown in \textbf{Fig. \ref{Sphere}(c)}. Since
$|\boldsymbol{ob}|=R=|\boldsymbol{oa_3}+\boldsymbol{a_3b}|$, we have $d_{link}=|\boldsymbol{a_3b}|=2R\left|\sin\psi_{a_3}\sin\beta_{\boldsymbol{v}}\cos(\theta_{a_3}-\alpha_{\boldsymbol{v}})+\cos\psi_{a_3}\cos\beta_{\boldsymbol{v}}\right|$. The whole LD can be calculated by 
\begin{equation}\label{Whole LD}
	T^{wl}=\frac{2R}{v}\left|\sin\psi_{a_3}\sin\beta_{\boldsymbol{v}}\cos(\theta_{a_3}-\alpha_{\boldsymbol{v}})+\cos\psi_{a_3}\cos\beta_{\boldsymbol{v}}\right|.
\end{equation}
Thus, the general distributed LD \cite{Zheng}, namely the service duration, is affected by relative velocity and the entrance of $\Omega_c$.

\subsubsection{Distribution of Customers}

Assuming that the position of the UAVs follow a 3-D Poisson Point Process (PPP) with density $\rho$, that is, the probability that there are $n$ nodes in a 3-D space \textbf{S} with a volume of $V$ is defined as
\begin{equation}\label{3-D PPP}
	Prob\{n\ nodes\ in\ \textbf{S}\}=\frac{(\rho V)^n\ e^{-\rho V}}{n!},
\end{equation}
where ${\rho}V$ equals to the expected number of nodes in \textbf{S}, and $\rho$ denotes the average network density. Assuming that both $\alpha_{\boldsymbol{v_k}}$ and $\beta_{\boldsymbol{v_k}}$ are uniformly distributed in $\left[-\pi,\pi \right]$, while $v_k$ has a similar distribution as them in $\left[v_l,v_u\right]$. Besides, the position, speed and direction of nodes are independent. 
\begin{table}
	\centering
	\caption{The Connections between CQS and Topology Changes Event.}
	\renewcommand{\arraystretch}{1}
	\begin{tabular}{m{3.3cm}<{\centering}|m{4.7cm}<{\centering}}
		\toprule[1pt]\toprule[0.5pt]
		\multicolumn{1}{c|} {Description of the CQS} & \multicolumn{1}{c}{Topology Changes Event} \\
		\midrule[0.5pt]
		The servers and customers & The UAV $S_c$ and its neighbors \\
		The service range & The communication range $\Omega_c$ of $S_c$ \\
		The start / end of the service & The time when nodes enter / leave $\Omega_c$  \\
		The service duration & The LD between $S_c$ and neighbors \\
		The customers in service & $M_1$: the nodes inside $\Omega_c$ \\
		The customers' source & $M_2$: the nodes outside $\Omega_c$ \\
		The arrival, departure and change of customers & The establishment, disconnection and change of neighbors \\
		\bottomrule[0.5pt]\bottomrule[1pt]
	\end{tabular}
	\label{Connections}
\end{table}

\subsection{Arrival of Customers}
\subsubsection{Customers' Arrival Rate}

\begin{figure}
	\centering
	\subfigure[]{\label{a}\includegraphics[width=4.2cm]{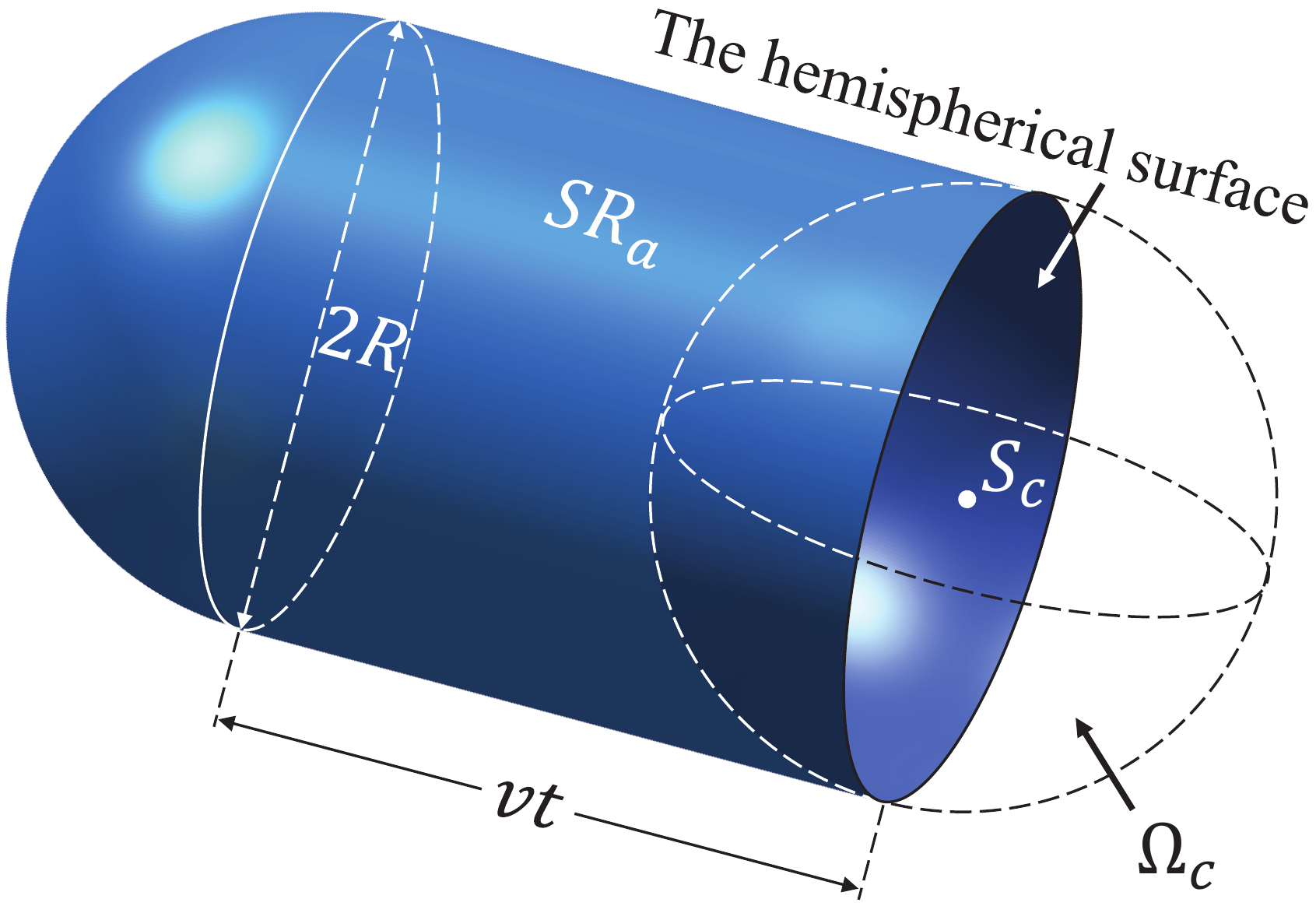}}\quad
	\subfigure[]{\label{b}\includegraphics[width=3.4cm]{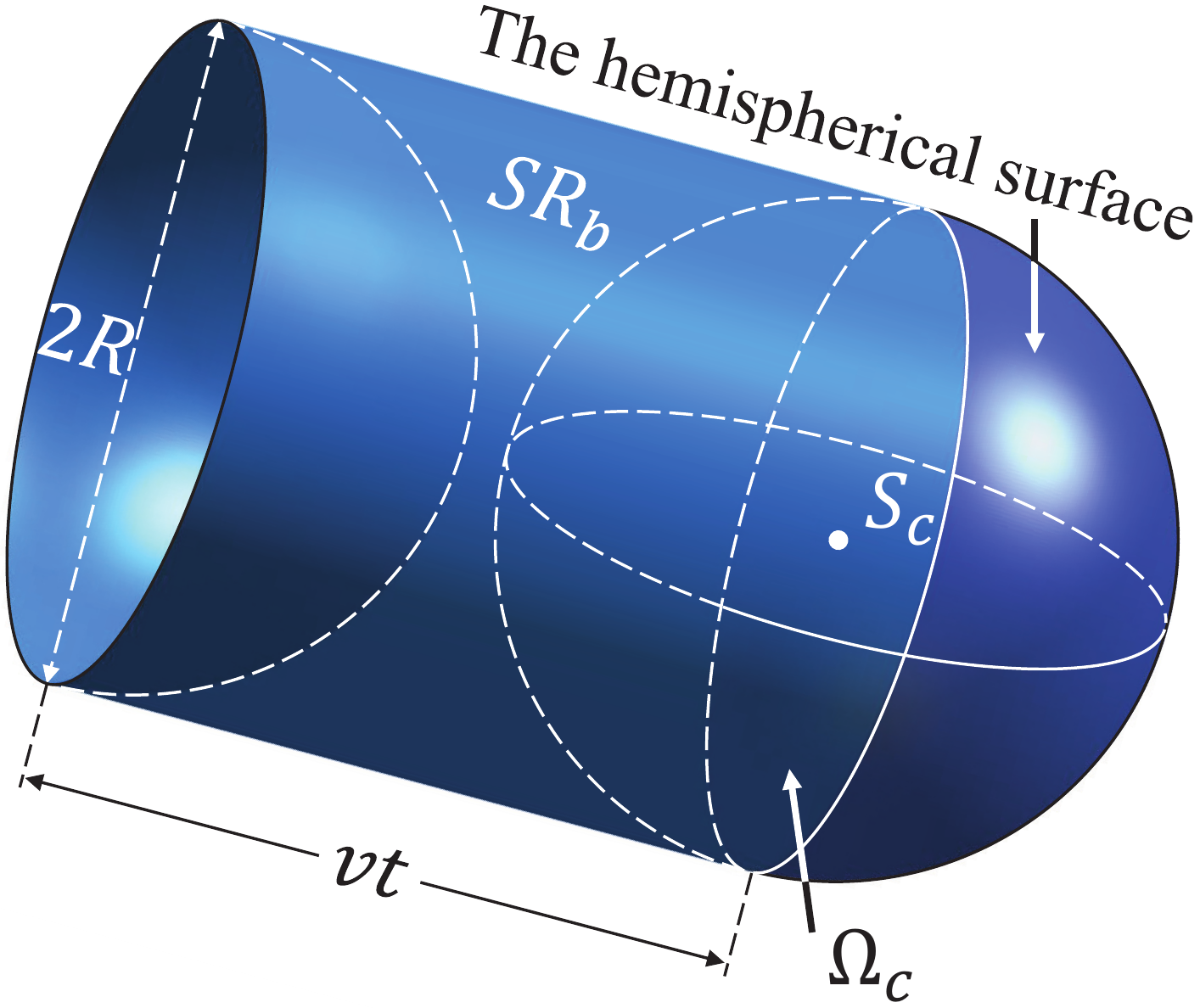}}
	\caption{The region where UAVs will (a) enter and (b) leave $\Omega_c$ from the hemispherical surface at velocity $\boldsymbol{v}$ within the upcoming $t$ seconds.}\label{Region_Enter}
\end{figure}

\textbf{Fig. \ref{Region_Enter}(a)} presents an intuitive illustration that a node can only enter $\Omega_c$ from a certain point on the hemispherical surface once its velocity is determined. The shaded region $SR_a$ shows the position of nodes that are moving at a relative velocity $\boldsymbol{v}$ and are about to enter $\Omega_c$, namely will be served by $S_c$, within the upcoming $t$ seconds. Considering all possibilities of the relative velocity, the expected number of customers arriving within the upcoming $t$ seconds is given by
\begin{equation}
	\eta_A=\int_0^\infty\int_{-\pi}^\pi\int_{-\pi}^\pi \rho V_1\cdot f(v,\alpha_{\boldsymbol{v}},\beta_{\boldsymbol{v}})d\beta_{\boldsymbol{v}}d\alpha_{\boldsymbol{v}}dv,
\end{equation}
where $V_1 = vt\pi R^2$, and  $f(v,\alpha_{\boldsymbol{v}},\beta_{\boldsymbol{v}})$ is 
derived in \textbf{Appendix A}. Thus, the expected number of customers arriving the CQS per second, namely the expected new customers' arrival rate ${\dot{\eta}}_A$ is given by
\begin{equation}\label{Arrival_Rate}
	\begin{aligned}
		{\dot{\eta}}_A = &\frac{R^2\rho}{v_u-v_l}
			\left[
				v_u^2 \boldsymbol{E}\left(\frac{v_c}{v_u}\right) - 
				2v_l^2 
				2\boldsymbol{E}\left(\frac{v_c}{v_l}\right)
				+
				v_l^2 \boldsymbol{E}\left(\beta_0,\frac{v_c}{v_l}\right)
				\right. \\ & \left. 
				+\frac{v_c^2}{4}
				\boldsymbol{M}_{0,v_c}^+ \left(\beta_{\boldsymbol{v}},v_u,v_c\right)		
				-\frac{v_c^2}{4}
				\boldsymbol{M}_{\beta_0,v_c}^- \left(\beta_{\boldsymbol{v}},v_l,v_l\right)
			\right],
	\end{aligned}
\end{equation}
where $\boldsymbol{E}\left(\cdot\right)$ and $\boldsymbol{E}\left(\cdot,\cdot\right)$ are the complete and incomplete elliptic integral of the second kind \cite{Yang} \cite{Fukushima}, respectively. And $\boldsymbol{M}_{a,b}^\pm(x,y,z)$ is defined as
 \begin{equation}\label{M}
 	\boldsymbol{M}_{a,b}^\pm(x,y,z) = 
 	\int_a^\pi\omega\left(x\right)ln\left|\frac{y + \sqrt{y^2-b^2sin^2x}}{z\pm\sqrt{z^2-b^2sin^2x}}\right|dx,
 \end{equation}
where $\omega\left(x\right)=1+3cos\left(x\right)$ and $\beta_0=\pi-sin^{-1}\left(v_l/v_c\right)$.

The conclusion that ${\dot{\eta}}_A$ is proportional to $\rho$ and $R^2$ can be drawn based on (\ref{Arrival_Rate}), which is consistent with intuition, that is, the number of nodes entering $\Omega_c$ per second will inevitably increase with $\rho$, and the nodes that were originally just passing by $S_c$ will be more likely to enter $\Omega_c$ due to the increase of $R$, namely ${\dot{\eta}}_A$ will be increased.

\begin{figure}
	\centering
	\subfigure[Range of node's speed: 5-20 m/s.]{\includegraphics[width=4.3cm]{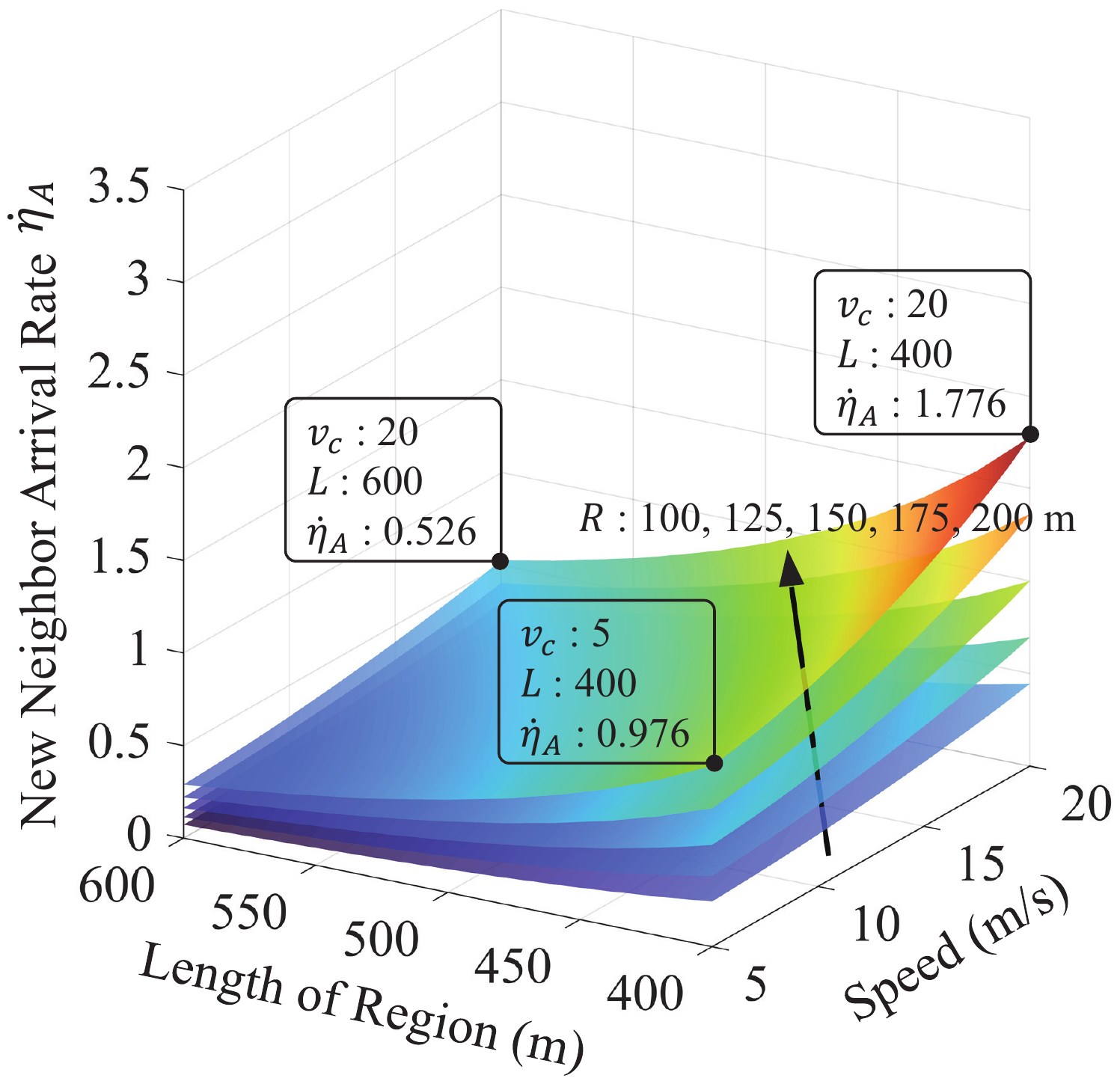}}
	\subfigure[Range of node's speed: 5-40 m/s.]{\includegraphics[width=4.3cm]{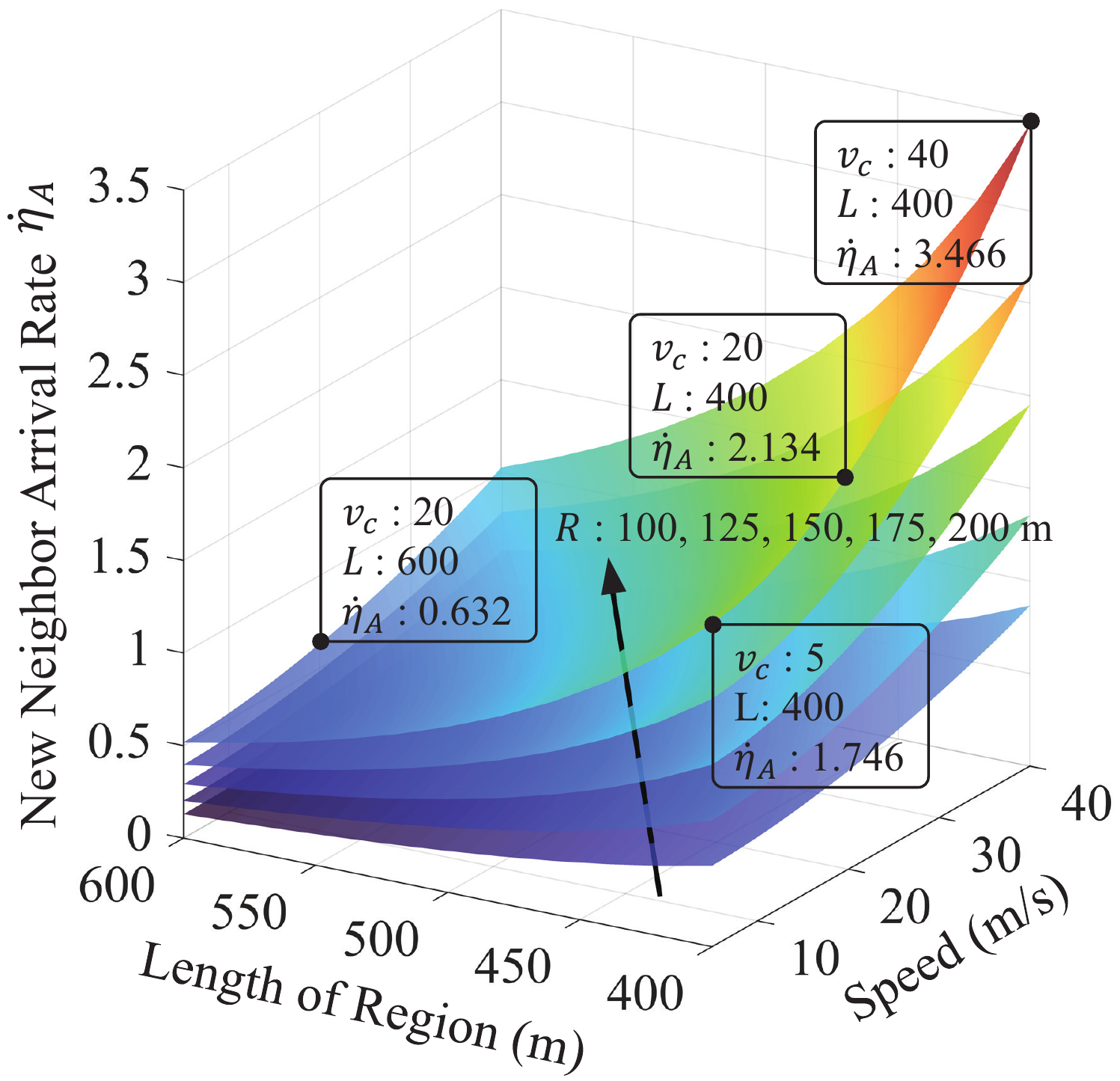}}
	\caption{New neighbor arrival rate of $S_c$ when moving with different speeds in various cases with 40 nodes, where the length of the region ranges from 400 m to 600 m, $v_l=5$ m/s while $R$ takes a value every 25 m from 100 m to 200 m.}\label{LAR}
\end{figure}

In \textbf{Fig. \ref{LAR}}, we plot ${\dot{\eta}}_A$ in various cases based on numerical integration results. The raise of $v_c$ will cause an increase in $v$ and thus more neighbors will arrive. It is worth noting that ${\dot{\eta}}_A$ and $v_c$ are not simply proportional, namely the increase of $v_c$ will promote the growth rate of ${\dot{\eta}}_A$. In addition, the expansion of the speed range will also increase ${\dot{\eta}}_A$, which can be verified by the difference of \textbf{Fig. \ref{LAR}(a)} and \textbf{Fig. \ref{LAR}(b)}.

\subsubsection{Customers' Arrival Time Distribution}

Given the relative velocity, the probability that the neighbors' inter-arrival time (NIT) is not greater than $t$ is equivalent to the probability that there is at least one node moving in $SR_a$ with relative velocity $\boldsymbol{v}$, thus we have
\begin{equation}
	\begin{aligned}
		&Prob\left\{{\rm NIT}\le t|v,\alpha_{\boldsymbol{v}},\beta_{\boldsymbol{v}}\right\}\\
		&=Prob\left\{{\rm At\ least\ one\ node\ in\ }SR_a|v,\alpha_{\boldsymbol{v}},\beta_{\boldsymbol{v}}\right\}\\
		&=1-e^{-\rho V_1}.
	\end{aligned}
\end{equation}
Therefore, the cumulative distribution function (CDF) of NIT is given by
\begin{equation}\label{NIT_CDF}
	\begin{aligned}
		&F_A\left(t\right)=Prob\left\{{\rm NIT}\le t\right\} \\
		&=\int_{0}^{\infty}\int_{-\pi}^{\pi}\int_{-\pi}^{\pi}\left(1-e^{-\rho V_1}\right)f\left(v,\alpha_{\boldsymbol{v}},\beta_{\boldsymbol{v}}\right)d\beta_{\boldsymbol{v}}d\alpha_{\boldsymbol{v}}dv.
	\end{aligned}
\end{equation}
Substituting for $f\left(v,\alpha_{\boldsymbol{v}},\beta_{\boldsymbol{v}}\right)$ from (\ref{V_angle_JPDF_simple}), we have
\begin{equation}\label{NIT_CDF_2}
	\begin{aligned}
		&F_A\left(t\right)=\\&1-\frac{1}{\pi\left(v_u-v_l\right)}\int_{0}^{\pi}{\int_{0}^{\infty}{v\cdot e^{-\rho V_1}g\left(v,\alpha_{\boldsymbol{v}},\beta_{\boldsymbol{v}}\right)}dv}d\beta_{\boldsymbol{v}},
	\end{aligned}
\end{equation}
where $g\left(v,\alpha_{\boldsymbol{v}},\beta_{\boldsymbol{v}}\right)$ is defined by (\ref{g}). By differentiating (\ref{NIT_CDF_2}) with respect to time $t$, the probability density function (PDF) of NIT is given by 
\begin{equation}\label{NIT_PDF}
	f_A\left(t\right)=\frac{\rho R^2}{v_u-v_l}\int_{0}^{\pi}{\int_{0}^{\infty}{v^2e^{-\rho V_1}g\left(v,\alpha_{\boldsymbol{v}},\beta_{\boldsymbol{v}}\right)}dv}d\beta_{\boldsymbol{v}}.
\end{equation}
It is worth noting that (\ref{NIT_PDF}) can be accurately approximated by the exponential distribution with the parameter ${\dot{\eta}}_A$, namely
\begin{equation}\label{NIT_PDF_Exponential}
	f_A\left(t\right)\approx{\dot{\eta}}_Ae^{-{\dot{\eta}}_At}.
\end{equation}

\subsection{Departure of Customers}

Assuming that the customers in CQS have changed during the period of $t_p$, and the number of customers entering, leaving and staying in the CQS is denoted as $\eta_A(t_p)$, $\eta_D(t_p)$ and $N(t_p)$, respectively. Hence, we have
\begin{equation}\label{Limit_1}
	\eta_A\left(t_p\right)-\eta_D\left(t_p\right)=N\left(t_p\right).
\end{equation}
By dividing both sides by $t_p$ and taking a limit, we have
\begin{equation}
	\lim\limits_{t_p\to\infty}{\left(\frac{\eta_A\left(t_p\right)}{t_p}-\frac{\eta_D\left(t_p\right)}{t_p}\right)}
	=\lim\limits_{t_p\to\infty}{\frac{N\left(t_p\right)}{t_p}},
\end{equation}
where $\eta_A(t_p)/t_p$ and $\eta_D(t_p)/t_p$ are respectively the average number of neighbors arriving and leaving during $t_p$, and they will become the expected arrival and departure rate, namely ${\dot{\eta}}_A$ and ${\dot{\eta}}_D$, when $t_p\rightarrow\infty$. Since the number of customers of $S_c$ is limited,\footnote{This is the case for any practical FANETs since the number of neighbors of a UAV is bounded.} thus $N\left(t_p\right)/t_p\rightarrow0$ when $t_p\rightarrow\infty$, and finally we have ${\dot{\eta}}_D={\dot{\eta}}_A$.

As mentioned earlier, the service duration follows a general distribution and the number of servers and system capacity is infinite, and we can infer from (\ref{NIT_PDF_Exponential}) that the inter-arrival time of customers is approximately exponentially distributed. Thus, the above CQS belongs to the $M/G/\infty/\infty$ type. Based on Theorem 4.14 in \cite{Su}, the departure of customers follows a Poisson process. We have $F_D\left(t\right)=F_A\left(t\right)$ and $f_D\left(t\right)=f_A\left(t\right)$ due to ${\dot{\eta}}_D={\dot{\eta}}_A$ and flow conservation, where $f_D\left(t\right)$ and $F_D\left(t\right)$ are the PDF and CDF of neighbor departure inter-arrival time, respectively.

\subsection{Change of Customers}

\subsubsection{Customers' Change Rate}

The arrival or departure of neighbors will lead to topology changes. Thus, the NCR is defined as the sum of ${\dot{\eta}}_A$ and ${\dot{\eta}}_D$, namely we have
\begin{equation}\label{NCR}
{\dot{\eta}}_C={\dot{\eta}}_A+{\dot{\eta}}_D=2{\dot{\eta}}_A,
\end{equation}
where ${\dot{\eta}}_A$ is calculated in (\ref{Arrival_Rate}). It provides a theoretical basis for calculating dynamic ${\dot{\eta}}_E$ for equations (\ref{Sensing_Delay_CDF_2})$\sim$(\ref{Expected_Sensing_Delay}) and finally contributes to the resilient perception strategy.

\subsubsection{Customers' Change Inter-arrival Time Distribution}

\begin{figure}
	\centering
	\subfigure[$L=400m, R=150m$.]{\label{fig:subfig:a}\includegraphics[width=4.2cm]{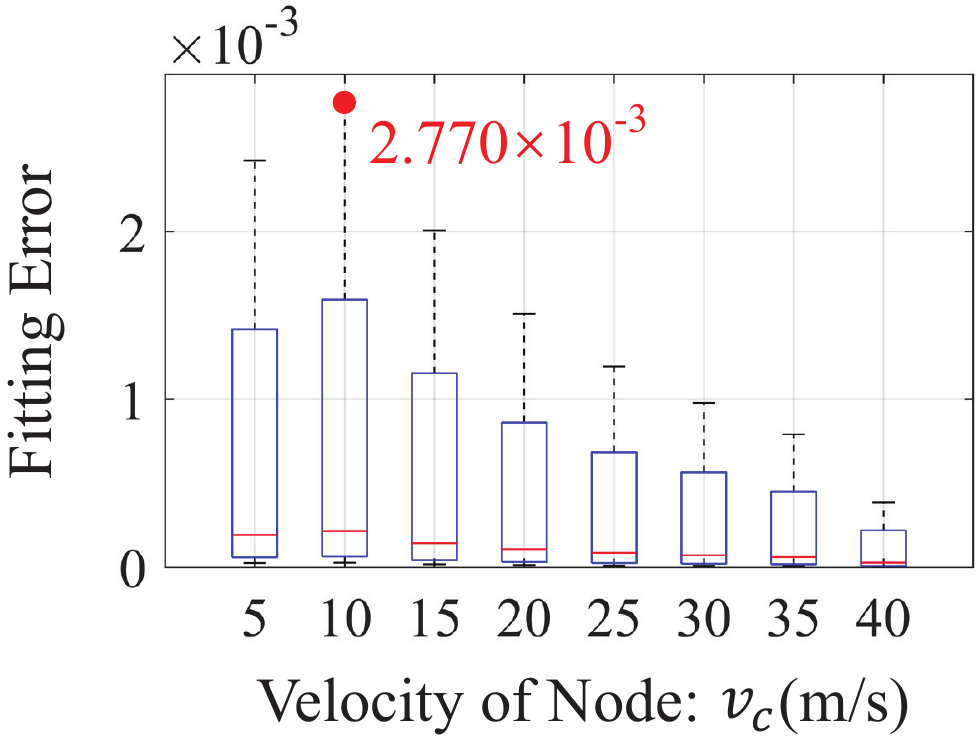}}
	\subfigure[$L=400m, R=200m$.]{\label{fig:subfig:b}\includegraphics[width=4.2cm]{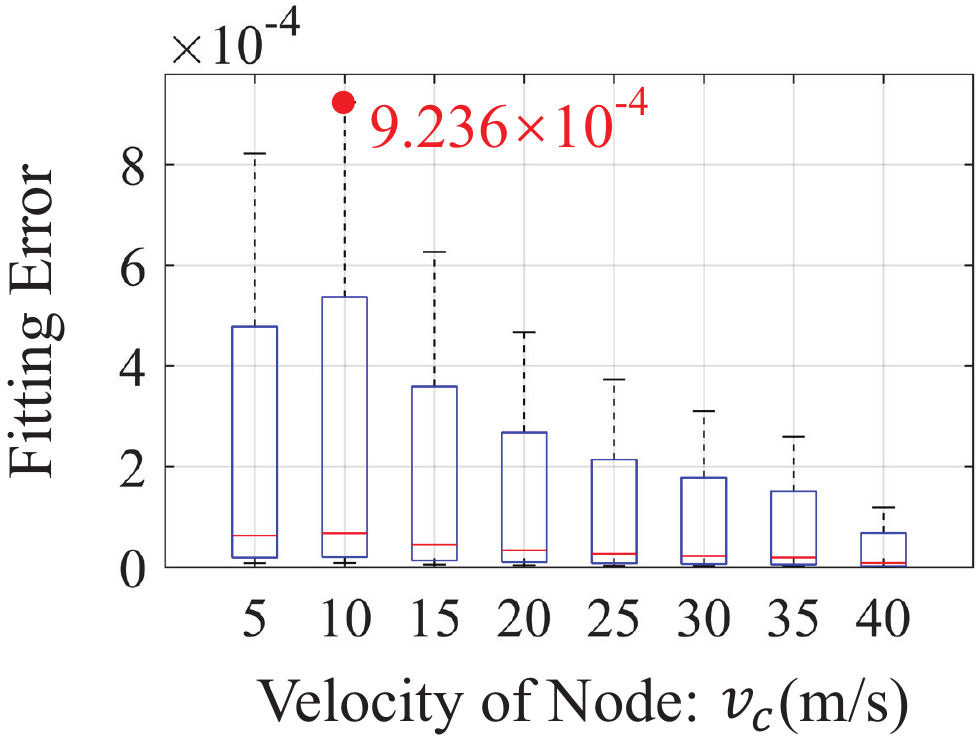}}
	\vfill
	\subfigure[$L=600m, R=150m$.]{\label{fig:subfig:a}\includegraphics[width=4.2cm]{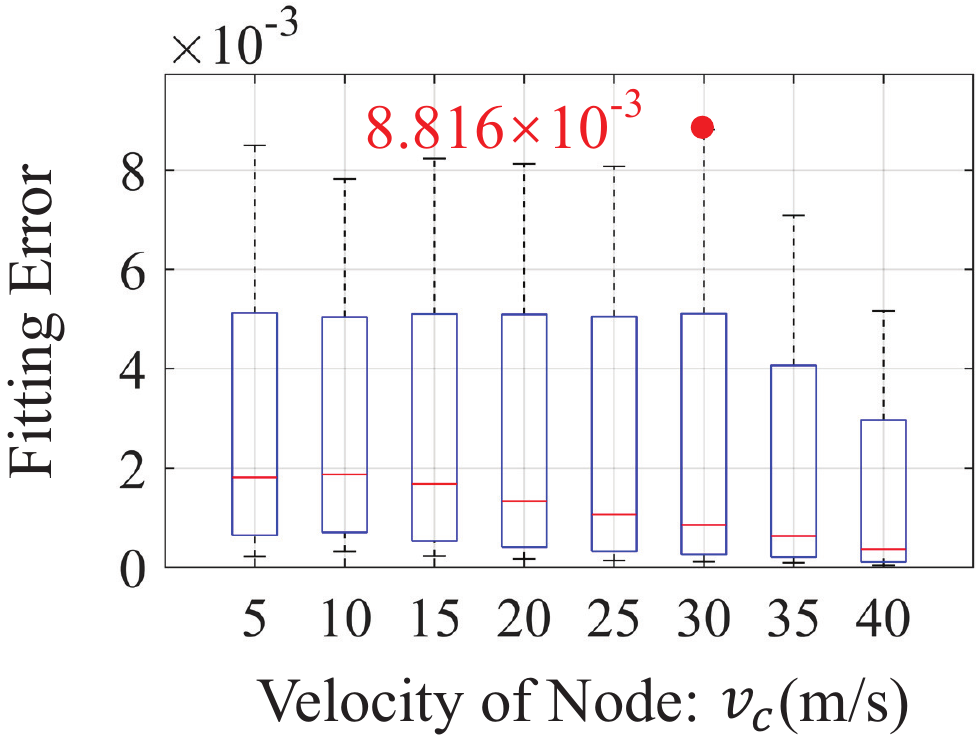}}
	\subfigure[$L=600m, R=200m$.]{\label{fig:subfig:b}\includegraphics[width=4.2cm]{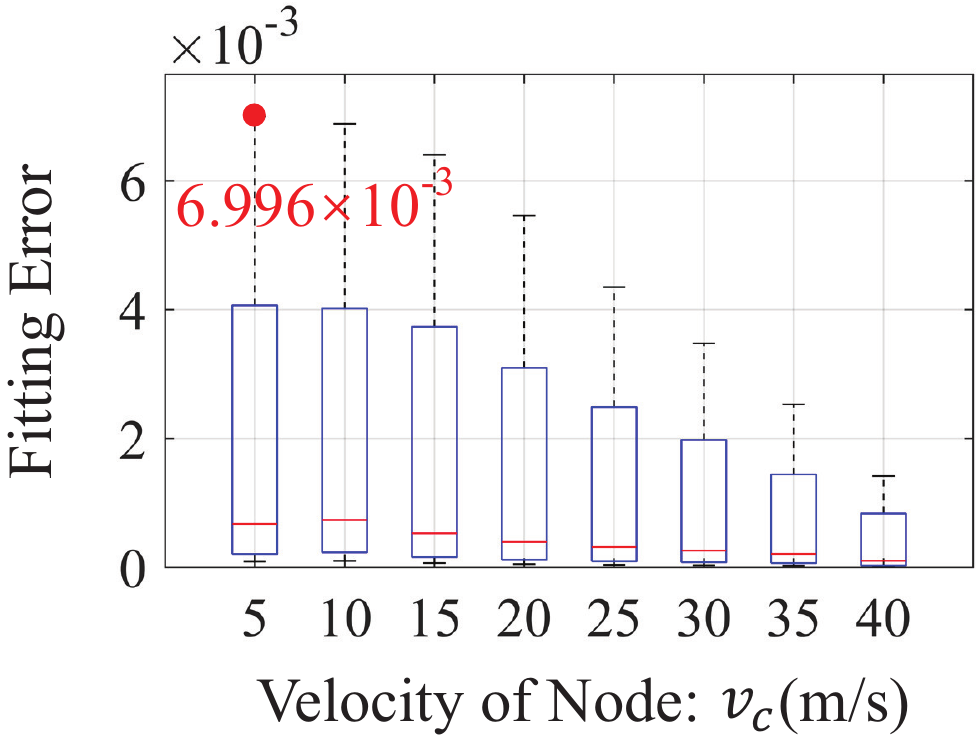}}
	\caption{The fitting error of equation (\ref{NCIT_PDF_Exponential}) in various cases with 40 UAVs. The speed range is limited to 5-40m/s.}\label{Fit_Error_NCIT}
\end{figure}

As shown in \textbf{Fig. \ref{Region_Enter}(a)} (or \textbf{(b)}), given the relative velocity $\boldsymbol{v}$, the nodes located in $SR_a$ (or $SR_b$) will arrive (or leave) $\Omega_c$ within the upcoming $t$ seconds, hence the nodes located in the union of the two shaded regions, ${SR}_u={SR}_a\cup{SR}_b$, will cause the customer changes within the upcoming $t$ seconds. The probability that the NCIT is not greater than $t$ is equivalent to the probability that there is at least one node in $SR_u$. Thus, we have
\begin{equation}
	Prob\left\{{\rm NCIT}\le t|v,\alpha_{\boldsymbol{v}},\beta_{\boldsymbol{v}}\right\}=1-e^{-\rho V_u},
\end{equation}
and the CDF of NCIT is given by
\begin{equation}\label{NCIT_CDF}
\begin{aligned}
	&F_C\left(t\right)=P\left\{{\rm NCIT}\le t\right\}\\
	&=\int_{-\pi}^{\pi}\int_{-\pi}^{\pi}\int_{0}^{\infty}\left(1-e^{-\rho V_u}\right)f\left(v,\alpha_{\boldsymbol{v}},\beta_{\boldsymbol{v}}\right)d\beta_{\boldsymbol{v}}d\alpha_{\boldsymbol{v}}dv.
\end{aligned}
\end{equation}
The volume of ${SR}_u$, namely $V_u$, varies with different $v,t$ and $R$, which are derived in \textbf{Appendix B}. Substituting for $f\left(v,\alpha_{\boldsymbol{v}},\beta_{\boldsymbol{v}}\right)$ from (\ref{V_angle_JPDF_simple}), $F_C\left(t\right)$ can be further solved by
\begin{equation}\label{NCIT_CDF_SIMP}
\begin{aligned}
	&F_C\left(t\right) = \\
	&1-\frac{1}{\pi\left(v_u-v_l\right)}\int_{0}^{\pi}\left[\int_{0}^{\frac{2R}{t}}{e^{-\rho V_3\left(v,t\right)}vg\left(v,v_c,\beta_{\boldsymbol{v}}\right)}dv
\right. \\ & \left. 
	+ \int_{\frac{2R}{t}}^{\infty}{e^{-\rho V_2\left(v,t\right)}vg\left(v,v_c,\beta_{\boldsymbol{v}}\right)}dv\right]d\beta_{\boldsymbol{v}}.
\end{aligned}
\end{equation}
By differentiating (\ref{NCIT_CDF_SIMP}) with respect to time $t$, the PDF of NCIT is given by
\begin{equation}\label{NCIT_PDF}
\begin{aligned}
	&f_C\left(t\right) = \frac{R^2\rho}{v_u-v_l}\int_{0}^{\pi}\left[\int_{\frac{2R}{t}}^{\infty}{e^{-\rho V_2\left(v,t\right)}v^2g\left(v,v_c,\beta_{\boldsymbol{v}}\right)}dv
\right. \\ & \left.
	+ \int_{0}^{\frac{2R}{t}}{\left(2-\frac{v^2t^2}{4R^2}\right)e^{-\rho V_3\left(v,t\right)}v^2g\left(v,v_c,\beta_{\boldsymbol{v}}\right)}dv\right]d\beta_{\boldsymbol{v}},
\end{aligned}
\end{equation}
where $V_2\left(v,t\right)$ and $V_3\left(v,t\right)$ are derived in \textbf{Appendix B}.
Note that (\ref{NCIT_PDF}) can be accurately approximated by
\begin{equation}\label{NCIT_PDF_Exponential}
f_C\left(t\right)\approx{\dot{\eta}}_Ce^{-{\dot{\eta}}_Ct},
\end{equation}
namely the exponential distribution with the parameter ${\dot{\eta}}_C$. By fitting the PDF of NCIT with (\ref{NCIT_PDF_Exponential}), the boxplot graphs of fitting errors in various cases are revealed in \textbf{Fig. \ref{Fit_Error_NCIT}}. The error decreases with the increase of $R$ and $\rho$. Besides, the maximum error remains below 0.0088 and even lower with higher speed, larger communication range and smaller flight area, which means that the fitting of $f_C\left(t\right)\approx{\dot{\eta}}_Ce^{-{\dot{\eta}}_Ct}$ and $F_C\left(t\right)\approx1-e^{-{\dot{\eta}}_Ct}$ will be more accurate with the increase of $\dot{\eta}_C$. Equation (\ref{NCIT_PDF_Exponential}) provides a theoretical basis for calculating $F_{EIT}\left(t\right)$ and the resilient SI (see equation (\ref{Sensing_Delay_CDF}) for detail).

\section{Topology-Aware Resilient Routing based on Adaptive Q-Learning}

In this section, we present the TARRAQ protocol, which includes the following phases: neighbor discovery, neighbor maintenance and relay selection. 
During the first two phases, nodes will find and update the available neighbors to obtain an accurate action space for Q-learning-based routing decisions, and the fragmentation of FANETs will be detected. When it comes to the relay selection phase, the information of neighbors' states determines the next hop, and the problem of route interruption caused by fragmentation will be solved.

\subsection{Neighbor Discovery Phase}

\textit{1) DEWMA Scheme for NCR}

Generally, the conventional methods tend to obtain the NCR by monitoring the changes in the neighbor list. However, the accuracy is greatly affected by the SI. And an inaccurate NCR will also result in the inability to set the best SI since the latter is usually calculated by the former. 
To avoid the above problems, we calculate the NCR by capturing the factors that affect topology changes, namely $R$, $\rho$, $v_c$, $v_u$ and $v_l$ discussed in (\ref{NCR}) and (\ref{Arrival_Rate}). Each node calculates the dynamic NCR according to (\ref{NCR}) via the following DEWMA Scheme.

Given that the speed of the neighbor node, network density and other characteristics may be accidental and time-varying, namely, $\rho$, $v_l$ and $v_u$ in (\ref{NCR}) may be much higher or lower than the average value. However, their instantaneous outliers cannot represent the real situation within a period of time. Thus, a DEWMA scheme is introduced, and the smoothed values are estimated by increasing or decreasing the old and new variables sequentially to adapt to accidental changes.

Specifically, the neighbor list is checked and compared periodically, and the current variables are smoothed by
\begin{equation}\label{DEWMA}
	{var}_{est}^t=\tau_{var}\times{var}_{est}^{t-1}+\left(1-\tau_{var}\right)\times{var}_{sam}^t,
\end{equation}
where ${var}_{est}^t$ and ${var}_{sam}^t$ are the values of $\rho$, $v_l$ and $v_u$ estimated previously and collected currently, respectively. The weight of short-term and long-term data is controlled by $\tau_{var}$, which is hoped to have the following functions: when the variable fluctuates significantly, the latest ones are expected to be concerned particularly to ensure a dynamic reaction and decision, hence the current value needs to be smoothed with greater weight, and $\tau_{var}$ is defined by
\begin{equation}\label{Tau}
\tau_{var}=min\left({{var}_{est}^{t-1}}/{{var}_{sam}^t},{{var}_{sam}^t}/{{var}_{est}^{t-1}}\right).
\end{equation}
Taking the network density $\rho$ as an example, $\tau_{\rho}$ is set to ${\rho}_{est}^{t-1}/{\rho}_{sam}^t$ (or ${\rho}_{sam}^t/{\rho}_{est}^{t-1}$) if ${\rho}_{sam}^t$ is greater (or less) than ${\rho}_{est}^{t-1}$, thus the weight of ${\rho}_{sam}^t$ in (\ref{DEWMA}) increases with the increase of the difference between ${\rho}_{sam}^t$ and ${\rho}_{est}^{t-1}$. Similar analysis can also be applied to $v_l$ and $v_u$, which won't be reiterated here. In addition, given that the $R$ and $v_c$ are individual rather than overall parameters, which means, the randomness of the network will not bring fluctuation to them, and their real-time value represents the real situation, hence it should be obtained directly without taking the impact of historical data into account.

\textit{2) Resilient Perception Strategy based on Sensing Delay}

In addition to the changes in topology, the characteristics of network events also vary with scenarios, which can be divided into the following two categories. \textbf{Case 1}: The NCIT is greater than the traffic inter-arrival time. 
There is no need to perceive the network before each traffic arrives, since the neighbor nodes may not change in the short term. \textbf{Case 2}: The NCIT is smaller than the traffic inter-arrival time. 
There is no need to perceive the network whenever link changes, since the repeated perception is wasted owing to the infrequent traffic.

Therefore, defining SI according to the event with the lowest arrival rate $\dot{\eta}_E=\min\{{\rm NCR,TAR}\}$, namely NCR in case 1 and TAR in case 2, is the best solution for minimizing the overhead without sacrificing accuracy.

Note that different performance requirements should be satisfied since there is always a compromise between overhead and other performance indicators. Thus, the concept of sensing delay is proposed to characterize and control the trade-off relationship.
\begin{figure}
\centering
	{\includegraphics[width=8.5cm]{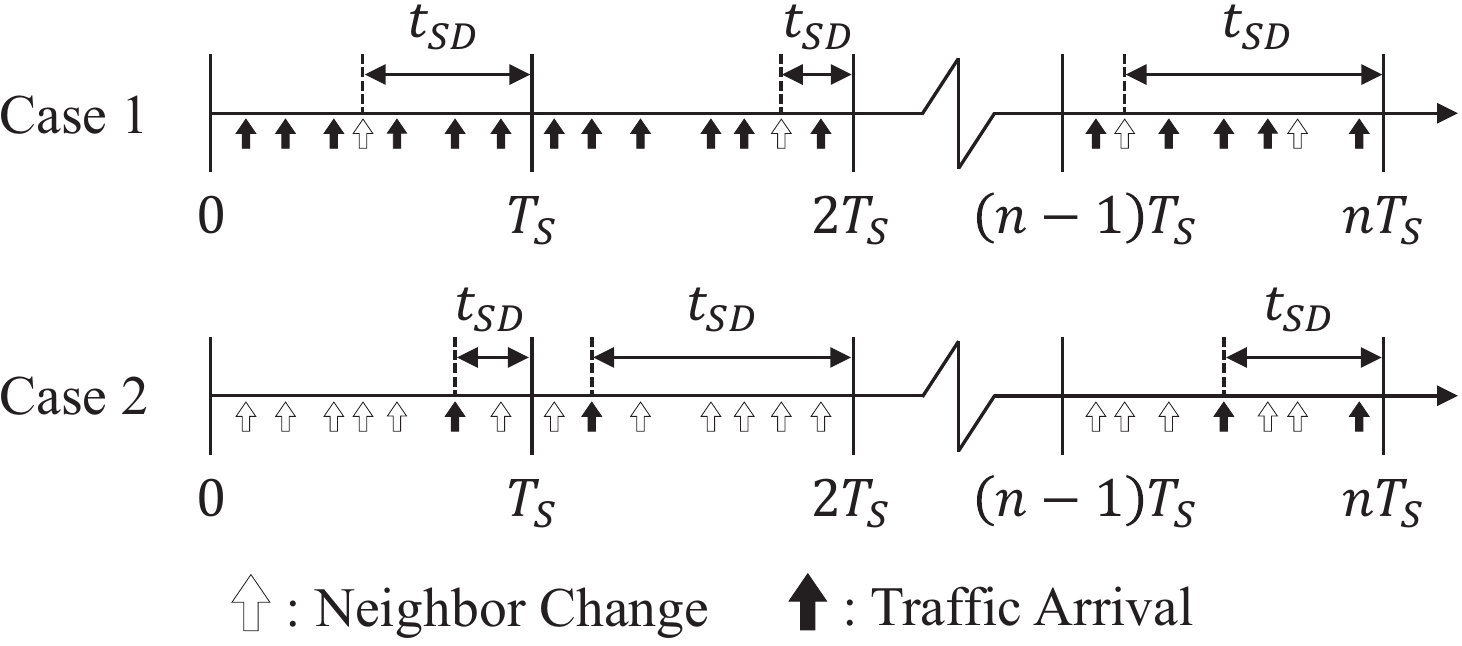}}
\caption{Sensing delay of network events in various cases.}\label{Event}
\end{figure}

We define sensing delay $t_{SD}$ as the time difference between perception behavior and the last event, and it is ubiquitous due to stochastic network events. As shown in \textbf{Fig. \ref{Event}}, periodic perception is performed after every $T_S$. In case 1, the sensing delay reflects the degree of compromise between perception overhead and accuracy, that is, the link change can be captured swiftly if the sensing delay is small enough, hence the accuracy increases with the overhead increasing, and vice versa. Thus, it is expected to find the maximum $T_S$ while minimizing the link errors caused by large sensing delay. In case 2, the sensing delay reflects the time difference between the arrival of the last traffic and the latest perception. Within a certain range, the larger it is, the closer the perception is to the arrival of the next traffic, hence the overhead is smaller and the neighbor information is more accurate, and vice versa. Thus, it is expected to find the maximum $T_S$ while avoiding link information expiration caused by a large sensing delay. The resilient perception can be achieved by presetting the expected sensing delay. Therefore, a resilient perception strategy based on sensing delay is proposed as follows.

Assuming that the network is initialized at $t=0$ and a network event occurs at time $t_{EO}$, which lies between the $n-$1 th and the $n$ th perception, $n=\left\lceil{t_{EO}/T_S}\right\rceil$, the sensing delay is given by $t_{SD}=nT_S-t_{EO}$, $t_{SD}\in\left[0,T_S\right]$, where $T_S$ symbolizes the SI of perception. Thus the CDF of $t_{SD}$ can be calculated by
\begin{equation}
\begin{aligned}\label{Sensing_Delay_CDF}
&F_{SD}\left(t\right)=Prob\left(t_{SD}\le t\right)
\\&=Prob\left(t_{EO}\in\left\{\left[T_S-t,T_S\right]\cup\left[2T_S-t,2T_S\right]\cup\cdots\right\}\right)
\\&=\sum_{k=1}^{\infty}\left(F_{EIT}\left(kT_S\right)-F_{EIT}\left(kT_S-t\right)\right).
\end{aligned}
\end{equation}

Since the traffic arrival generally follows the Poisson process \cite{Baltaci} and the NCIT can be accurately approximated by the exponential distribution according to (\ref{NCIT_PDF_Exponential}), the CDF of events' inter-arrival time (EIT) can be given by $F_{EIT}\left(t\right)=1-e^{-{\dot{\eta}}_Et}$, and (\ref{Sensing_Delay_CDF}) can be further simplified as
\begin{equation}
\begin{aligned}\label{Sensing_Delay_CDF_2}
&F_{SD}\left(t\right)=\sum_{k=1}^{\infty}\left(e^{-{\dot{\eta}}_E\left(kT_S-t\right)}-e^{-{\dot{\eta}}_EkT_S}\right)
\\&=\frac{e^{-{\dot{\eta}}_ET_S}\left(e^{{\dot{\eta}}_Et}-1\right)}{1-e^{-{\dot{\eta}}_ET_S}},
\end{aligned}
\end{equation}
where ${\dot{\eta}}_E$ is the incidence rate of event, which represents the NCR and TAR in cases 1 and 2, respectively. The PDF of EIT can be calculated by differentiating (\ref{Sensing_Delay_CDF_2}) with respect to $t$,
\begin{equation}
f_{SD}\left(t\right)=\frac{e^{-{\dot{\eta}}_ET_S}}{1-e^{-{\dot{\eta}}_ET_S}}{\dot{\eta}}_Ee^{{\dot{\eta}}_Et}\ ,\ 0\le t\le T_S.
\end{equation}
Thus, the expected sensing delay can be calculated by
\begin{equation}\label{Expected_Sensing_Delay}
\begin{aligned}
E_{SD}\left(t\right)&=\int_{0}^{T_S}{tf_{SD}\left(t\right)}dt
=\frac{e^{-{\dot{\eta}}_ET_S}}{1-e^{-{\dot{\eta}}_ET_S}}{\dot{\eta}}_E\int_{0}^{T_S}{te^{{\dot{\eta}}_Et}}dt
\\&=\frac{T_S}{1-e^{-{\dot{\eta}}_ET_S}}-\frac{1}{{\dot{\eta}}_E}\le\delta T_S,
\end{aligned}
\end{equation}
where $\delta\in\left[0,1\right]$ since $t_{SD}\in\left[0,T_S\right]$. 

It provides a method for capturing the topology changes on demand, that is, $T_S$ can be dynamically determined based on ${\dot{\eta}}_H$ and the performance requirements, which can be met by giving different $\delta$. If the accuracy of the neighbor is more valued, a smaller sensing delay namely a smaller $\delta$ should be set to achieve a more accurate perception of topology changes. If the overhead is more valued, a larger $T_S$ can be obtained by setting a larger $\delta$ to reduce the network overhead.

\subsection{Neighbor Maintenance Phase}

As one of the most crucial variables during the neighbor maintenance phase, the expiration timer reflects the residual LD of available links, which can be calculated from the position and velocity embedded in the Hello messages. Note that for high-mobility communication scenarios, it is not sufficient to only use the information in Hello messages. In fact, the node should have the capability to predict the status of neighbors to meet the critical timeliness requirement. Therefore, our TARRAQ calculates the adaptive expiration timer by predicting the residual LD via KF.

\textit{1) Mobility Prediction:} Aiming at predicting the mobility of neighbors, KF model is introduced during the neighbor maintenance phase, which is mainly composed of estimation and correction. According to the law of RWP mobility model, the discrete state evolution model and measurement model can be expressed by 

\begin{equation}\label{Station and measurement model}
	\left\{
	\begin{aligned}
	&\boldsymbol{x}[t]=\textbf{F}\boldsymbol{x}[t-1]+\boldsymbol{u}[t-1]\\
	&\boldsymbol{y}[t]=\textbf{H}\boldsymbol{x}[t-1]+\boldsymbol{z}[t-1]
	\end{aligned}\ \ ,
\right.
\end{equation}
where $\boldsymbol{x}[t]=[\,\textbf{p}^T[t],\boldsymbol{v}^T[t]]^T$ denotes the predicted mobility state of node $S_j$, $\textbf{p}[t]=[p_{j,1}[t],p_{j,2}[t],p_{j,3}[t]]^T$ and $\boldsymbol{v}[t]=[v_{j,1}[t],v_{j,2}[t],v_{j,3}[t]]^T$. $\boldsymbol{x}[t-1]$ denotes the state information embedded in the Hello messages or predicted at last discrete time. $\boldsymbol{y}[t]$ denotes the measurement vector obtained by GPS. $\textbf{F}\in\mathbb{R}^{6\times6}$ and $\textbf{H}\in\mathbb{R}^{3\times6}$ are the matrices of state transition and observation. $\boldsymbol{u}$ and $\boldsymbol{z}$ are zero-mean Gaussian distributed noise with covariance matrices as $\textbf{Q}$ and $\textbf{R}$.

Based on the RWP mobility model, the state transition and measurement matrix should be initialized as $\textbf{F}=\left[\textbf{I}_{3\times3},{\rm SI}\cdot\textbf{I}_{3\times3};\textbf{0}_{3\times3},\textbf{I}_{3\times3}\right]$ and $\textbf{H}=\left[\textbf{I}_{3\times3},\textbf{0}_{3\times3}\right]$, respectively. According to \cite{Liu}, we determine the initial values for covariance matrix by $\textbf{M}[0]=10^4\textbf{I}_{6\times6}$, $\textbf{Q}=10^{-3}\textbf{I}_{6\times6}$ and $\textbf{R}=\textbf{I}_{3\times3}$, where $\textbf{I}_{n\times n}$ is unit matrix. The initial state $\boldsymbol{x}[0]$ denotes the first measurement value since no prior information can be obtained.
Following the standard procedure of KF, the state prediction and tracking are summarized as follows.

\textit{Step 1. }State Prediction.  
\begin{equation}\label{Kalman_1}
	\hat{\boldsymbol{x}}[t|t-1]=\textbf{F}\hat{\boldsymbol{x}}[t-1].
\end{equation}

\textit{Step 2. Covariance Matrix Prediction}.  
\begin{equation}\label{Kalman_2}
	\textbf{M}[t|t-1]=\textbf{F}\textbf{M}[t-1]\textbf{F}^T+\textbf{Q}
\end{equation}

\textit{Step 3. Kalman Gain Calculation}.  
\begin{equation}\label{Kalman_3}
	\textbf{K}[t]=\textbf{M}[t|t-1]\textbf{H}^T\left(\textbf{R}+\textbf{H}\textbf{M}[t|t-1]\textbf{H}^T\right)
\end{equation}

\textit{Step 4. State Tracking}.  
\begin{equation}\label{Kalman_4}
	\hat{\boldsymbol{x}}[t]=\hat{\boldsymbol{x}}[t|t-1]+\textbf{K}[t]\left(\boldsymbol{y}[t]-\textbf{H}\hat{\boldsymbol{x}}[t|t-1]\right)
\end{equation}

\textit{Step 5. Covariance Matrix Update}. 
\begin{equation}\label{Kalman_5}
	\textbf{M}[t]=\left(\textbf{I}-\textbf{K}[t]\textbf{H}\right)\textbf{M}[t|t-1]
\end{equation}

\textit{2) Adaptive Expiration Timer:} After obtaining the predicted positions of available neighbors, the residual LD can be estimated by 
\begin{equation}\label{Residual LD}
	\hat{T}^{rl}_{i,j} = {T}_{i,j}^{wl} - \hat{T}_{i,j}^{el},
\end{equation}
where $T_{i,j}^{wl}$ is defined by (\ref{Whole LD}),
\begin{equation}\label{LD Elapsed}
	\hat{T}_{i,j}^{el} = \frac{1}{v_{i,j}}\sqrt{\sum\nolimits_{k=1}^3\left|\,p_{a_j,k}-\hat{p}_{j,k}\right|^2}
\end{equation}
represents the LD elapsed. Here $v_{i,j}$ denotes the relative velocity between $S_i$ and $S_j$, and $\hat{p}_{j,k}$ represents the 3D position of UAV $S_j$ predicted by (\ref{Kalman_1})$\sim$(\ref{Kalman_5}). As shown in \textbf{Fig. \ref{Sphere}(a)}, $\textbf{p}_{a_j}\in\mathbb{R}^{3\times1}$ denotes the position of $a_j$ where $S_j$ enters $\Omega_i$, which is obtained by their position and relative velocity.

Finally, for UAV $S_i$, we regard $S_j$'s ET as its residual LD, namely $\hat{T}^{rl}_{i,j}$. Once $S_i$ receives messages from $S_j$, ET will be set immediately if $S_j$ is not within the neighbor list, and it will be refreshed with the periodical KF procedure. If new messages are not received after ET expires, then $S_j$ is considered to have left $\Omega_i$ and should be deleted from the neighbor list of $S_i$.

\subsection{Relay Selection Phase}

In our TARRAQ, an adaptive Q-learning approach is proposed to make distributed and autonomous decisions for relay selection, where metrics of link, neighbor and distance are considered in the reward function. Besides, the Q-learning parameters are adjusted adaptively in the dynamic environment.

\subsubsection{Reward Function}
In order to obtain a stable route with low delay, the metrics of link, neighbor and distance are jointly considered as follows.

\textit{The link metric:}
The link quality should be taken into consideration since the route will be more reliable if links are stable. Thus we consider the link metric by $F^{[1]}_{i,j} = \hat{T}^{rl}_{i,j}$, where the predicted residual LD $\hat{T}^{rl}_{i,j}$ is defined in (\ref{Residual LD}).

\textit{The neighbor metric:}
The next hop with more useful neighbors close to the destination will be preferred since there will be more choices and a lower possibility of packet loss. Thus the neighbor metric is given by $F^{[2]}_{i,j} = {|N^{u}_{j}|}/{{\rm NCR}_{j}}$, where $N^{u}_{j}=N_j-(N_i\cap N_j)$ represents the useful neighbors of $S_j$, and $N_i$ denotes the neighbor set of $S_i$. If $S_j$ has a higher degree and smaller NCR, and owns more neighbors that are not in the neighbor list of $S_i$, it yields a larger $F^{[2]}_{i,j}$.

\textit{The distance metric:}
Note that the distant neighbors usually have shorter link lifetime and fewer hops to destination, while the adjacent ones have more stable links with worse E2ED, the distance metric is defined by
\begin{equation}\label{distance metric}
	F^{[3]}_{i,j} = \frac{z_{i,j}\cdot \Delta d_{i,j}}{\sigma^2}\exp\left(-\frac{z^2_{i,j}}{2\sigma^2}\right),
\end{equation}
where $z_{i,j}=(R/d_{i,j})^2-1,z_{i,j}\in(0,\infty)$, $d_{i,j}$ and $R$ can be calculated by (\ref{Euclidean Dis}) and (\ref{R}), respectively. $\Delta d_{i,j}=d_{i,D}-d_{j,D}$ is the difference between the distance of $S_i$ and $S_j$ to the destination. Generally, if $S_j$ has a greater $\Delta d_{i,j}$ and adjacent to our preset distance, it yields a larger $F^{[3]}_{i,j}$. Note that a larger (smaller) $\sigma$ means more preference for close (distant) neighbors, we set $\sigma=1$ by default.

Based on the above metric, the reward function is given by
\begin{equation}\label{Reward Function}
	R(s_t,a_t)=\left\{
	\begin{aligned}
		&R_{max}\ ,\ s_{t+1}\ {\rm is\ the\ destination}\\
		&R_{min}\ ,\ s_{t+1}\ {\rm is\ the\ local\ minimum}\\
		&\sum\nolimits_{k=1}^{3}\frac{\varphi_kF^{[k]}_{i,j}}{\sum\nolimits_{c=1}^{|N_i|}F^{[k]}_{i,c}}\ ,\ {\rm otherwise}
	\end{aligned},
	\right.
\end{equation}
where $R_{max}$ is the maximum reward once next hop is the destination. $R_{min}$ is the minimum value when there is no neighbor closer to the destination, which can avoid the routing holes caused by network fragmentation. The normal reward will be calculated by the last item when taking action from $s_t$ to $s_{t+1}$, namely transferring packets from node $S_i$ to $S_j$. $\varphi_k\in(0,1)$ is the weight for various $F^{[k]}_{i,j}$. There will be more reward if a node has larger factors of link, neighbor and distance. In this way, nodes that may become network fragments can be excluded from the best relay.

\subsubsection{Adaptive Exploitation and Exploration}
In highly dynamic FANETs, the action space will be invalidated once the residual duration of a link gradually drops to zero. In order to adapt to the dynamic of FANETs, the residual LD, namely the system service time, is considered as the key metric for balanced and adaptive exploitation and exploration.

First, residual LD-based softmax is designed to achieve a combination of exploitation and exploration. The probability that the agent selects action $a_t$ in state $s_t$ is given by
\begin{equation}\label{Softmax}
	\pi(s_t,a_t)=\frac{e^{{\hat{T}^{rl}_{i,j}}/{\tau}}}{\sum_{c=1}^{|N_i|}e^{{\hat{T}^{rl}_{i,c}}/{\tau}}},
\end{equation}
where $\tau$ is a positive parameter of temperature, and a higher value means that the agent selects more random actions while a lower one means less exploration. And $\tau$ is updated by $\tau(t)=\tau(0)/\log_2(1+t)$ during iteration. Thus $\pi(s_t,a_t)$ for different actions is almost equal at the beginning with a larger $\tau$, and after it decreases with time step increases, there will be a larger gap for different $\pi(s_t,a_t)$. Based on (\ref{Softmax}), the agents prefer the action with the largest LD whereas other actions are ranked instead of randomly chosen.

In addition, the residual LD is also used to control the update process of $\alpha$ and $\gamma$ in (\ref{Q}). Note that Q-values need to be updated faster when the action space is more unstable, thus a larger learning rate is preferred if link is about to break. Thus an adaptive $\alpha_{i,j}$ associated with the residual LD from $S_i$ to $S_j$ is designed by $\alpha_{i,j}={\rm exp}(-\hat{T}^{rl}_{i,j})$.
This makes $Q_{i,j}$ substantially updated if $S_j$ will soon disappear from $N_i$.
Similarly, if the residual LD from $S_j$ to $S_k$ is shorter and $S_k$ is the best relay for $S_j$, a smaller discount factor $\gamma_{i,j} $ is required since $S_j$ is probably not the best choice for $S_i$. Thus we have $\gamma_{i,j}=1-{\rm exp}(-\hat{T}^{rl}_{j,k})$.

\begin{algorithm}[h]  
	\caption{TARRAQ}  
	\label{Pseudo Code}
	\begin{algorithmic}[1]  
		\Require   $\textbf{p}_i$, $\delta$, $\boldsymbol{v_i}$, $v_l$, $v_u$, R, $\rho$, $K_{max}$, $\epsilon$, TAR and destination
		\Ensure The best relay for UAV $S_i$
		\State Initial $T_i^{\rm SI}$, KF parameters, Q-table and neighbor table
		\For{$t=1:\Delta t:t_{max}$}
			\State /*****\ \textbf{\textit{Neighbor Discovery and Maintenance}}\ *****/  
			\State Obtain the real-time value of $\textbf{p}_i$, $\boldsymbol{v}_i$ and $R$
			\State Calculate the ${\rm NCR}$ by (\ref{NCR}) and (\ref{Arrival_Rate})
			\State Set $\dot{\eta}_E$ as the minimum of ${\rm NCR}$ and ${\rm TAR}$
			\State Set $E_{SD}$ for different performance requirement
			\State Calculate the resilient SI by (\ref{Expected_Sensing_Delay})
			\If{Hello message is received (e.g. from $S_j$)}
				\State Calculate the actual $T_{i,j}^{rl}$ by (\ref{Residual LD})
				\State Perform DEWMA for NCR
				\State Update neighbor table and content of Hello
				\State Reply Hello, reset $T_i^{\rm SI}=0$
			\ElsIf{no Hello from any UAV is received}
				\If{$T_i^{\rm SI} \le$ SI} $T_i^{\rm SI}=T_i^{\rm SI}+\Delta t$ 
				\Else 
					\State Send Hello message and reset $T_i^{\rm SI}=0$
				\EndIf
				\For{$j=1:1:|N_i|$}
				\State Perform the KF procedure by (\ref{Kalman_1}) to (\ref{Kalman_5})
				\State Estimate $\hat{T}^{rl}_{i,j}$ by (\ref{Residual LD})
				\If{$\hat{T}^{rl}_{i,j} \le$ 0} remove $S_j$ from $N_i$
				\EndIf
				\EndFor
			\EndIf			
			\State /************\ \textbf{\textit{Next Hop Selection}}\ ************/
			\If{the destination is within one-hop} 
				\State The maximum reward $R_{max}$ is obtained
				\State Transmit the traffic to the destination
			\Else
			\State Update the action space via available neighbors
				\While{$k\le K_{max}$}
					\State Select action by (\ref{Softmax}), and adjust $\alpha$ and $\gamma$
					\State Measure the reward by (\ref{Reward Function})
					\State Update the Q-value by (\ref{Q})
					\If{$\max\limits_{j\in N_i}|Q_k(S_i,S_j)-Q_{k-1}(S_i,S_j)|\le\epsilon$}
						\State Break
					\EndIf
				\EndWhile
				\State $m = \mathop{\arg\max}\limits_{j\in N_i} Q_t(s_i,a_j)$
				\State Select UAV $S_m$ as the best relay
			\EndIf
		\EndFor
	\end{algorithmic}  
\end{algorithm}  

\subsection{TARRAQ Protocol}
We are now ready to present the complete TARRAQ protocol. \textbf{Algorithm \ref{Pseudo Code}} provides the pseudo-code of TARRAQ run by each UAV (taking UAV $S_i$ for example), which is mainly composed of the following two parts.

\subsubsection{Neighbor Discovery and Maintenance}
First, UAVs calculate the NCR dynamically by performing DEWMA. 
Then they preset the expected sensing delay for various performance demands and calculate the resilient SI based on the smallest of NCR and TAR.
Once they receive Hello messages from other nodes, the transmitters will be counted in the neighbor list if they are newly arrived, and the expected LD will be calculated instantly. 
If no message from any UAV is received, there will be an increment for SI's timer $T_i^{\rm SI}$, and it will not be reset namely Hello message will not be broadcast until $T_i^{\rm SI}>$ SI.
In addition, the KF procedure for mobility state will be performed for all neighbors to estimate the residual LD $\hat{T}^{rl}_{i,j}$, and the neighbor that has not exchanged valid information will be deleted once $\hat{T}^{rl}_{i,j}$ drops to zero.
\subsubsection{Next Hop Selection}
The traffic packets will be transmitted to the destination if it's within one-hop range. Otherwise, the distributed Q-learning process will be performed. The Q-table is first initialized and then updated periodically when a Hello message is received. The action is selected based on (\ref{Softmax}), and then the adaptive learning rate and discount function are adjusted. Multi metrics are considered in (\ref{Reward Function}) when calculating the award. Q-values are updated via (\ref{Q}) to choose the best relay, and it receives the minimum reward $R_{min}$ when reaching the local minimum. The iteration will be stopped when at least one of the following conditions is satisfied: 1) Maximum number $K_{max}$ of iterations is reached; 2) $\max\limits_{j\in N_i}|Q_k(S_i,S_j)-Q_{k-1}(S_i,S_j)|\le\epsilon$.
\begin{remark}
It should be emphasized that the conclusions of section \textbf{\uppercase\expandafter{\romannumeral4}} play a crucial role in TARRAQ. That is, as the premise of the dynamic calculation of NCR, (\ref{NCR}) together with (\ref{NCIT_PDF_Exponential}) provide a theoretical basis for resilient SI in (\ref{Expected_Sensing_Delay}), and the system service duration derived in (\ref{Whole LD}) is used to set a timer for link validity. Besides, they are also valuable in the reward function and action selection strategy, and the details have been offered in the last subsection thus won't be reiterated here.

\end{remark}

\section{Simulation and Discussion}

\subsection{Simulation for Dynamic Topology Feature}
The distribution of NIT and NCIT is analyzed and approximated by exponential distribution in section \uppercase\expandafter{\romannumeral4}. Here, Monte Carlo simulations are performed to verify their accuracy.

\begin{figure}
	\centering
	\subfigure[CDF of NIT]{\includegraphics[width=4.2cm]{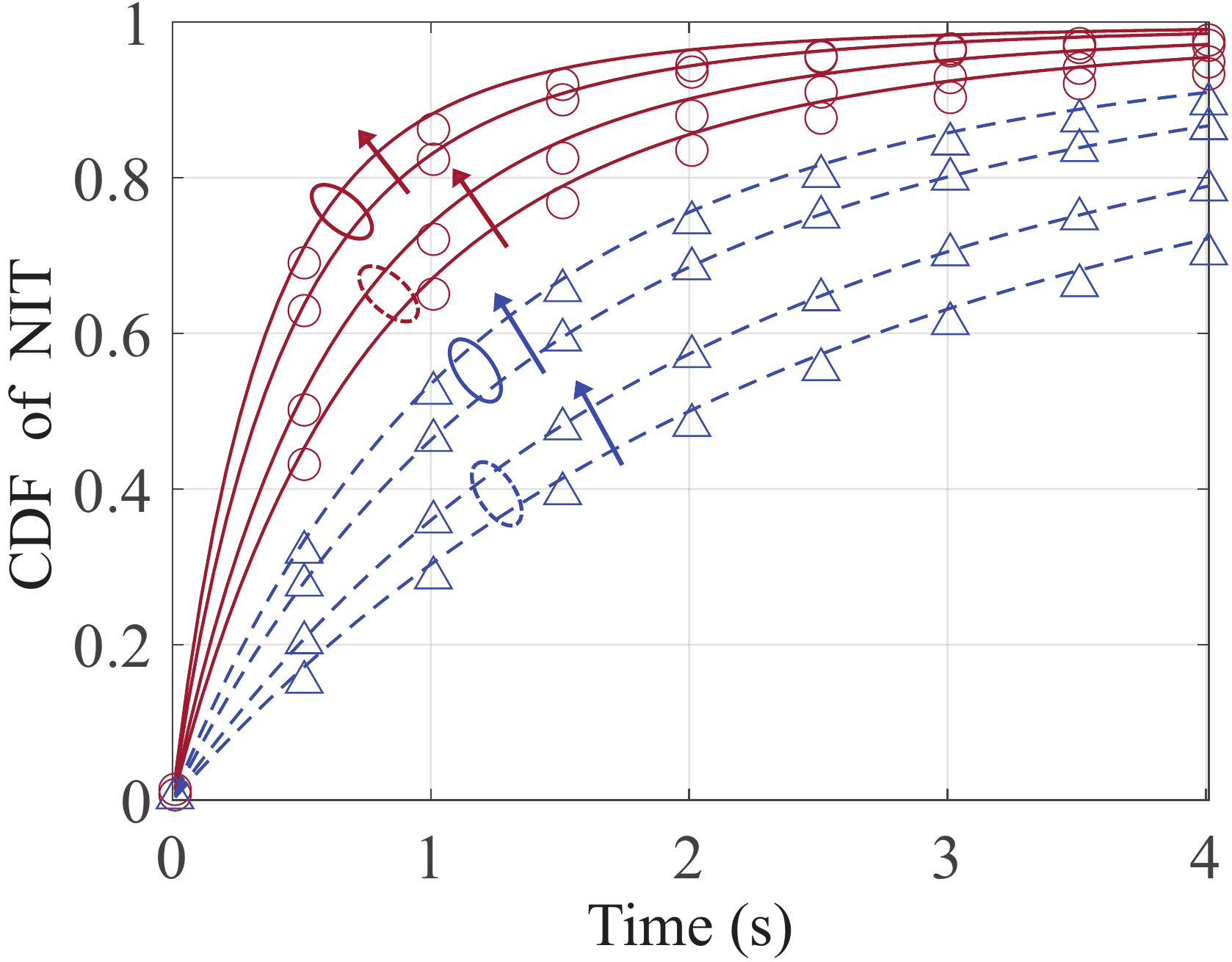}}
	\subfigure[CDF of NCIT]{\includegraphics[width=4.2cm]{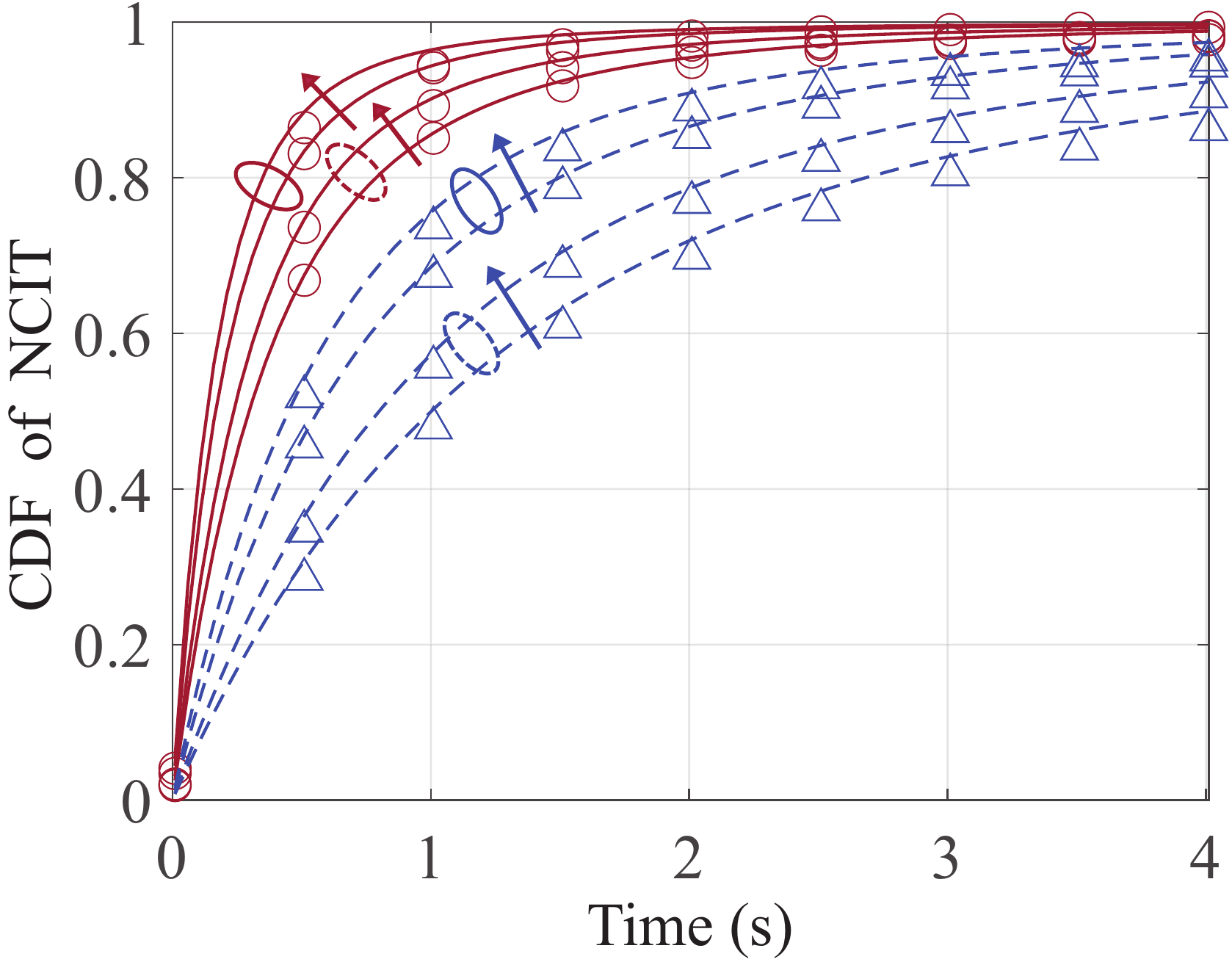}}
	\subfigure{\includegraphics[width=8.5cm]{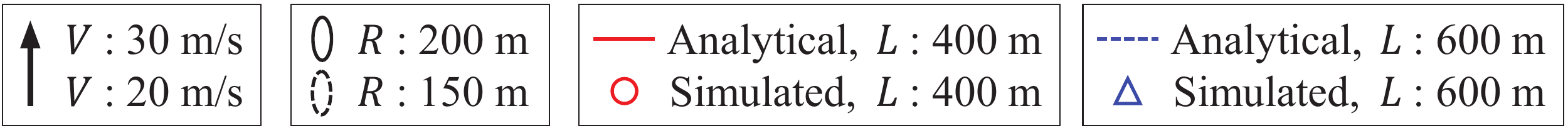}}
	\caption{Comparison between theoretical analysis and simulation statistics: the CDF of NIT and NCIT in various cases.} \label{FIG_CDF_PDF_NIT_NCIT}
\end{figure}

The closed-form solutions for the distribution of NIT and NCIT, namely equations (\ref{NIT_CDF_2}) and (\ref{NCIT_CDF_SIMP}), are analyzed by numerical integration, which is shown as the dotted lines in \textbf{Fig. \ref{FIG_CDF_PDF_NIT_NCIT}}. The simulation experiment was performed in a cube region with a side length of $L$, which is 400 m or 600 m for a dense or sparse scenario, respectively. Besides, 40 UAVs are randomly deployed and moving with the RWP mobility model, and their speed range is limited to 5-40 m/s while the communication range is set as 150 m or 200 m. The arrival and change of neighbors during the 3600 s experiment are recorded for each node, and the average results from 500 random simulations are shown as the scatter points in \textbf{Fig. \ref{FIG_CDF_PDF_NIT_NCIT}}. 

The numerical integration results from theoretical analysis in section \uppercase\expandafter{\romannumeral4} perfectly match the Monte Carlo simulations, and their slight difference is caused by the boundary effects. Specifically, when a UAV moves close to the boundary of the 3-D region, it will bounce back into the region soon. In addition, the probability that other nodes enter or leave their communication range from the direction of boundary will be very small or even non-existent, which makes the actual ${\dot{\eta}}_A$ and ${\dot{\eta}}_C$ lower than expected. Note that this error will be reduced as the network scale increases since it reduces the occurrence of the above case.

In addition, \textbf{Fig. \ref{FIG_CDF_PDF_NIT_NCIT}} also demonstrates that the CDF of NIT and NCIT increase sharply with the increase of time and present a long-tail effect. Since ${\dot{\eta}}_A$ and ${\dot{\eta}}_C$, which are related to the slope of CDF, are depending on $R$, $\rho$, $v_c$, $v_u$ and $v_l$. Thus the probability that neighbors will arrive or change in a short time becomes higher with faster mobility, a larger communication range and a smaller network scale. Interestingly, the CDF of NCIT is similar to the two times magnified and normalized version of NIT, due to that the distribution of NCIT can be accurately approximated by the exponential distribution with the parameter $\dot{\eta}_C$ and $\dot{\eta}_A$, and there is a relationship of ${\dot{\eta}}_C=2{\dot{\eta}}_A$.

\subsection{Simulation for Routing Protocol}
In order to test the performance of TARRAQ, we constructed a scenario by MATLAB R2021a. According to the parameters in \cite{Mahmud}, with the initial position generated randomly, 40 UAVs are moving in a region of 600 m$\times$600 m$\times$150 m, and the lowest speed is 5 m/s while the highest speed varies within 10$\sim$60 m/s. For the MAC layer, we utilize the IEEE 802.11b protocol, which is suitable for long-range communication with high data rate. The source node is randomly selected during 300 s simulation. The first 10 s are the initialization stage, and then 1 Mbps Rate of CBR traffics that follows the Poisson process with an average interval of 1 s are generated \cite{Mahmud} \cite{Tan} \cite{QTAR}. In our simulation, the maximum cache time is 5 s. The free-space path loss model is used to characterize propagation in our simulation, and each node is configured with an omni-directional antenna and has the same transmission power of 1 $W$ \cite{MPVR} \cite{QTAR}. According to \cite{Energy_Consumption}, the path loss exponent is set as $\alpha=2$, and we also have $E_{elec}=50\ nJ/bit$ and $E_{fs}=10\ pJ/bit/m^2$. The detailed parameters are summarized in Table \textbf{\ref{Simulation Parameters}}. 

\begin{table}
	\centering
	\caption{Simulation Parameters.}
	\renewcommand{\arraystretch}{1.2}
	\begin{tabular}{m{4.4cm}<{\centering} m{3.7cm}<{\centering}}
		\toprule[1pt]\toprule[0.5pt]
		Parameters & Value \\
		\midrule[0.5pt]
		Region Size & $600\ m \times 600\ m \times 150\ m$\\
		Number of UAVs & 40 \\
		Mobility Model & 3-D RWP\\
		Minimum Speed & 5 m/s\\
		Maximum Speed & 10 - 60 m/s\\
		Transmission Power & $1.0\ W$\\
		Coefficient of Transmitter or Receiver & $50\ nJ/bit$\\
		Coefficient of Power Amplifier &  $10\ pJ/bit/m^2$\\
		Path Loss Exponent & 2\\
		Traffic Type & CBR\\
		CBR Rate & 1 Mbps\\
		PHY/MAC Protocol & 802.11b\\
		Propagation Model & Free-space\\
		Antenna Type & Omni-directional\\
		\bottomrule[0.5pt]\bottomrule[1pt]
	\end{tabular}
	\label{Simulation Parameters}
\end{table}

We considered three state-of-art protocols for the performance comparison, namely GPSR-EE-Hello \cite{Mahmud}, MPVR \cite{MPVR} and QTAR \cite{QTAR}, where GPSR-EE-Hello presents the improved GPSR protocol based on Mahmud's scheme. Our TARRAQ is further divided into two cases: $\delta=0.55$ and $\delta=0.65$ for the sensing delay-sensitive and overhead-sensitive scenes, respectively. We performed 50 simulations and the results with 90\% confidence interval are presented as follows. 

\textit{1) Comparison Under Different SINR Thresholds}

\begin{figure*}
	\centering
	\subfigure[Packet delivery ratio]{\includegraphics[width=4.45cm]{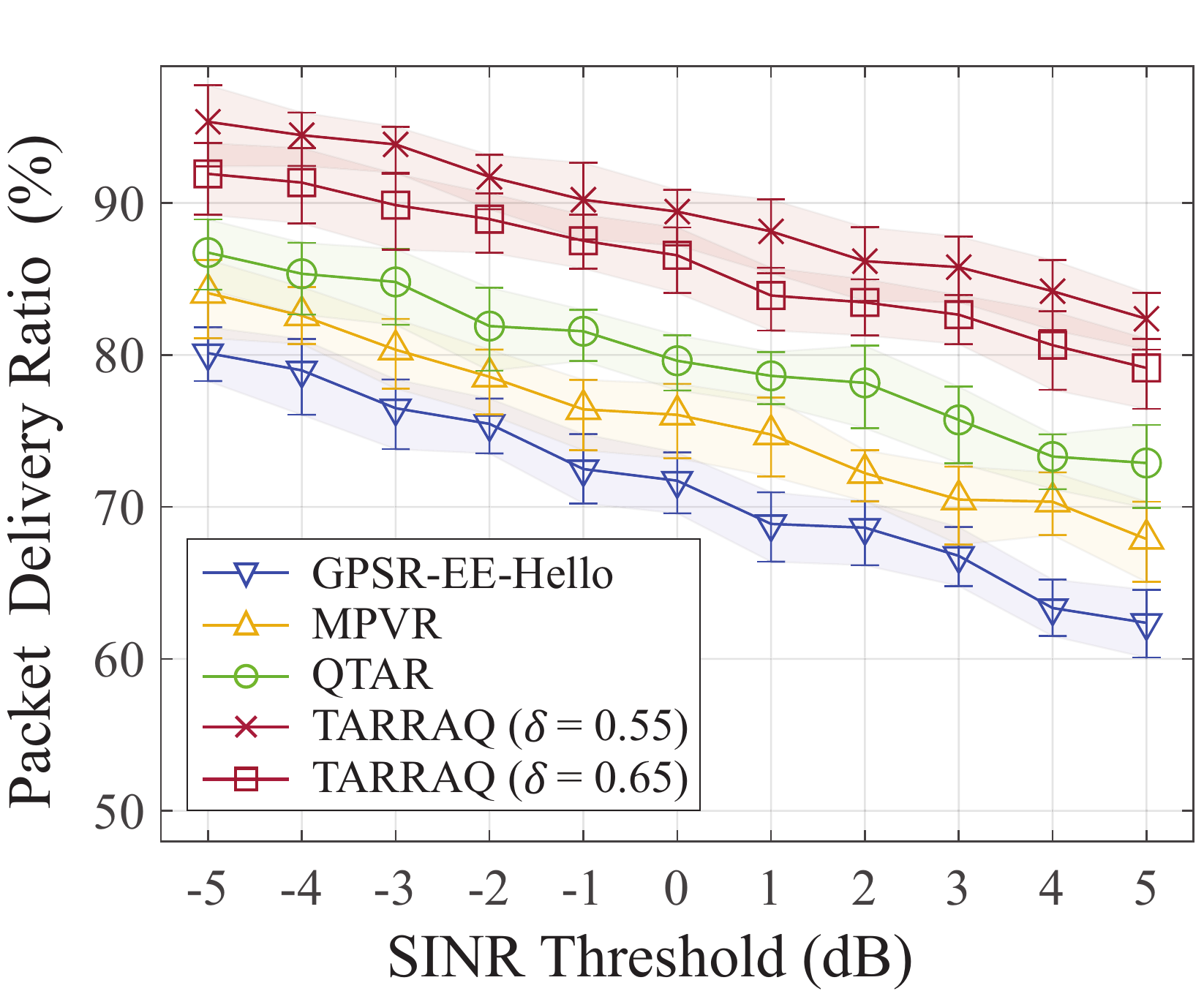}}
	\subfigure[End-to-end delay]{\includegraphics[width=4.45cm]{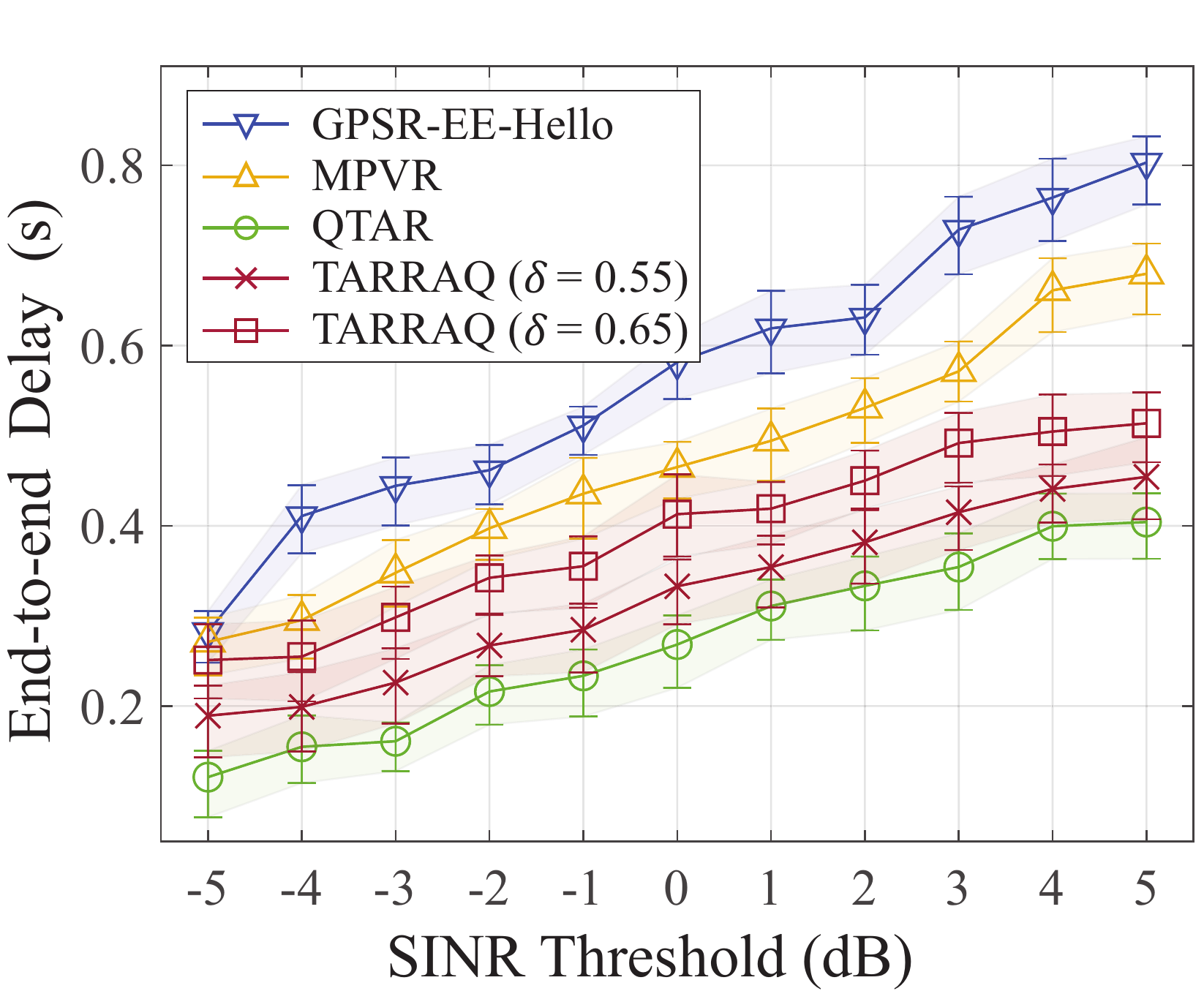}}
	\subfigure[Overhead]{\includegraphics[width=4.45cm]{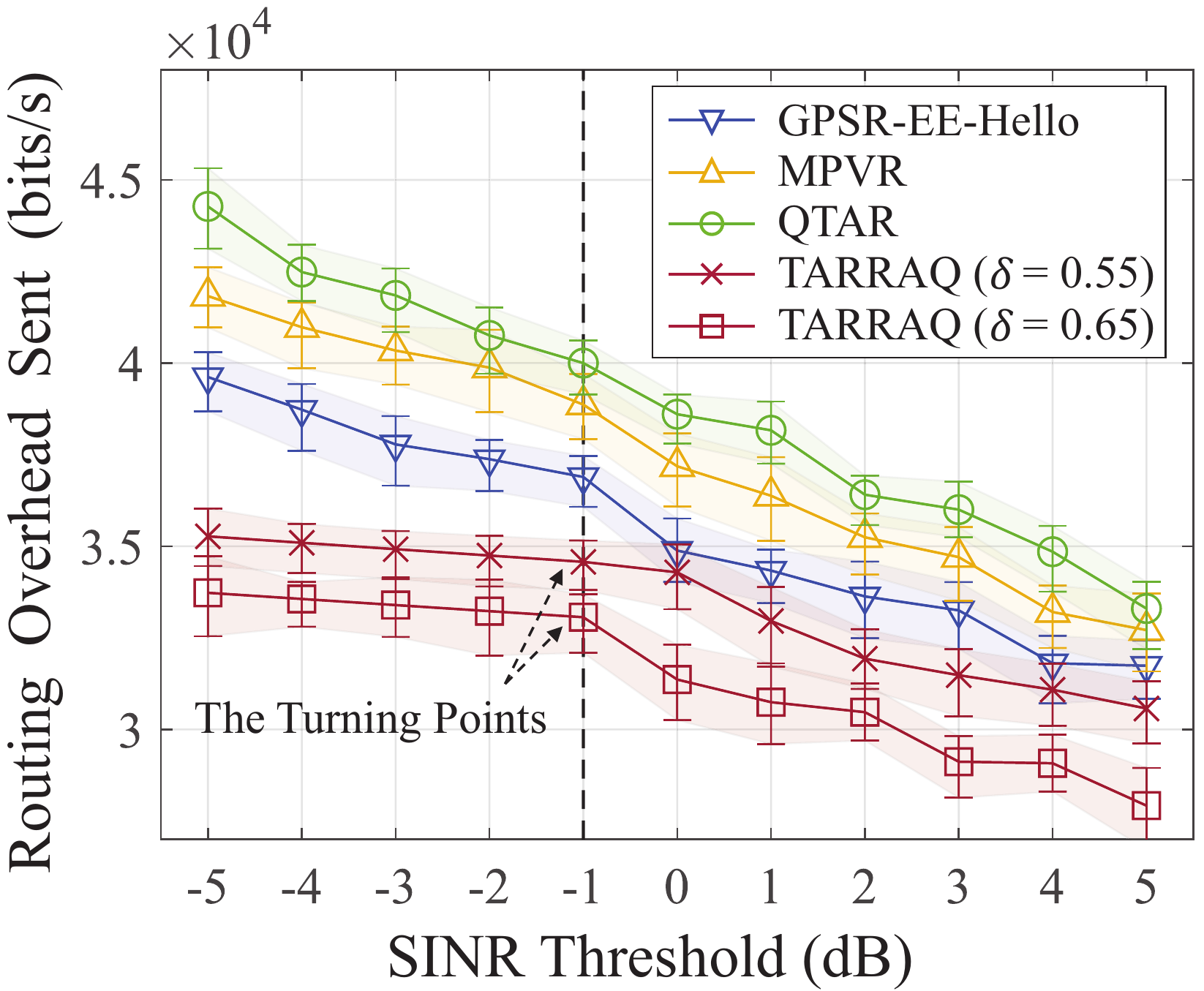}}
	\subfigure[Energy consumption]{\includegraphics[width=4.45cm]{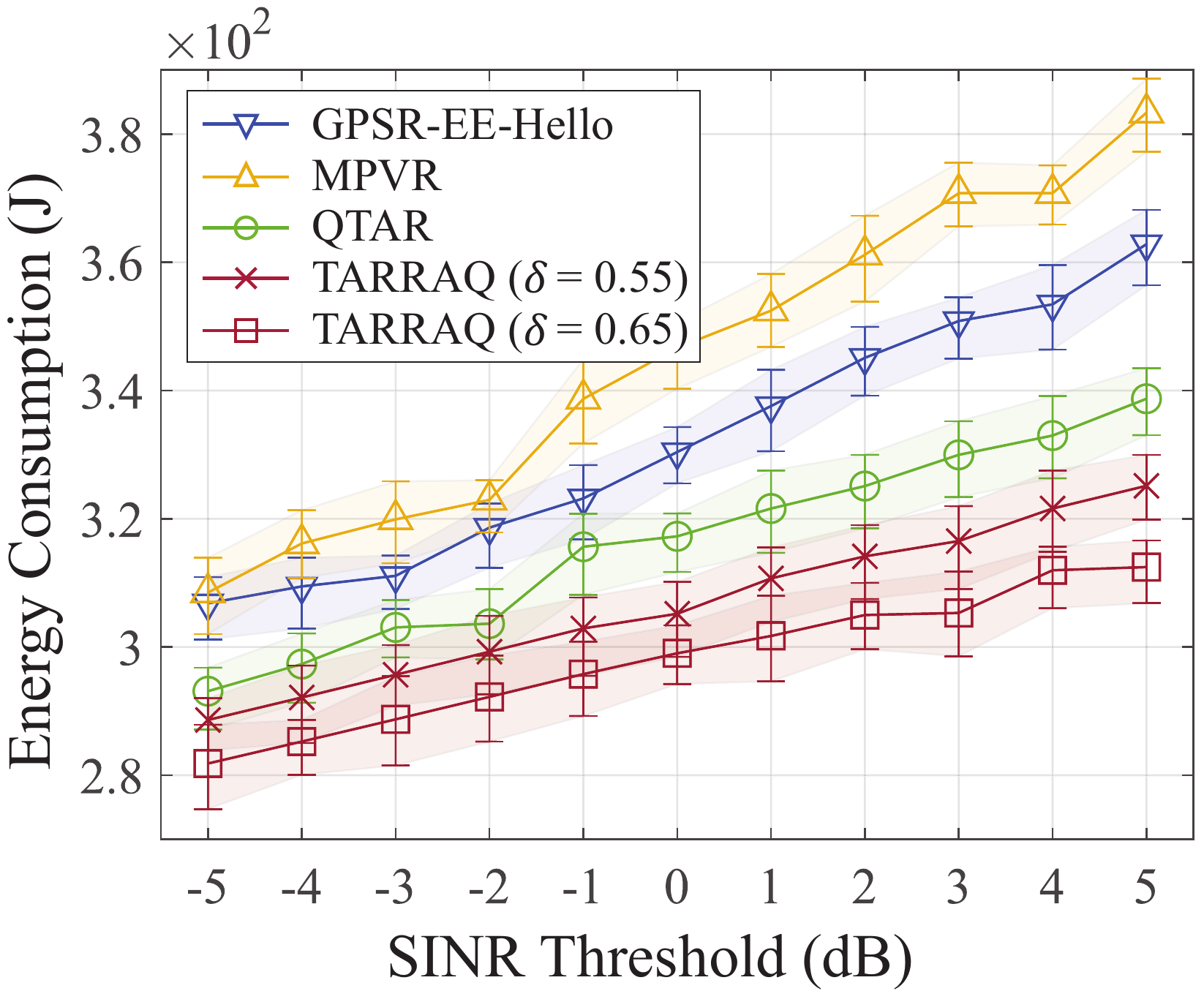}}
	\caption{Comparison of the routing performance under different SINR thresholds (speed: 5-20 m/s).}  \label{SINR_Sim}
\end{figure*}

\begin{figure*}
	\centering
	\subfigure[Packet delivery ratio]{\includegraphics[width=4.45cm]{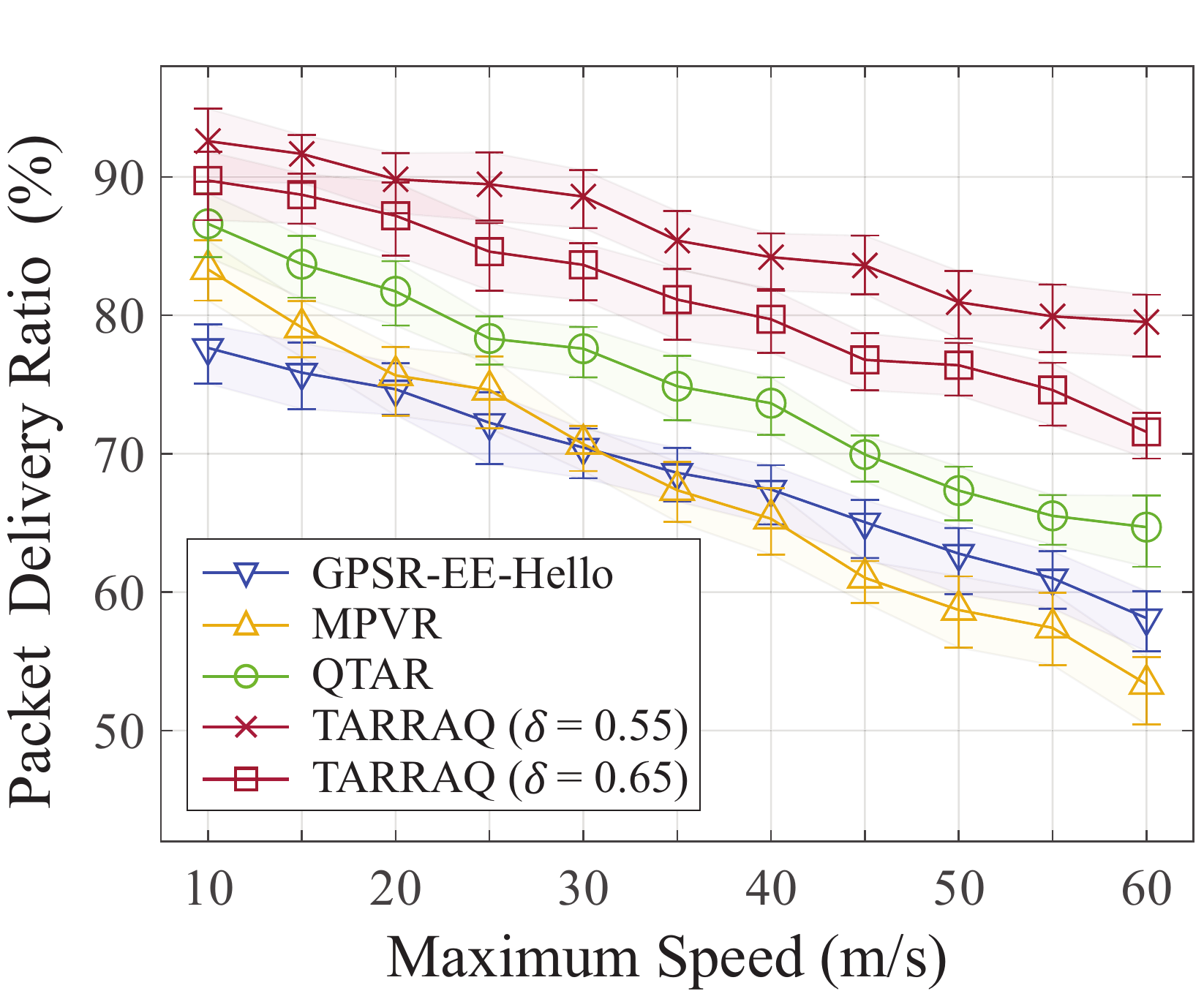}}
	\subfigure[End-to-end delay]{\includegraphics[width=4.45cm]{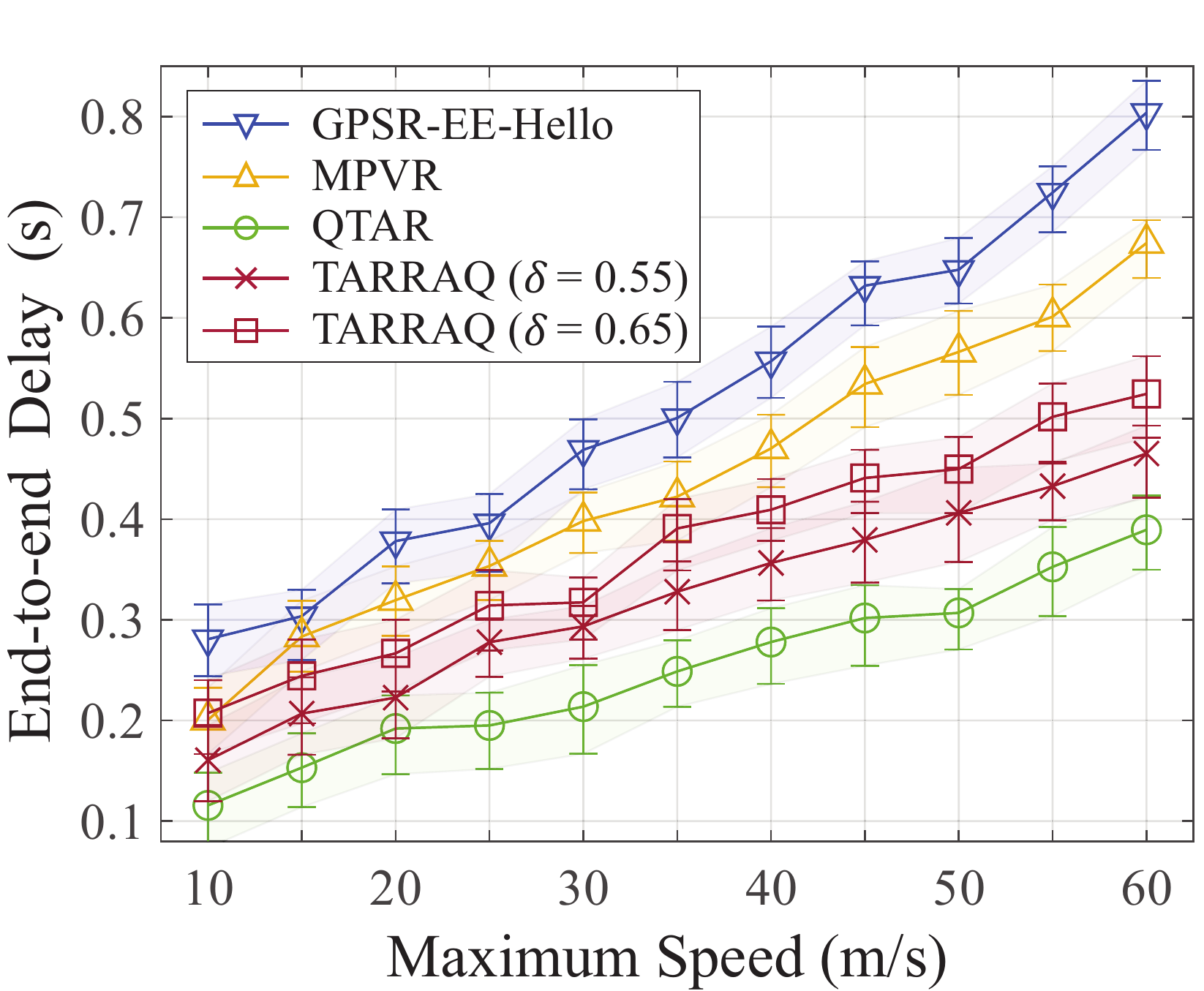}}
	\subfigure[Overhead]{\includegraphics[width=4.45cm]{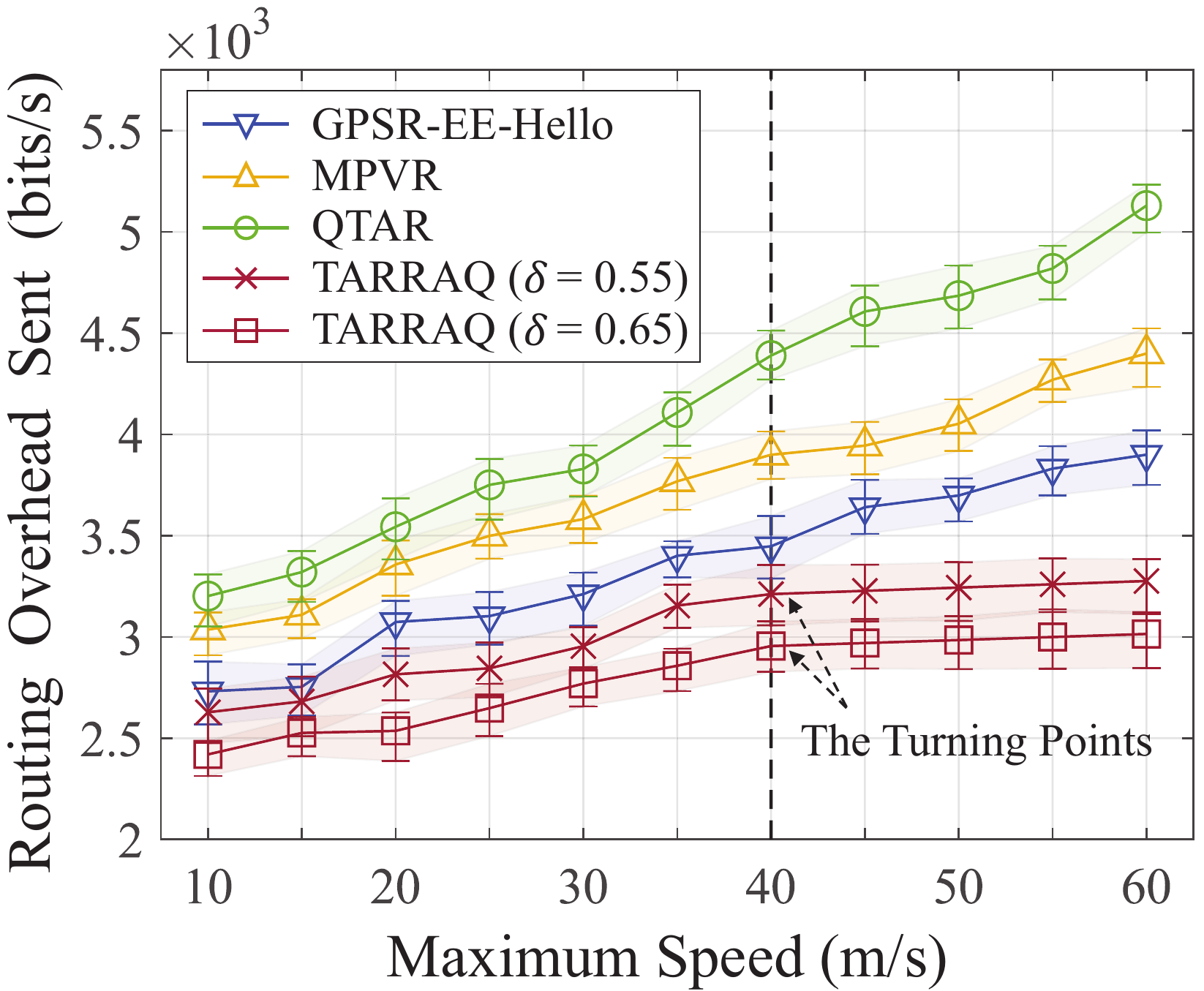}}
	\subfigure[Energy consumption]{\includegraphics[width=4.45cm]{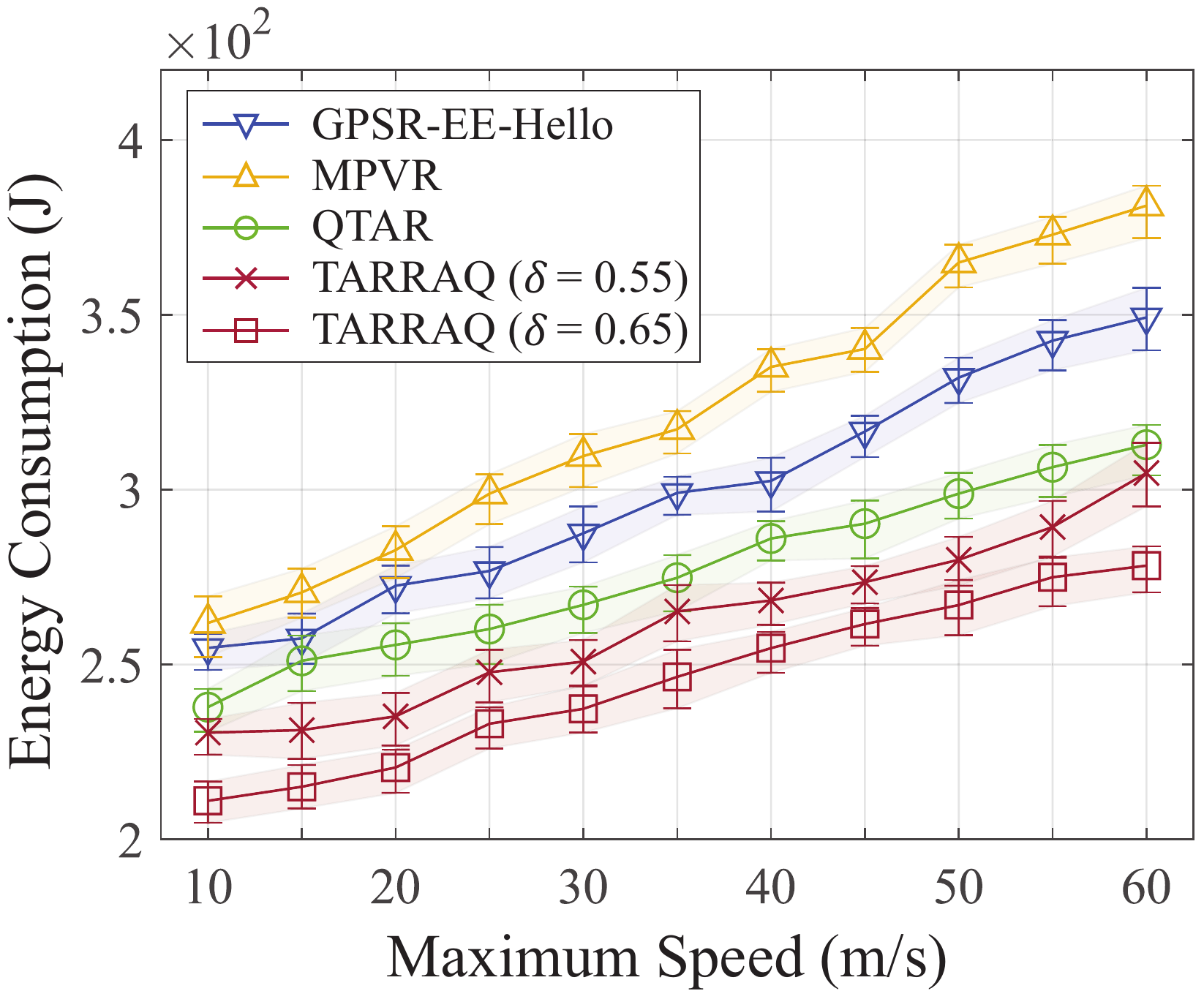}}
	\caption{Comparison of the routing performance under different moving speeds (SINR threshold: -3 dB).}  \label{Velocity_Sim}
\end{figure*}

As shown in \textbf{Fig. \ref{SINR_Sim}}, there will be better overheads whereas worse PDR, E2ED\footnote{The cache time caused by network fragmentation is excluded in E2ED.} and energy consumption with the increase of SINR threshold $\gamma_{th}$, owing to a lower successful probability and smaller effective range of transmission. 

\textbf{Fig. \ref{SINR_Sim}(a)} presents that PDR becomes lower with a larger $\gamma_{th}$, which is mainly caused by poor connectivity. The PDR of GPSR-EE-Hello degrades because it performs forwarding with link stability ignored. MPVR is slightly better since the LD is considered additionally. QTAR has a better performance than MPVR and GPSR-EE-Hello owing to the two-hop link information used in the reward function. However, our TARRAQ shows a significant advantage in PDR thanks to the KF prediction for residual LD and the multi metrics considered in the reward function, which enables the protocol to be aware of broken links in advance.

\textbf{Fig. \ref{SINR_Sim}(b)} illustrates that E2ED increases with a larger $\gamma_{th}$, due to the reduction of $R$ and the number of available neighbors, which probably causes detour. The worst E2ED occur on GPSR-EE-Hello owing to the perimeter forwarding mechanism used. MPVR preferentially selects the nodes close to the virtual trunk for the next hop to ensure fewer hops, thus its performance is better than that of GPSR-EE-Hello. However, they are inferior to our TARRAQ. For one thing, the topology change analysis based on queuing theory provides a theoretical basis for SI, which enables TARRAQ to capture neighbor change accurately and yields shorter routes. For another, the metrics of neighbor and distance in the reward function ensure fewer hops and reasonable distances to forwarding. Whereas, it creates more E2ED than QTAR since the latter considers the two-hop neighbors and the UAV's velocity to meet the packet delivery deadline. However, our TARRAQ has a significant advantage in PDR, overhead and energy consumption over QTAR since the residual LD is predicted and the resilient perception is achieved. Therefore, the E2ED increases to ensure the link quality, and thus a stable path with more hops may be selected by TARRAQ.

\textbf{Fig. \ref{SINR_Sim}(c)} shows that the overhead is inversely proportional to $\gamma_{th}$, because a larger $\gamma_{th}$ means a smaller $R$, thus fewer control packets will be exchanged. TARRAQ has an absolute advantage over the other three protocols, since that the closed-form solutions of LD, NCR and NCIT provide a rigorous theory for the preset of SI, which can perceive the network environment accurately with low overhead. GPSR-EE-Hello is slightly inferior due to the empirical SI calculation method. MPVR needs more overhead since the nodes are configured with fixed SI. QTAR has the worst overhead even though dynamic SI is preset, which is owing to a large number of control packets during two-hop interaction.

As indicated in \textbf{Fig. \ref{SINR_Sim}(d)}, as the SINR threshold increases, there will be more energy consumption since the number of hops and forwarding soared as the effective transmission range shrinks. MPVR has the highest energy consumption due to a large number of control packets. Although the GPSR-EE-Hello has a significant advantage of overhead over MPVR, its energy consumption is only slightly better since that there will be more transmission under the perimeter forwarding mechanism. Note that QTAR has fewer hops whereas worse overheads, it has superiority than GPSR-EE-Hello in terms of energy consumption. Our TARRAQ performs better than the other protocols owing to the lowest overhead and the distance metric in the reward function, which ensures fewer forwarding when transmitting traffic packets.

\textit{2) Comparison Under Different Moving Speeds}

In addition to the SINR threshold mentioned above, the UAV's speed is another key affecting the routing performance. As demonstrated in \textbf{Fig. \ref{Velocity_Sim}}, the overhead, PDR, E2ED and energy consumption deteriorate sharply as speed increases.

\textbf{Fig. \ref{Velocity_Sim}(a)} illustrates the impact of moving speed on PDR, that is, all protocols have worse PDR at higher mobility due to the frequent link disconnection. The PDR of GPSR-EE-Hello is the worst among the four protocols, while MPVR is slightly better since it considers link stability. However, as the speed increases, GPSR-EE-Hello is improved owing to the accurate information of neighbors obtained by a smaller SI. Thanks to the consideration of two-hop neighbor information in Hello messages, QTAR has a higher PDR than MPVR and GPSR-EE-Hello at both low and high speeds. Given that increasing speed makes a worse link condition, the KF prediction of residual LD enables a better link quality with fewer control packets. Thus TARRAQ presents the best PDR performance.

It can be seen from \textbf{Fig. \ref{Velocity_Sim}(b)} that there will be a worse E2ED as the moving speed increases owing to the unstable links, which cause more routing errors. Compared to GPSR-EE-Hello and MPVR, Our TARRAQ shows a notable performance since it selects the path with a lower E2ED in two ways: 1) The neighbor with lower NCR and more useful options will be preferred as a relay to avoid local minimum and high delay; 2) The adaptive adjustment of exploitation and exploration ensures adaptation of routing decisions in high mobility environment. Unfortunately, TARRAQ creates more E2ED than QTAR since the former prefers link stability and energy efficiency while the latter pays more attention to E2ED, which can be seen from the required and offered velocity it defined. Nevertheless, our TARRAQ performs better than QTAR in terms of PDR, overhead and energy efficiency owing to the cost of E2ED.

\textbf{Fig. \ref{Velocity_Sim}(c)} indicates that as the speed increases, there will be a larger NCR and more link errors, and thus the number of control packets namely overhead increases. Two-hop neighbor information increases the packet header size, and thus QTAR has the highest overhead. MPVR needs to transmit about two times more control packets than GPSR-EE-Hello since the latter adjusts SI according to node's mobility. Obviously, the resilient perception strategy in TARRAQ provides excellent performance in terms of overhead. That is, the node can calculate NCR accurately and preset appropriate SI based on network condition and performance demand. In addition, the KF procedure can accurately predict the residual LD and reduce the link error, thus a lower overhead appears.

\textbf{Fig. \ref{Velocity_Sim}(d)} proves that the energy consumption is presented as a function of UAV speeds. When UAVs are moving at high speed, more Hello messages will be exchanged to discover the neighbor's change and more transmission may be performed to ensure the successful routing, namely more energy will be consumed. In addition to the energy optimization obtained by low overhead, TARRAQ also benefits from the neighbor metric of the reward function. Besides, the NCR is calculated based on DEWMA to adapt to rapid changes. QTAR presents a worse performance than TARRAQ owing to the non-optimal SI and the transmission of two-hop messages. GPSR-EE-Hello consumes more energy than QTAR since that there may be more transmission caused by the perimeter forwarding mechanism or route failure when nodes move rapidly. MPVR performs the worst since it requires a large number of control packets when topology changes rapidly.

\textit{3) Discussion on Routing Performance}

MPVR: Based on the default sensing scheme, the Hello interval in MPVR is uniformly distributed in [0.5SI,1.5SI] and the expiration timer is three times the maximum Hello interval. All nodes broadcast Hello messages at a fixed interval regardless of the network density, speed and transmission range. Thus, it can only maintain a good result when link change occurs approximately once per second since SI$=$1000 ms, and the overheads may be out of control owing to the unnecessary Hello messages exchanged in low dynamic networks, and the frequent routing errors will appear in high dynamic cases. According to statistics, the NCR varies from 0.1 to 5, which means that the best path for fast-moving UAVs is difficult to find since the neighbor changes may not be captured accurately. And it also results in higher E2ED, energy consumption and lower PDR, and the overhead will also increase with the communication links frequently disconnected. 

GPSR-EE-Hello: The GPSR-EE-Hello is worse than our TARRAQ in all aspects, although the low overhead is known as the specialty of EE-Hello. The SI is about 2 to 5 times larger than the actual NCIT according to statistics, and the greater the upper limit of speed, the greater the difference. Even if the number of Hello messages is reduced owing to the deliberately increased SI, a large number of valuable links information will be lost, which results in more route errors. That is the reason why its energy consumption is worse than QTAR even though the Hello messages are reduced. Due to the perimeter forwarding mechanism of GPSR, the number of hops may increase and accordingly, resulting in poor E2ED. Note that the PDR of GPSR-EE-Hello gradually becomes better than that of MPVR as the speed increases, since it adjusts SI to ensure more accurate neighbor information. However, the residual LD, which is a crucial factor to ensure a better PDR in a rapidly changing network, is not considered, thus its PDR is worse than QTAR and our TARRAQ.

QTAR: As proved in \cite{QTAR}, QTAR indeed has an overwhelming E2ED advantage owing to the consideration of required velocity and velocity offered by the two-hop potential forwarding pairs. Besides, The simulation results demonstrate that QTAR outperforms GPSR-EE-Hello and MPVR in all metrics except overhead since it works via the two-hop neighbor information regardless of scenarios. However, there is still a little gap with our TARRAQ in terms of PDR and energy consumption. As inherited from \cite{Hong}, QTAR presents an intuitive approach for calculating LD and SI, which is designed for UAV swarm networks rather than FANETs. It may not apply to FANETs since the topology changes faster. Besides, there will be a large calculation error of LD and SI once the displacement of two nodes is large yet their relative distance changes little, and thus resulting in an inaccurate neighbor relationship and poor PDR. In addition, LD is ignored when designing the reward function, which means that the relay will be selected without considering the path duration, and thus a poor PDR appears.

TARRAQ: Our protocol shows the best performance advantages in terms of PDR, energy consumption and overhead, and is only slightly worse than QTAR in E2ED.
First of all, the topology dynamic analysis model based on queuing theory in section \uppercase\expandafter{\romannumeral4} guides the proper setting of SI, and the coupling relationship between NCR and dynamic features is described accurately. Thus a resilient SI is determined and the lowest overhead can be achieved owing to the proposed expected sensing delay, which means that the most accurate neighbor information can be obtained with the lowest overhead.
Besides, thanks to the service duration time derived in (\ref{Whole LD}), the accurate residual LD can be predicted by KF procedure via (\ref{Residual LD}) and (\ref{LD Elapsed}), which provides an expiration timer for available links and ensures a better PDR than other protocols.
Furthermore, the predicted residual LD is working as a link metric to guarantee stable and reliable links, and the node that has more useful neighbors and lower NCR is preferred to avoid frequent link disconnection. By giving more rewards to the neighbors who are closer to the destination and at the proper location, our TARRAQ achieves a lower energy consumption owing to the limited forwarding distance.

It is worth emphasizing that TARRAQ has strong adaptability in a rapidly changing environment. The DEWMA process of NCR not only guarantees sensitivity to dynamic topology but also prevents nodes from making abnormal decisions due to fluctuations in network conditions, thus TARRAQ can adapt to changes in SINR threshold and speed. More importantly, the self-adaptation ability is also reflected in different traffic and topology changes, which is intuitively shown by the turning points marked in \textbf{Fig. \ref{SINR_Sim}} and \textbf{\ref{Velocity_Sim}}. Specifically, NCR will be greater than TAR once the SINR threshold drops or speed increases to a certain value, and the minimum of NCR and TAR is used to determine SI. Thus the overhead and energy degradation of TARRAQ is the slowest since it undoubtedly reduces the spread of useless Hello messages. In addition, the exploitation and exploration scheme of Q-learning is adaptively adjusted based on the estimated residual LD, which improves the sensitivity of routing decisions to the rapidly changing environment.

As one of the most crucial contributions, the on-demand performance of TARRAQ is verified by simulation results. The different expected sensing delays are realized by various $\delta$. A smaller overhead with inferior link accuracy is achieved via lager $\delta$, and vice versa. Thus, it provides a method to achieve resilient routing performance. The two red curves in \textbf{Fig. \ref{SINR_Sim}} and \textbf{\ref{Velocity_Sim}} demonstrate that the TARRAQ with $\delta = 0.55$ has a poor overhead and energy level, while the PDR and E2ED are improved. Conversely, the TARRAQ with $\delta=0.65$ sacrifices PDR and E2ED in exchange for overhead and energy efficiency. No matter what kind of performance is compromised, our TARRAQ is better than the comparison schemes on most metrics. The average comparison results are presented in Table \textbf{\ref{Improvement}}. 

\begin{table}
	\centering
	\caption{The performance improvement of TARRAQ when compared with GPSR-EE-Hello, MPVR and QTAR.}
	\renewcommand{\arraystretch}{1}
	\begin{tabular}{m{1.9cm}<{\centering}|m{1.15cm}<{\centering}m{1cm}<{\centering}m{1cm}<{\centering}m{1.6cm}<{\centering}}
		\toprule[1pt]\toprule[0.5pt]
		Protocol & Overhead & E2ED  & PDR & Energy Consumption\\
		\midrule
		GPSR-EE-Hello  & 13.73\% & 40.32\% & 16.70\% & 11.31\% \\
		MPVR  & 20.24\% & 28.72\% & 14.77\% & 15.65\% \\
		QTAR & 25.23\% & -22.57\% & 9.41\% & 5.12\% \\
		\bottomrule[0.5pt]\bottomrule[1pt]
	\end{tabular}%
	\label{Improvement}
\end{table}%

\subsection{Computation Complexity Comparison}

As shown in Table \textbf{\ref{Computation Complexity}}, we present a comparative analysis of the computation complexity by comparing TARRAQ to other state-of-art protocols. For each UAV of the active route, the best decision should be made from $N$ available neighbors by each round. During the neighbor discovery and maintenance phase, it should be emphasized that only (\ref{Expected_Sensing_Delay}) and (\ref {LD Elapsed}) need to be calculated additionally in our TARRAQ. Thus a complexity of $O(N)$ is necessary, which can be considered the same as other algorithms except for QTAR, since it considers two-hop neighbors and has a complexity of $O(N^2)$. During the relay selection phase, the MPVR and GPSR-EE-Hello choose the best hop among $N$ candidate relays according to a certain criterion, thus $O(N)$ appears. For QTAR and our TARRAQ, $K_{max}$ iterations should be performed in the worst case, which is generally larger than MPVR and GPSR-EE-Hello since there are probably hundreds of iterations for learning before convergence. But it is affordable and worthwhile since the learning-based decentralized methods that adapt to dynamic networks without the need for global knowledge are more suitable for UAV networks. Overall, our protocol constantly adapts to abrupt changes hence leading to a higher PDR and efficiency as well as maximal connectivity. Therefore, compared with MPVR and GPSR-EE-Hello, the slightly larger complexity of TARRAQ is justified by the increased network performance. Besides, it provides better performances while having a lower complexity with respect to QTAR.

\begin{table}
	\centering
	\caption{The comparison of computation complexity.}
	\renewcommand{\arraystretch}{1}
	\begin{tabular}{m{1.4cm}<{\centering}|m{1.8cm}<{\centering}||m{1.9cm}<{\centering}|m{2cm}<{\centering}}
		\toprule[1pt]\toprule[0.5pt]
		Protocol & Complexity & Protocol & Complexity \\
		\midrule[0.5pt]
		TARRAQ & $O(N+K_{max})$ & QTAR & $O(N^2+K_{max})$ \\
		MPVR & $O(2N)$ & GPSR-EE-Hello & $O(2N)$ \\
		\bottomrule[0.5pt]\bottomrule[1pt]
	\end{tabular}
	\label{Computation Complexity}
\end{table}

\section{Conclusion}
In this work, we propose a novel protocol called TARRAQ to accurately capture topology changes with the lowest overhead and make routing decisions in a distributed and autonomous way. To reveal the mapping relationship between NCR and dynamic behavior of UAVs, we analyze the topology change characteristics of FANETs via queuing theory, and the closed-form solutions of NCR and neighbor change inter-arrival time distribution are derived. Based on this, TARRAQ defines the expected sensing delay, performs a DEWMA procedure on the real-time NCR, and determines the validity period of links by predicting the residual link duration. The proposed TARRAQ can make distributed and autonomous routing decisions via an adaptive Q-learning approach, where the reward function is designed to find a stable path. It can also achieve adaptive learning from the variable network environment owing to the dynamic adjustment of action selection, learning rate and discount factor.

As our future work, we will extend TARRAQ to various architectures and scenarios, e.g. UAV swarm networks, and pay more attention to the traffic and aerial channel characteristics. In addition, designing different queuing models and RL techniques (DQN and DDPG, etc.) for more UAV networks (cluster-based and multi-layer networks, etc.) will be another interesting future work.
\appendices
\section{Joint Probability Density of Relative Speed}
Here, we derive the joint PDF of $v$, $\alpha_{\boldsymbol{v}}$ and $\beta_{\boldsymbol{v}}$, namely $f\left(v,\alpha_{\boldsymbol{v}},\beta_{\boldsymbol{v}}\right)$ for the UAVs that enter $\Omega_c$, as illustrated in \textbf{Fig. \ref{Sphere}(a)}. The PDF of moving direction and speed of $S_o$ are given by $f_{\beta_{\boldsymbol{v_o}}}\left(\beta_{\boldsymbol{v_o}}\right) = f_{\alpha_{\boldsymbol{v_o}}}\left(\alpha_{\boldsymbol{v_o}}\right)=1/{2\pi}$ and
\begin{equation}\label{V_PDF}
	f_{v_o}\left(v_o\right) = \frac{u\left(v_o-v_l\right)-u\left(v_o-v_u\right)}{v_u-v_l}.
\end{equation}
Thus, their joint PDF is given by
\begin{equation}\label{Vo_angle_JPDF}
	f_{v_o,\beta_{\boldsymbol{v_o}},\alpha_{\boldsymbol{v_o}}}\left(v_o,\beta_{\boldsymbol{v_o}},\alpha_{\boldsymbol{v_o}}\right)=\frac{u\left(v_o-v_l\right)-u\left(v_o-v_u\right)}{4\pi^2\left(v_u-v_l\right)},
\end{equation}
and the joint PDF of $v$, $\alpha_{\boldsymbol{v}}$ and $\beta_{\boldsymbol{v}}$ can be calculated as
\begin{equation}\label{V_angle_JPDF}
	f_{v,\alpha_{\boldsymbol{v}},\beta_{\boldsymbol{v}}}\left(v,\alpha_{\boldsymbol{v}},\beta_{\boldsymbol{v}}\right)=\frac{f_{v_o,\beta_{\boldsymbol{v_o}},\alpha_{\boldsymbol{v_o}}}\left(v_o,\beta_{\boldsymbol{v_o}},\alpha_{\boldsymbol{v_o}}\right)}{\left|J\left(v_o,\beta_{\boldsymbol{v_o}},\alpha_{\boldsymbol{v_o}}\right)\right|},
\end{equation}
where $J\left(v_o,\beta_{\boldsymbol{v_o}},\alpha_{\boldsymbol{v_o}}\right)$ is the Jacobian matrix, and its determinant is calculated by
\begin{equation}\label{Jacobian}
	\begin{aligned}
		\left|J\left(v_o,\beta_{\boldsymbol{v_o}},\alpha_{\boldsymbol{v_o}}\right)\right|
		&=\left|\frac{\partial\left(v,\beta_{\boldsymbol{v}},\alpha_{\boldsymbol{v}}\right)}{\partial\left(v_o,\beta_{\boldsymbol{v_o}},\alpha_{\boldsymbol{v_o}}\right)}\right|\vspace{1ex}
		\\&=\frac{v_o}{\sqrt{v_o^2+v_c^2-2v_cv_ocos\beta_{\boldsymbol{v_o}}}}.
	\end{aligned}
\end{equation}
We have $v_o=\sqrt{v^2+v_c^2+2vv_ccos\beta_{\boldsymbol{v}}}$ by solving equations (\ref{speed}) and (\ref{beta_v}), and equation (\ref{Jacobian}) can be further converted to
\begin{equation}\label{Jacobian_2}
	\begin{aligned}
		&\left|J\left(v_o,\beta_{\boldsymbol{v_o}},\alpha_{\boldsymbol{v_o}}\right)\right|=\frac{\sqrt{v^2+v_c^2+2vv_ccos\beta_{\boldsymbol{v}}}}{v}.
	\end{aligned}
\end{equation}
Thus, equation (\ref{V_angle_JPDF}) can be further simplified as
\begin{equation}\label{V_angle_JPDF_simple}
	f\left(v,\alpha_{\boldsymbol{v}},\beta_{\boldsymbol{v}}\right)=\frac{vg\left(v,v_c,\beta_{\boldsymbol{v}}\right)}{4\pi^2\left(v_u-v_l\right)},
\end{equation}
where 
\begin{equation}\label{g}
	g\left(v,v_c,\beta_{\boldsymbol{v}}\right)=\frac{u\left(h\left(v,v_c,\beta_{\boldsymbol{v}}\right)-v_l\right)-u\left(h\left(v,v_c,\beta_{\boldsymbol{v}}\right)-v_u\right)}{h\left(v,v_c,\beta_{\boldsymbol{v}}\right)},
\end{equation}
\begin{equation}\label{h}
	h\left(v,v_c,\beta_{\boldsymbol{v}}\right)=\sqrt{v^2+v_c^2+2vv_ccos\beta_{\boldsymbol{v}}},
\end{equation}
and $u\left(\cdot\right)$ is the unit step function. 

\section{The Calculation of $V_u$}
Here, we derive the volume of ${SR}_u$, namely $V_u$, in the following two cases. 

\begin{figure}
	\centering
	\subfigure[$vt\textgreater2R$.]{\includegraphics[width=4.0cm]{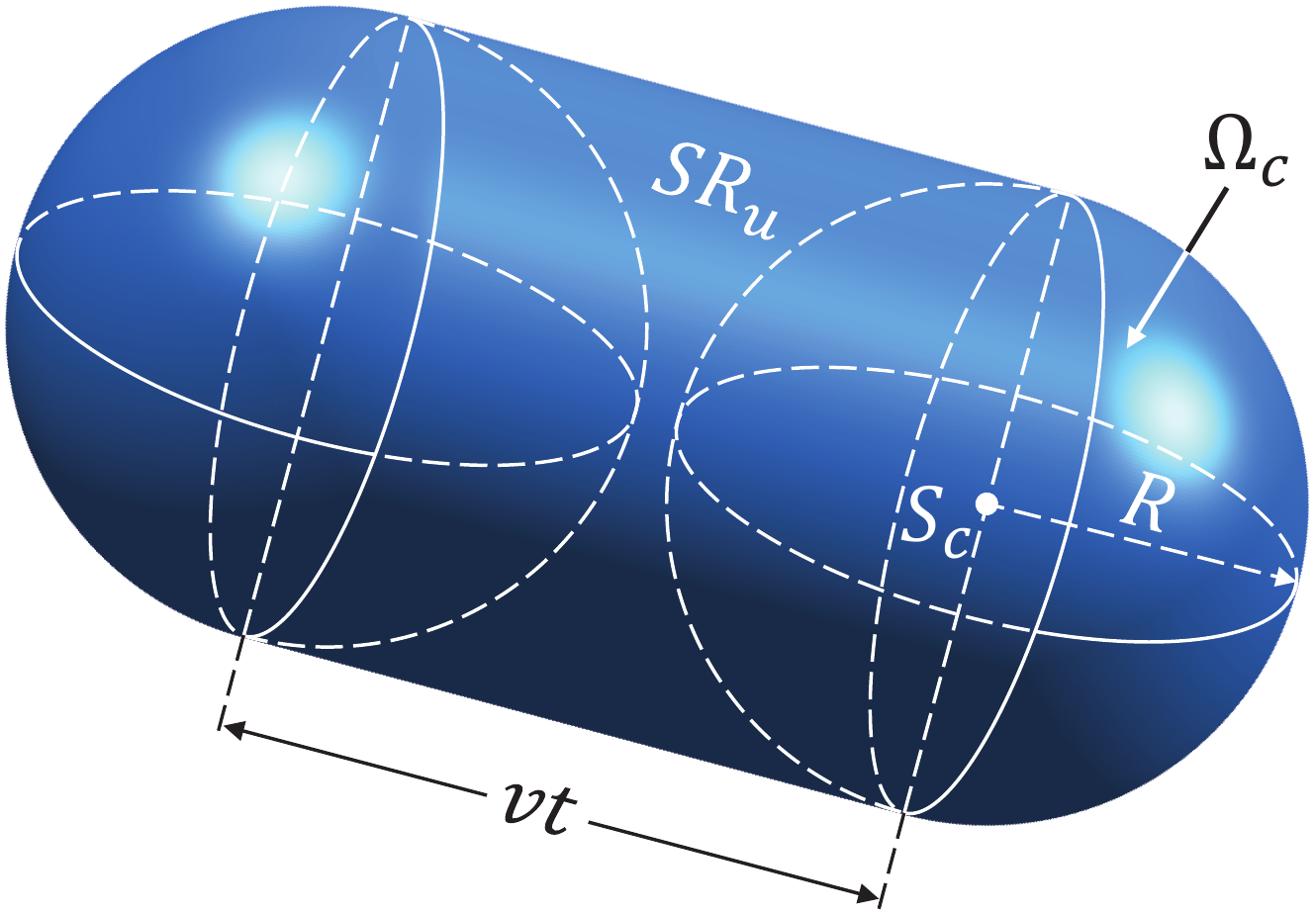}}\quad\quad
	\subfigure[$vt\leq2R$.]{\includegraphics[width=3.2cm]{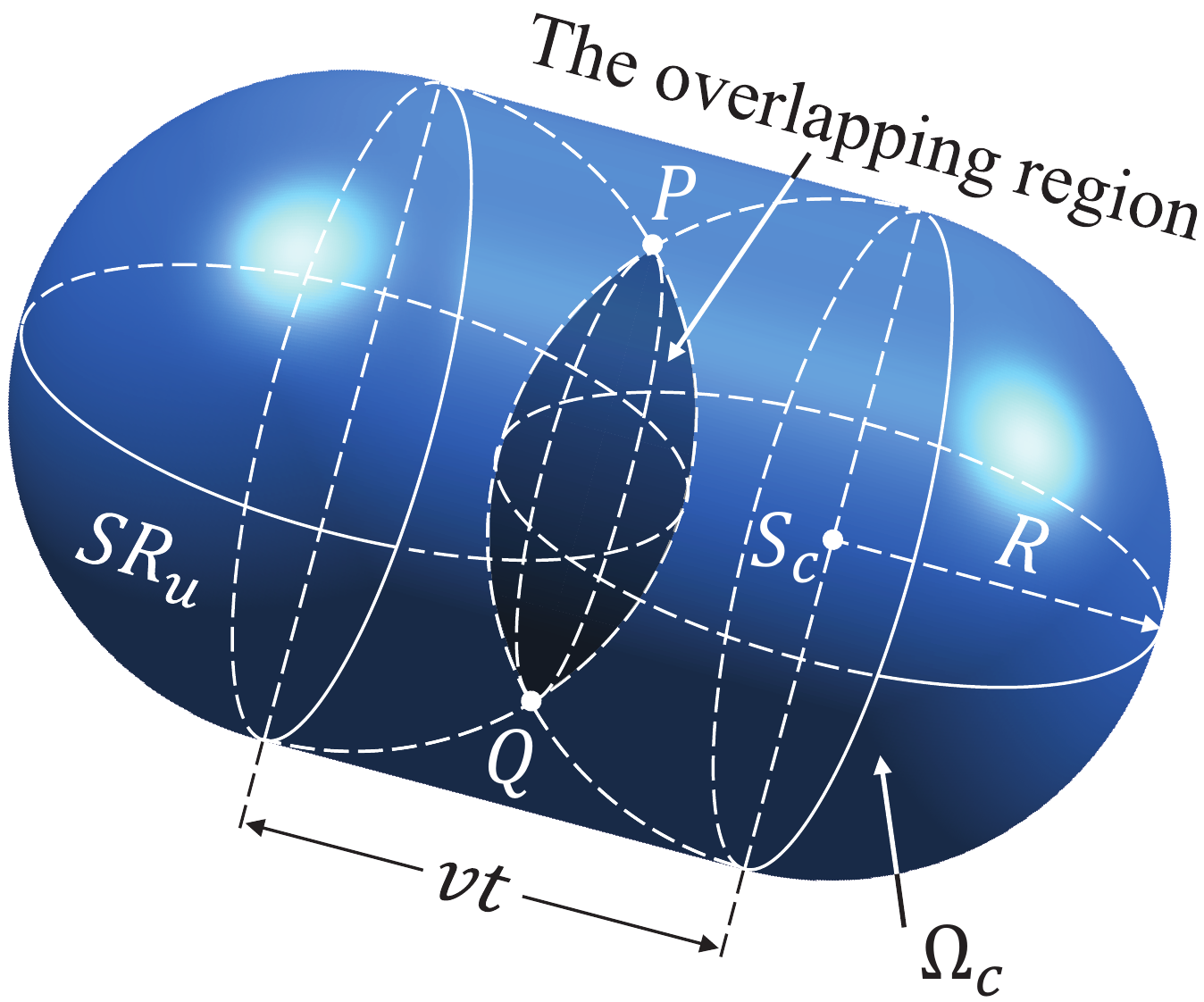}}
	\caption{The region where UAVs will enter or leave $\Omega_c$ with a velocity $\boldsymbol{v}$ within the upcoming $t$ seconds.}\label{Region_4_Change}
\end{figure}

Case \uppercase\expandafter{\romannumeral1}: As shown in \textbf{Fig. \ref{Region_4_Change}(a)}, there is no overlapping region when $vt\textgreater2R$. Hence the volume of ${SR}_u$ is given by
\begin{equation}\label{V1}
	V_2\left(v,t\right)=\frac{\pi R^2\left(4R+3vt\right)}{3}.
\end{equation}

Case \uppercase\expandafter{\romannumeral2}: As shown in \textbf{Fig. \ref{Region_4_Change}(b)}, there is an overlapping region that does not belong to ${SR}_u$, when $vt\leq2R$. Hence the volume of the overlapping region can be calculated by
\begin{equation}
	\begin{aligned}
		V_{overlap}\left(v,t\right)&=2\int_{0}^{R-\frac{vt}{2}}\pi\left[R^2-\left(\frac{vt}{2}+x\right)^2\right]dx\\
		&=\frac{2\pi}{3}\left(R-\frac{vt}{2}\right)^2\left(2R+\frac{vt}{2}\right),
	\end{aligned}
\end{equation}
and the volume of ${SR}_u$ is calculated by
\begin{equation}\label{V2}
	\begin{aligned}
		V_3\left(v,t\right)&=\frac{\pi R^2\left(4R+3vt\right)}{3}-V_{overlap}\left(v,t\right)
		\\&=\frac{\pi vt\left(24R^2-v^2t^2\right)}{12}.
	\end{aligned}
\end{equation}

\ifCLASSOPTIONcaptionsoff
	\newpage
\fi

\begin{IEEEbiography}[{\includegraphics[width=1.1in,height=1.4in,clip,keepaspectratio]{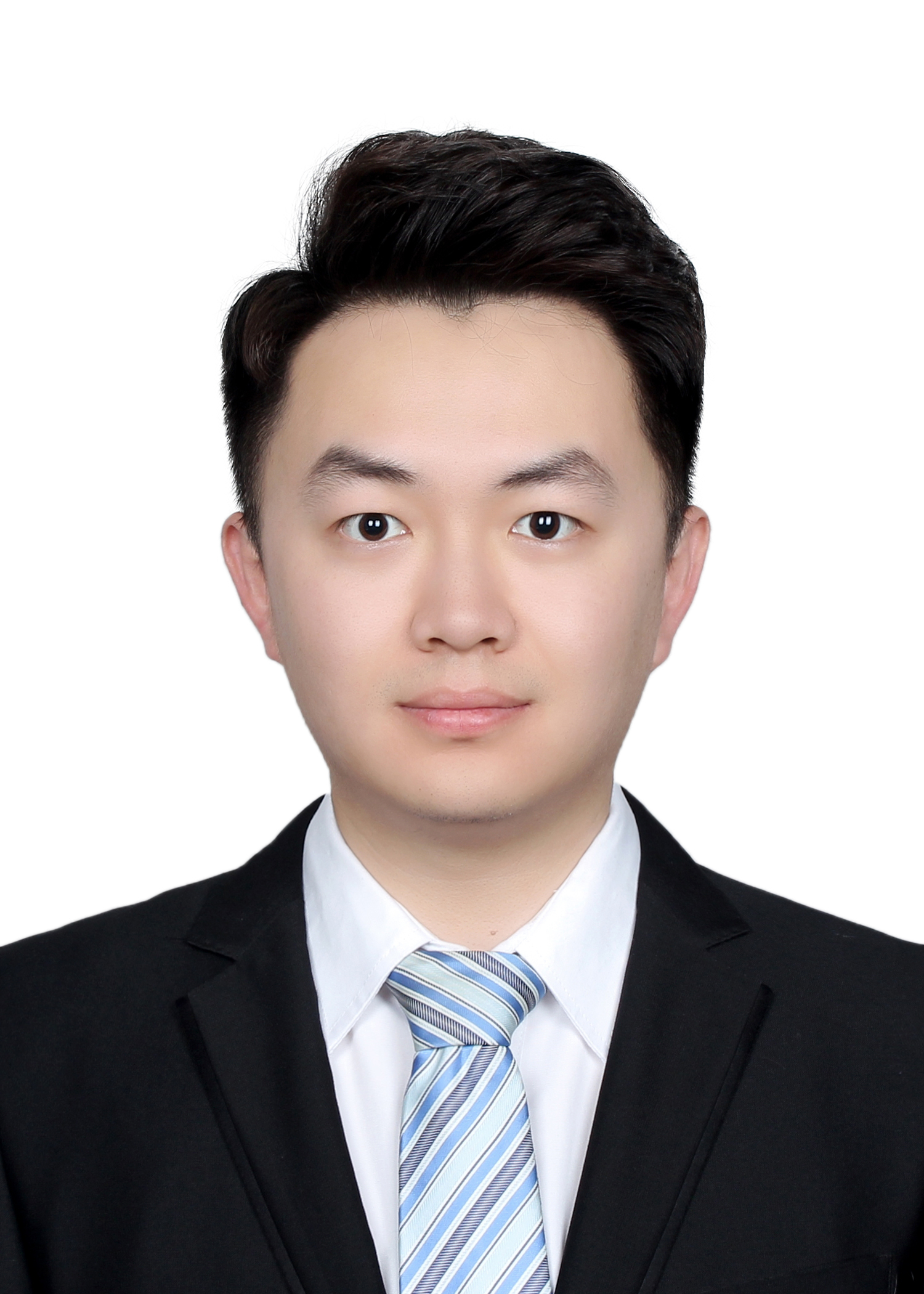}}]{Yanpeng Cui}
(S’20) received the B.S. degree from the Henan University of Technology, Zhengzhou, China, in 2016, and the M.S. degree from the Xi’an University of Posts and Telecommunications, Xi’an, China, in 2020. He is currently pursuing the Ph.D. degree with the School of Information and Communication Engineering, Beijing University of Posts and Telecommunications (BUPT), Beijing, China. His current research interests include the Flying ad hoc networks, and integrated sensing and communication for UAV networks.
\end{IEEEbiography}

\begin{IEEEbiography}[{\includegraphics[width=1.1in,height=1.4in,clip,keepaspectratio]{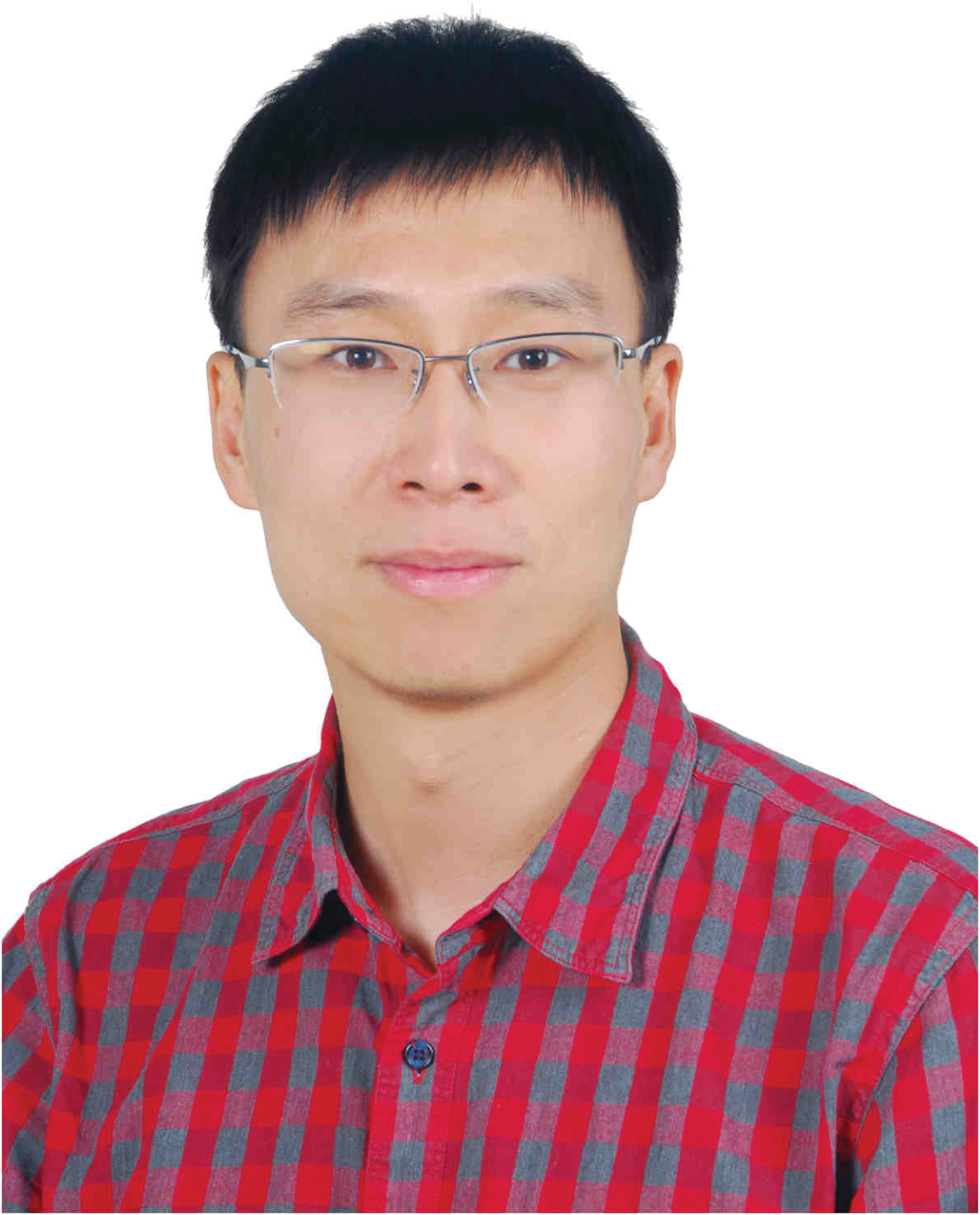}}]{Qixun Zhang}
(M’12) received the B.E. and the Ph.D. degree from BUPT, Beijing, China, in 2006 and 2011, respectively. From Mar. to Jun. 2006, he was a Visiting Scholar at the University of Maryland, College Park, Maryland. From Nov. 2018 to Nov. 2019, he was a Visiting Scholar in the Electrical and Computer Engineering Department at the University of Houston, Texas. He is a Professor with the Key Laboratory of Universal Wireless Communications, Ministry of Education, and the School of Information and Communication Engineering, BUPT. His research interests include 5G mobile communication system, integrated sensing and communication for autonomous driving vehicle, mmWave communication system, and unmanned aerial vehicles (UAVs) communication. He is a member of IEEE and active in ITU-R WP5A/5C/5D standards.
\end{IEEEbiography}

\begin{IEEEbiography}[{\includegraphics[width=1.1in,height=1.4in,clip,keepaspectratio]{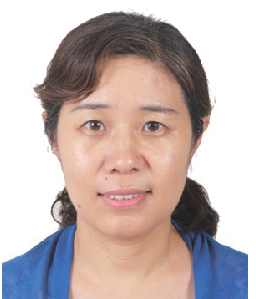}}]{Zhiyong Feng}
(M’08-SM’15) received her B.S., M.S., and Ph.D. degrees from BUPT, Beijing, China. She is a Professor with the School of Information and Communication Engineering, BUPT, and the director of the Key Laboratory of Universal Wireless Communications, Ministry of Education, China. Her research interests include wireless network architecture design and radio resource management in 5th generation mobile networks (5G), spectrum sensing and dynamic spectrum management in cognitive wireless networks, universal signal detection and identification, and network information theory. She is a senior member of IEEE and active in standards development, such as ITU-R WP5A/5C/5D, IEEE 1900, ETSI, and CCSA.
\end{IEEEbiography}

\begin{IEEEbiography}[{\includegraphics[width=1.1in,height=1.4in,clip,keepaspectratio]{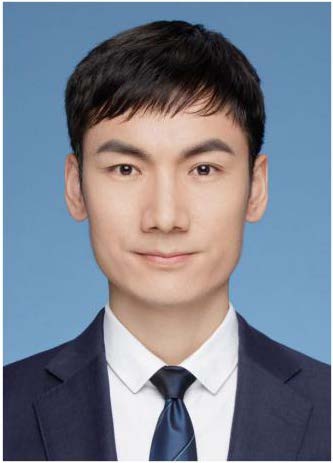}}]
{Zhiqing Wei}
(S'12-M'15) received his B.E. and Ph.D. degrees from BUPT in 2010 and 2015. Now he is an associate professor at BUPT. He was granted the Exemplary Reviewer of IEEE Wireless Communications Letters in 2017, the Best Paper Award of International Conference on Wireless Communications and Signal Processing 2018. He was the Registration Co-Chair of IEEE/CIC International Conference on Communications in China (ICCC) 2018 and the publication Co-Chair of IEEE/CIC ICCC 2019. His research interest is the performance analysis and optimization of mobile ad hoc networks.
\end{IEEEbiography}

\begin{IEEEbiography}[{\includegraphics[width=1.1in,height=1.4in,clip,keepaspectratio]{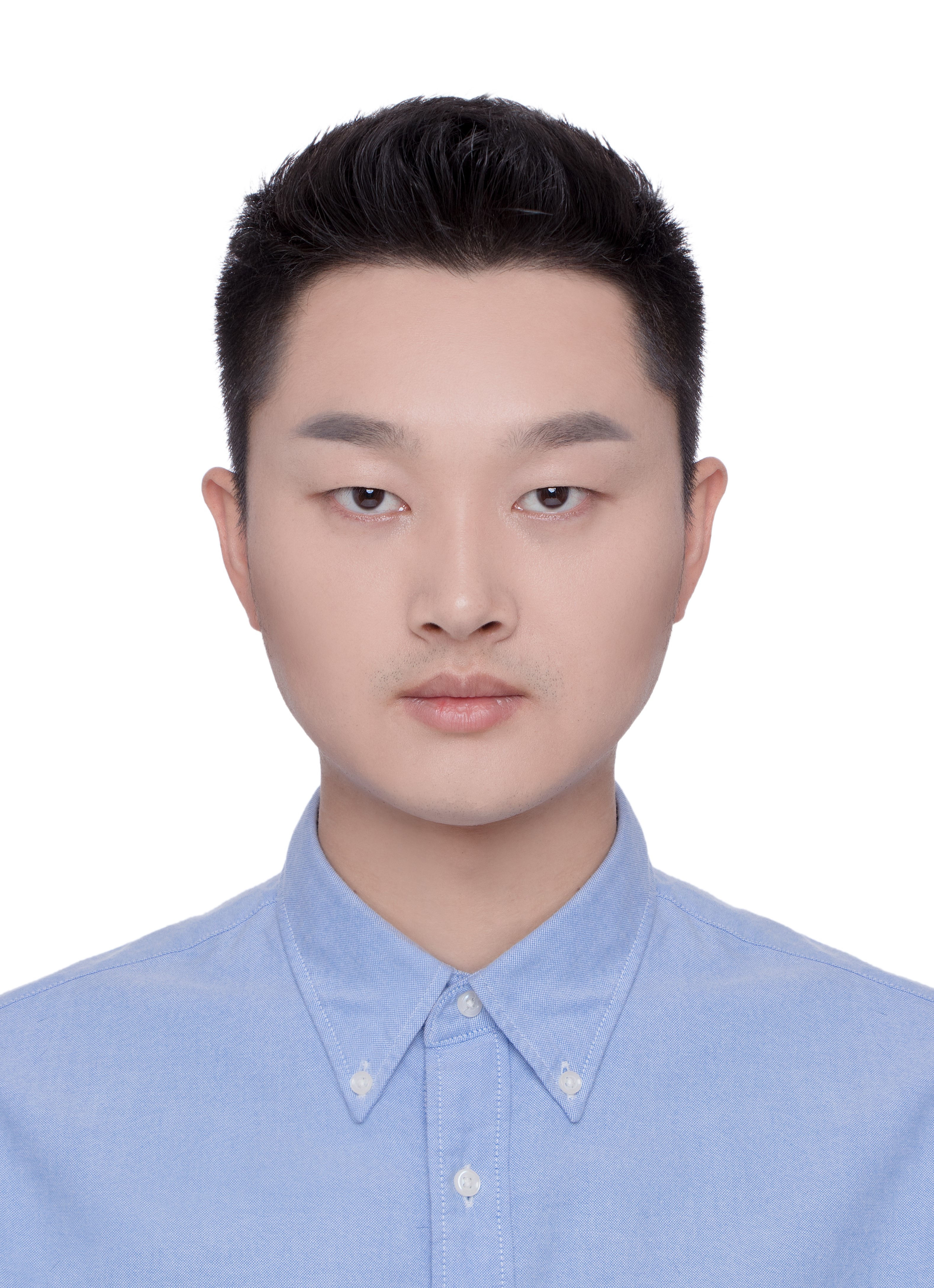}}]
{Ce Shi} received the B.S. degree from the Chongqing Technology and Business University, Chongqing, China, in 2017, and the M.S. degree from the Chongqing University of Posts and Telecommunications (CQUPT), Chongqing, China, in 2020. He is currently pursuing the Ph.D. degree with the School of Information and Communication Engineering, BUPT, Beijing, China. His current research interests include the interference elimination at the physical layer and orthogonal time frequency space modulation.
\end{IEEEbiography}

\begin{IEEEbiography}[{\includegraphics[width=1.5in,height=1.3in,clip,keepaspectratio]{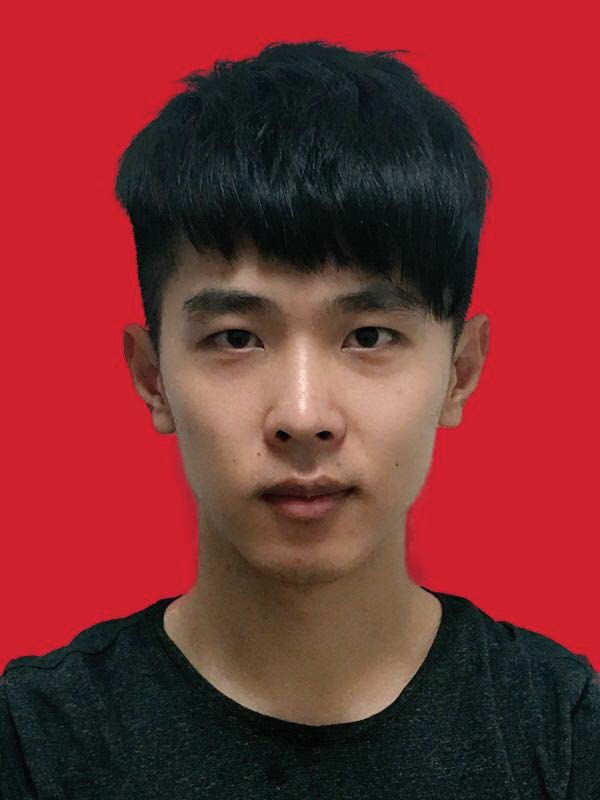}}]
{Heng Yang}
(S'20) received the B.Sc. degree and the M.S. degree from CQUPT, Chongqing, China, in 2016 and 2019. He is currently pursuing the Ph.D. degree with the School of Information and Communication Engineering, BUPT, Beijing, China. His current research interests include radio resource management, vehicular network and joint communication and sensing network.
\end{IEEEbiography}


\begin{thebibliography}{1}

\bibitem{Cui}
X. Zhao, Y. Cui, C. Gao, Z. Guo and Q. Gao, ``Energy-Efficient Coverage Enhancement Strategy for 3-D Wireless Sensor Networks Based on a Vampire Bat Optimizer,'' \emph{IEEE Internet Things J.}, vol. 7, no. 1, pp. 325-338, Jan. 2020.

\bibitem{WildFire}
O. M. Bushnaq, A. Chaaban and T. Y. Al-Naffouri, ``The Role of UAV-IoT Networks in Future Wildfire Detection,'' \emph{IEEE Internet Things J.}, vol. 8, no. 23, pp. 16984-16999, Dec. 2021.

\bibitem{Arafat}
M. Y. Arafat and S. Moh, ``Localization and Clustering Based on Swarm Intelligence in UAV Networks for Emergency Communications,'' \emph{IEEE Internet Things J.}, vol. 6, no. 5, pp. 8958-8976, Oct. 2019.

\bibitem{B. Alzahrani}
B. Alzahrani, O. S. Oubbati, A. Bernawi, A. Atiquzzaman, and D. Alghazzawi, ``UAV assistance paradigm: State-of-the-art in applications and challenges,'' \emph{J. Netw. Comput. Appl.}, vol. 166, pp. 1–44, 2020, Art. no. 102706.

\bibitem{O. S. Oubbati}	
O. S. Oubbati, M. Atiquzzaman, P. Lorenz, M. H. Tareque, and M. S. Hossain, ``Routing in flying ad hoc networks: Survey, constraints, and future challenge perspectives,'' \emph{IEEE Access}, vol. 7, pp. 81057–81105, 2019.

\bibitem{Arafat_Survey_Cluster}
M. Y. Arafat and S. Moh, ``A Survey on Cluster-Based Routing Protocols for Unmanned Aerial Vehicle Networks,'' \emph{IEEE Access}, vol. 7, pp. 498-516, 2019.

\bibitem{J. Jiang}
J. Jiang and G. Han, ``Routing Protocols for Unmanned Aerial Vehicles,'' \emph{IEEE Commun. Mag.}, vol. 56, no. 1, pp. 58-63, Jan. 2018.

\bibitem{A. Bujari}
A. Bujari, C. E. Palazzi and D. Ronzani, ``A Comparison of Stateless Position-based Packet Routing Algorithms for FANETs,'' \emph{IEEE Trans. Mobile Comput.}, vol. 17, no. 11, pp. 2468-2482, Nov. 2018.

\bibitem{Reinforcement_Survey}
R. A. Nazib and S. Moh, ``Reinforcement Learning-Based Routing Protocols for Vehicular Ad Hoc Networks: A Comparative Survey,'' \emph{IEEE Access}, vol. 9, pp. 27552-27587, 2021.

\bibitem{L. Lin}
L. Lin, Q. Sun, S. Wang and F. Yang, ``A geographic mobility prediction routing protocol for Ad Hoc UAV Network,'' in \emph{Proc. IEEE Globecom Workshops (GC Wkshps)}, Anaheim, CA, USA, Dec. 2012, pp. 1597-1602.

\bibitem{Lakew}
D. S. Lakew, U. Sa’ad, N. Dao, W. Na, and S. Cho, ``Routing in flying ad hoc networks: A comprehensive survey,'' \emph{IEEE Commun. Surv. Tuts.}, vol. 22, no. 2, pp. 1071-1120, 2nd Quart., 2020.

\bibitem{Jovel}
F. Jovel, J. McCartney, P. J. Teller, E. Ruiz and M. P. McGarry, ``Neighbor discovery message hold times for MANETs,'' \emph{Comput. Commun.}, vol. 112, pp. 38-46, Nov. 2017.

\bibitem{OLSR}
T. Clausen and P. Jacquet, \emph{Optimized link state routing protocol (OLSR)}, IETF Standard RFC 3626, Oct. 2003 [Online]. Available: https://datatracker.ietf.org/doc/rfc7181.

\bibitem{Oliveira}
R. Oliveira, M. Luis, L. Bernardo, R. Dinis and P. Pinto, ``The impact of node's mobility on link-detection based on routing hello messages,'' in \emph{IEEE Wireless Commun. Networking. Conf. (WCNC)}, Sydney, NSW, Australia, Apr. 2010, pp. 1-6.

\bibitem{Xiao}
Z. Xiao, H. Dong, L. Bai, D. Wu, and X. Xia, ``Unmanned aerial vehicle base station (UAV-BS) deployment with millimeter wave beamforming,'' \emph{IEEE Internet Things J.}, vol. 7, no. 2, pp. 1336-1349, Feb. 2020.

\bibitem{Lin}
Z. Lin, H. H. T. Liu and M. Wotton, ``Kalman filter-based large-scale wildfire monitoring with a system of UAVs,'' \emph{IEEE Trans. Ind. Electron.}, vol. 66, no. 1, pp. 606-615, Jan. 2019.

\bibitem{Chen}
Y. Chen, D. Chang and C. Zhang, ``Autonomous tracking using a swarm of UAVs: A constrained multi-agent reinforcement learning approach,'' \emph{IEEE Trans. Veh. Technol.}, vol. 69, no. 11, pp. 13702-13717, Nov. 2020.

\bibitem{Giruka}
V. Giruka and M. Singhal, ``Hello protocols for ad-hoc networks: overhead and accuracy tradeoffs,'' in \emph{Proc. 6th IEEE Int. Symp. World Wireless Mobile Multimedia Netw. (WoWMoM)}, Taormina-Giardini Naxos, Italy, Jun. 2005, pp. 354-361.

\bibitem{Park}
N. Park, J. Nam and Y. Cho, ``Impact of node speed and transmission range on the hello interval of MANET routing protocols,'' in \emph{Int. Conf. Inf. Commun. Technol. Converg. (ICTC)}, Jeju, Korea (South), Oct. 2016, pp. 634-636.

\bibitem{Mahmud}
I. Mahmud and Y. Cho, ``Adaptive hello interval in FANET routing protocols for green UAVs,'' \emph{IEEE Access}, vol. 7, pp. 63004-63015, 2019.

\bibitem{Shah}
S. K. Shah and D. D. Vishwakarma, ``Development and simulation of artificial neural network based decision on parametric values for performance optimization of reactive routing protocol for MANET using Qualnet,'' in \emph{Proc. Int. Conf. Comput. Intell. Commun. Netw. (CICN)}, Bhopal, India, Nov. 2010, pp. 167-171.

\bibitem{Bisen}
D. Bisen and S. Sharma, ``An energy-efficient routing approach for performance enhancement of MANET through adaptive neuro-fuzzy inference system,'' \emph{Int. J. Fuzzy Syst.}, vol. 20, no. 8, pp. 2693-2708, Aug. 2018.

\bibitem{Daas}
M. S. Daas, Z. Benahmed and S. Chikhi, ``An optimized energy-efficient mission-based routing protocol for unmanned aerial vehicles,'' in \emph{Int. Symp. Model. Implementation Complex Syst.}, Batna, Algeria, Sept. 2020, pp. 62-76.

\bibitem{Liu}
C. Liu, G. Zhang, W. Guo, and R. He, ``Kalman prediction-based neighbor discovery and its effect on routing protocol in vehicular ad hoc networks,'' \emph{IEEE Trans. Intell. Transp. Syst.}, vol. 21, no. 1, pp. 159-169, Jun. 2020.

\bibitem{Ernst}
R. Ernst and P. Martini, ``Adaptive HELLO for the neighborhood discovery protocol,'' in \emph{37th Annu. IEEE Conf. Local Comput. Netw. (LCN)}, Clearwater Beach, FL, USA, Oct. 2012, pp. 470-478.

\bibitem{Hernandez-Cons}
N. Hernandez-Cons, S. Kasahara, and Y. Takahashi, ``Dynamic hello/timeout timer adjustment in routing protocols for reducing overhead in MANETs,'' \emph{Comput. Commun.}, vol. 33, no. 15, pp. 1864-1878,
Sep. 2010.

\bibitem{Han}
S. Y. Han and D. Lee, ``An adaptive hello messaging scheme for neighbor discovery in on-demand MANET routing protocols'', \emph{IEEE Commun. Lett.}, vol. 17, no. 5, pp. 1040-1043, May 2013.

\bibitem{Tan}
X. Tan, Z. Zuo, S. Su, X. Guo, X. Sun and D. Jiang, ``Performance Analysis of Routing Protocols for UAV Communication Networks'', \emph{IEEE Access}, vol. 8, pp. 92212-92224, 2020.

\bibitem{P-OLSR}
S. Rosati, K. Krużelecki, G. Heitz, D. Floreano and B. Rimoldi, ``Dynamic Routing for Flying Ad Hoc Networks,''  \emph{IEEE Trans. Veh. Technol.}, vol. 65, no. 3, pp. 1690-1700, Mar. 2016.

\bibitem{OLSR_PMD}
M. Song, J. Liu and S. Yang, ``A Mobility Prediction and Delay Prediction Routing Protocol for UAV Networks,'' in \emph{Proc. IEEE 10th Int. Conf. Wireless Commun. Signal Process. (WCSP)}, Hangzhou, China, Oct. 2018, pp. 1-6.

\bibitem{ECaD}
O. Oubbati, M. Mozaffari, N. Chaib, P. Lorenz, M. Atiquzzaman, and A. Jamalipour, “ECaD: Energy-efficient routing in flying ad hoc networks,” Int. J. Commun. Syst., vol. 32, no. 18, 2019, Art. no. e4156.

\bibitem{Hong}
L. Hong, H. Guo, J. Liu and Y. Zhang, ``Toward swarm coordination: topology-aware inter-UAV routing optimization,'' \emph{IEEE Trans. Veh. Technol.}, vol. 69, no. 9, pp. 10177-10187, Sept. 2020.

\bibitem{Jian}
X. Jian, P. Leng, Y. Wang, M. Alrashoud and M. S. Hossain, ``Blockchain-empowered trusted networking for unmanned aerial vehicles in the B5G era,'' \emph{IEEE Network}, vol. 35, no. 1, pp. 72-77, Feb. 2021.

\bibitem{MPVR}
M. Jiang, Q. Zhang, Z. Feng, Z. Han and W. Li, ``Mobility Prediction Based Virtual Routing for Ad Hoc UAV Network,'' in \emph{Proc. IEEE GLOBECOM}, Waikoloa, HI, USA, Dec. 2019, pp. 1-6.

\bibitem{QGEO}
W. Jung, J. Yim and Y. Ko, ``QGeo: Q-Learning-Based Geographic Ad Hoc Routing Protocol for Unmanned Robotic Networks,'' \emph{IEEE Commun. Lett.}, vol. 21, no. 10, pp. 2258-2261, Oct. 2017.

\bibitem{QMR}
J. Liu, Q. Wang, C. He, \textit{et al}, ``QMR:Q-learning based Multi-objective optimization Routing protocol for Flying Ad Hoc Networks,'' \emph{Comput. Commun.}, vol. 150, PP. 304-316, Jan. 2019.

\bibitem{QTAR}
M. Y. Arafat and S. Moh, ``A Q-Learning-Based Topology-Aware Routing Protocol for Flying Ad Hoc Networks,'' \emph{IEEE Internet Things J.}, Early Access. doi: 10.1109/JIOT.2021.3089759.

\bibitem{Z. Xiao}
Z. Xiao, \textit{et al.}, ``A Survey on Millimeter-Wave Beamforming Enabled UAV Communications and Networking,'' \emph{IEEE Commun. Surv. Tuts.}, Early Access. doi: 10.1109/COMST.2021.3124512.

\bibitem{Srinivasa}
S. Srinivasa and M. Haenggi, ``Distance Distributions in Finite Uniformly Random Networks: Theory and Applications,''  \emph{IEEE Trans. Veh. Technol.}, vol. 59, no. 2, pp. 940-949, Feb. 2010.

\bibitem{Energy_Consumption}
J. Baek, S. I. Han and Y. Han, ``Energy-Efficient UAV Routing for Wireless Sensor Networks,''  \emph{IEEE Trans. Veh. Technol.}, vol. 69, no. 2, pp. 1741-1750, Feb. 2020.

\bibitem{DQN}
D. Liu, J. Cui, J. Zhang, C. Yang and L. Hanzo, ``Deep Reinforcement Learning Aided Packet-Routing for Aeronautical Ad-Hoc Networks Formed by Passenger Planes,''  \emph{IEEE Trans. Veh. Technol.}, vol. 70, no. 5, pp. 5166-5171, May 2021.

\bibitem{DDPG}
A. Nahar and D. Das, ``SeScR: SDN-Enabled Spectral Clustering-Based Optimized Routing Using Deep Learning in VANET Environment,'' in \emph{Proc. IEEE Int. Symp. Netw. Comput. Appl. (NCA)}, Cambridge, MA, USA, Jan. 2020, pp. 1-9.

\bibitem{Zheng}
B. Zheng, G. Huang and H. Yang, ``Link dynamics in three-dimensional mobile ad hoc networks,'' \emph{J. Electron. Inf. Technol.}, vol. 33, no. 11, pp. 2605-2609, Nov. 2011.

\bibitem{Yang}
Z. Yang, Y. Chu and W. Zhang, ``Accurate approximations for the complete elliptic integral of the second kind,'' \emph{J. Math. Anal. Appl.}, vol. 438, no. 2, pp. 875-888, Jun. 2016.

\bibitem{Fukushima}
T. Fukushima, ``Precise and Fast Computation of Elliptic Integrals and Functions,'' in \emph{IEEE 22nd Symp. Comput. Arithmetic}, Lyon, France, Jun. 2015, pp. 50-57.

\bibitem{Su}
Z. Su, ``M/G/1 queuing,'' in \emph{Fundamentals of Queueing Theory.} Chengdu, China: UEST Press, 1998, ch.4, sec.12, pp. 289-290.

\bibitem{Baltaci}
A. Baltaci, \textit{et al.}, ``Experimental UAV Data Traffic Modeling and Network Performance Analysis,'' in \emph{Proc. IEEE Conf. Comput. Commun. (INFOCOM)}, Vancouver, BC, Canada, May 2021, pp. 1-10.

\end{thebibliography}
\end{document}